\documentclass[twocolumn, switch]{article} 

\usepackage{my_preprint}

\usepackage[numbers,square]{natbib}
\usepackage[utf8]{inputenc}	
\usepackage[T1]{fontenc}	
\usepackage{xcolor}		
\usepackage[colorlinks = true,
            linkcolor = purple,
            urlcolor  = blue,
            citecolor = cyan,
            anchorcolor = black]{hyperref}	
\usepackage{booktabs} 		
\usepackage{nicefrac}		
\usepackage{microtype}		
\usepackage{lineno}		
\usepackage{float}	

\usepackage{amsfonts,amsmath,amssymb,mathtools,amsthm,mathrsfs}
\usepackage[caption = false]{subfig}
\usepackage[algo2e,ruled,vlined]{algorithm2e}
\usepackage{pgfplots,caption,doi}

\newtheorem{definition}{Definition}
\newtheorem{corollary}{Corollary}
\newtheorem{proposition}{Proposition}
\newenvironment{proof-sketch}{\noindent{\bf Sketch of Proof. \\}\hspace*{1em}}{\qed\bigskip}

\usepackage{newfloat}
\DeclareFloatingEnvironment[name={Supplementary Figure}]{suppfigure}
\usepackage{sidecap}
\sidecaptionvpos{figure}{c}

\usepackage{titlesec}
\titlespacing\section{0pt}{12pt plus 3pt minus 3pt}{1pt plus 1pt minus 1pt}
\titlespacing\subsection{0pt}{10pt plus 3pt minus 3pt}{1pt plus 1pt minus 1pt}
\titlespacing\subsubsection{0pt}{8pt plus 3pt minus 3pt}{1pt plus 1pt minus 1pt}

\usepackage{tikz,xcolor,hyperref}

\definecolor{lime}{HTML}{A6CE39}
\DeclareRobustCommand{\orcidicon}{
	\begin{tikzpicture}
	\draw[lime, fill=lime] (0,0) 
	circle [radius=0.16] 
	node[white] {{\fontfamily{qag}\selectfont \tiny ID}};
	\draw[white, fill=white] (-0.0625,0.095) 
	circle [radius=0.007];
	\end{tikzpicture}
	\hspace{-2mm}
}
\foreach \x in {A, ..., Z}{\expandafter\xdef\csname orcid\x\endcsname{\noexpand\href{https://orcid.org/\csname orcidauthor\x\endcsname}
			{\noexpand\orcidicon}}
}

\title{Introducing an experimental distortion-tolerant speech encryption scheme for secure voice communication}

\usepackage{xwatermark}
\usepackage{authblk}

\author{%
  \orcidA{} Piotr Krasnowski,  \orcidB{} Jerome Lebrun, \orcidC{} Bruno Martin\\
  \normalfont{C\^ote d'Azur, 98, Boulevard Edouard Herriot, 06200 Nice-Cedex, France} \\
  \normalfont{I3S-CNRS, 2000, Route des Lucioles, BP 121, 06903 Sophia Antipolis-Cedex, France} \\
  \texttt{\{krasnowski,lebrun,bruno.martin\}@i3s.unice.fr}
}

\begin{document}

\twocolumn[ 
  \begin{@twocolumnfalse} 
  
\maketitle

\begin{abstract}
The current increasing need for privacy-preserving voice communications is leading to new ideas for securing voice transmission. This paper refers to a relatively new concept of sending encrypted speech as pseudo-speech in the audio domain over digital voice communication infrastructures, like 3G cellular network and VoIP. This setting is more versatile compared with secure VoIP applications because vocal communication does not depend on a specific transmission medium. However, it is very challenging to maintain simultaneously high security and robustness of communication over voice channels.  

This work presents a novel distortion-tolerant speech encryption scheme for secure voice communications over voice channels that combines the robustness of analog speech scrambling and elevated security offered by digital ciphers like AES-CTR. The system scrambles vocal parameters of a speech signal (loudness, pitch, timbre) using distance-preserving pseudo-random translations and rotations on a hypersphere of parameters. Next, scrambled parameters are encoded to a pseudo-speech signal adapted to transmission over digital voice channels equipped with voice activity detection. Upon reception of this pseudo-speech signal, the legitimate receiver restores distorted copies of the initial vocal parameters. Despite some deciphering errors, an integrated neural-based vocoder based on the LPCNet architecture reconstructs an intelligible speech.

The experimental implementation of this speech encryption scheme has been tested by simulations and sending an encrypted signal over FaceTime between two iPhones~6 connected to the same WiFi network. Moreover, speech excerpts restored from encrypted signals were evaluated by a speech quality assessment on a group of about 40 participants. The experiments demonstrated that the proposed scheme produces intelligible speech with a gracefully progressive quality degradation depending on the channel noise. Finally, the preliminary computational analysis suggested that the presented setting may operate on high-end portable devices in nearly real-time.
\end{abstract}
\vspace{0.35cm}
\end{@twocolumnfalse} 
] 


\section{Introduction}

The mainstreaming of mobile networks opens new possibilities for personal communication. However, the rising numbers of reported privacy violations and cyber-espionage undermine confidence in the communication infrastructure. Another issue is the inadequate security of many voice communication systems, such as GSM which encrypted voice traffic using the insecure A5/1 stream cipher with a 64-bit key \cite{biham2000cryptanalysis}. Low trust results in a growing need for alternative methods of securing vocal communication. 

\begin{figure}[h!]
\centering
 \subfloat[Key exchange and authentication with short authentication codes displayed on the devices.]{
\includegraphics[width=0.40\textwidth]{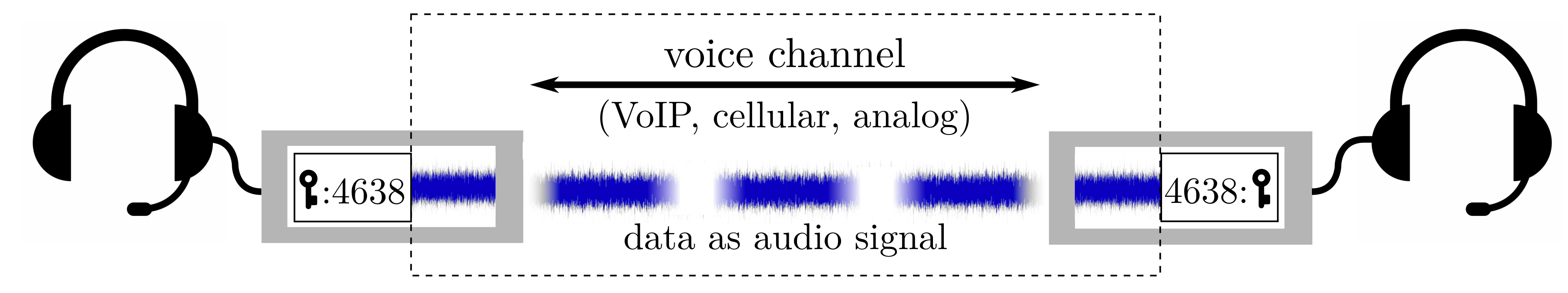}
 }
 \\
 \subfloat[Encrypted conversation.]{
\includegraphics[width=0.40\textwidth]{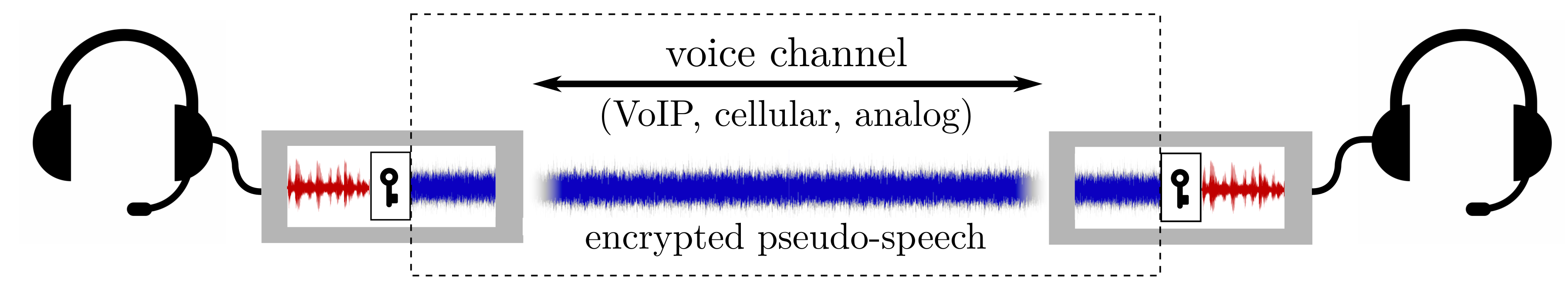}
 }
  \vspace{-4mm}
  \caption{Establishing a secure vocal link over a voice channel. \vspace{-8mm}}
  \label{setting1}
\end{figure}

This work addresses the issue of secure voice communications over untrusted voice channels. The procedure for establishing a secure vocal link is illustrated in Fig. \ref{setting1}. In the first step, two users carrying dedicated devices initiate an insecure call using a preferred communication technique, like cellular telephony, Voice over Internet Protocol (VoIP), or fixed-line telephone circuits. Then, the two devices securely acknowledge their cryptographic keys by sending binary messages over the voice channel, the same way as ordinary voice. Once the cryptographic key is computed and authenticated, the speakers can start a secure conversation. Each device encrypts speech in real-time into an incomprehensible noise and sends the encrypted signal over the channel. Upon reception, the paired device decrypts the signal and restores the initial voice.

The outlined communication scheme for secure voice communication revives the idea dating back to the late 30s \cite{kahn1996codebreakers}. First secure voice systems relied on analog signal scrambling in time, frequency and some mixed transform domains \cite{kak1980overview,kinnon80development,wyner1979analog,lee1984new,goldburg1993design}. Their role was to obscure a conversation by making the speech signal unintelligible for interceptors. The biggest advantage of analog scramblers, apart from simplicity, was their exceptional robustness against distortion introduced by telephone lines because transmission noise linearly mapped to reconstructed speech. However, due to insufficient security level \cite{goldburg1993cryptanalysis, zhao2007puzzle}, these analog and semi-analog scramblers were later superceded by fully-digital systems. Probably the first fully-digital secret telephony system was SIGSALY\footnote{\url{https://www.nsa.gov/about/cryptologic-heritage/historical-figures-publications/publications/wwii/sigsaly-story/}} (`X-System'), constructed in 1943 during WWII \cite{Sigsaly}. However, the proliferation of digital secure voice systems became possible when communications over fading telephone lines was replaced with packet networks \cite{forgie1975speech}. This change paved the way to modern secure VoIP, with Telegram and Signal being the iconic examples. \footnote{\url{https://signal.org}, \url{https://core.telegram.org}}

Nowadays, sending encrypted audio over voice channels appears to be more complicated than exchanging encrypted bits over digital packet networks. Modern voice channels aim at preserving speech intelligibility at an acceptable speech quality degradation. This goal is accomplished by applying perceptual speech processing, such as voice compression, voice activity detection, and adaptive noise suppression \cite{benesty2007springer,rabiner2011theory}. All these operations considerably modify the synthetic signal. As a result, the current golden standard for securing voice communication that consists of consecutive voice encoding and binary enciphering is considerably hindered by prohibitive error rates and extremely limited bandwidth.

On the other hand, secure communication over voice channels is supposed to be more versatile because the encrypted audio signal can be made compatible with arbitrary communication infrastructure. Furthermore, the encrypted audio signal is more likely to pass through firewalls without being blocked \cite{lee2017vulnerability}. Finally, the system can protect against spying malware installed on the portable device if speech encryption is done by an external unit \cite{krasnowski2020introducing}. The mentioned advantages suggest that the proposed setting could be especially useful for diplomatic and military services, journalists, lawyers, and traders who require secure communications in an unreliable environment and without confidential communication infrastructure. Consequently, the system should reflect high security requirements by elevating the level of secrecy, privacy, and authentication.

This work presents an experimental joint source-cryptographic enciphering scheme for secure voice communications over voice channels, which maintains the security level of digital enciphering, and to some extent enjoys a similar distortion-tolerant property of analog speech scramblers. The lossy enciphering unit scrambles perceptual speech parameters (loudness, pitch, timbre) \cite{huang2001spoken} of a recorded speech signal using distance-preserving \cite{tex2018towards} techniques, and produces a synthetic signal adapted for transmission over a voice channel. Upon reception, a recipient who owns a valid cryptographic key restores distorted copies of speech parameters and decodes speech with a trained neural vocoder.

The system architecture and its operation is thoroughly detailed, emphasizing security aspects, computational complexity, and robustness to distortion. The scheme operates on speech frames and produces an enciphered signal of equal duration, what can be seen as a strong advantage for making the system operation real-time. Moreover, it is proved that encrypted speech is computationally indistinguishable from random when enciphering is done using a secure pseudo-random number generator (PRNG) with a secret seed of a sufficient length. 

Simulations and real-world experiments follow the system description. Simulations confirmed the scheme's capability to decode mildly distorted signals. Furthermore, the encrypted speech signal was transmitted over FaceTime between two mobile phones and successfully decrypted. A speech quality assessment with about 40 listeners showed that the proposed encryption scheme produces intelligible speech and is robust against Gaussian noise at SNR = 15~dB and voice compression at bitrate 48 kbps with the Opus-Silk speech coder. Finally, the preliminary computational analysis suggests that an optimized system implementation may run on high-end portable devices. The experimental code used in simulations and speech samples evaluated in the speech quality assessment are available online.

This work is organized as follows. Section \ref{section-sphericalcodes} details the notion of distortion-tolerant encryption and introduces a scrambling technique of unit vectors on hyperspheres that is robust against channel noise. The technique is one of the main building blocks of a distortion-tolerant speech encryption scheme described in Section \ref{speech_encryption}. The scheme's properties are discussed in Section \ref{discussion}. Section \ref{evaluation} presents experimentation results and finally, Section \ref{conclusions} concludes the work. 

\section{Distortion-tolerant scrambling of unit vectors on N-spheres}
\label{section-sphericalcodes}

The perceptually-oriented processing of encrypted pseudo-speech in voice channels leads to inevitable transmission errors. This pessimistic property undermines the usefulness of many prominent cryptographic algorithms in the studied speech encryption system. The reason is their non-compliance to error, which is the property that prevents adversarial data manipulation and guarantees exact message decryption \cite{Bellare1998crypto,katz2014introduction}. In contrast, successful operation of speech coding and compression techniques proves that a good enough approximation of vocal parameters is sufficient to reconstruct intelligible speech \cite{benesty2007springer}. Consequently, some imperfect decryption in secure voice communication is acceptable if the lower decryption accuracy is somehow compensated by a higher robustness against noise.

Designing a secure cryptographic scheme that operates correctly despite encrypted data distortion could be achieved using unconventional enciphering techniques. This work is inspired by the notion of distance-preserving \cite{tex2018towards} encryption stated by Definition \ref{distance-preserving}. Intuitively speaking, an encryption scheme is distance-preserving when the distance between any two pieces of encrypted data is the same after decryption. This property seems to be useful for protecting audio media streams in real-time applications. For example, when some small channel noise degrades the enciphered vocal parameters, the original signal could still be approximately decrypted without disrupting the communication.
\begin{definition}{\cite{tex2018towards}} Let $\mathcal{M}$ be a data set, $\mathcal{K}$ be a key space, $\mathrm{d}$ be a distance measure and $\mathrm{Enc}$ be an encryption algorithm for data items in $\mathcal{M}$. Then, $\mathrm{Enc}$ is $\mathrm{d}$-distance preserving if:
\begin{align*}\forall m_0, m_1 &\in \mathcal{M} \ \mathrm{and} \ \forall k \in \mathcal{K} \ : \\ 
&\mathrm{d}(\mathrm{Enc}_k(m_0),\mathrm{Enc}_k(m_1)) = \mathrm{d}(m_1,m_2).
\end{align*}
\label{distance-preserving}
\end{definition}

Applying the distance-preserving encryption technique directly on speech parameters is far from being straightforward. Naively, we could scramble independently three perceptual speech components: pitch, loudness, and speech timbre \cite{huang2001spoken}. Pitch and loudness are associated with the fundamental frequency and signal energy \cite{rabiner2011theory}, both scalars. However, timbre is a multi-dimensional signal loosely related to the spectral envelope and with many possible representations. For instance, in the speech recognition domain, a spectral envelope is usually encoded by 13-19 Mel-Frequency Cepstral Coefficients (MFCC) \cite{davis1980comparison,benesty2007springer,rabiner2011theory}. 

On the other hand, exact distance preservation in perceptually-oriented speech data seems to be an overly strict property. It is because some small inaccuracies in speech representations are usually acceptable or perceptually irrelevant. It is especially true in real-time applications that prioritize robustness and efficiency over the quality of representation. Thus, we propose a considerable relaxation of distance-preserving encryption that, in our opinion, is better suited for protecting real-time voice communication. The new notion, coined distortion-tolerant encryption, is stated by Definition~\ref{distortion_tolerance}.

\begin{definition}
Let $\mathrm{KeyGen}$ be a key generation algorithm that produces a secret key over the keyspace $\mathcal{K}$, $\mathrm{Enc}$ be an encryption algorithm and $\mathrm{Dec}$ be a decryption algorithm. Moreover, let $\mathrm{d}_{\mathcal{M}}$ and $\mathrm{d}_{\mathcal{C}}$ denote respectively distance measures over the plaintext space $\mathcal{M}$ and the ciphertext space $\mathcal{C}$.  We say that the encryption scheme $\Pi = (\mathrm{KeyGen}, \mathrm{Enc},$ $\mathrm{Dec})$ is distortion-tolerant with respect to $\mathrm{d}_{\mathcal{M}}$ and $\mathrm{d}_{\mathcal{C}}$ if for every key $k \in \mathcal{K}$ produced by $\mathrm{KeyGen}$, any two ciphertexts $\mathit{c}_1,\mathit{c}_2 \in \mathcal{C}$, and $\delta > 0 $ not too large, there is $\tau > 0 $ such that:
\begin{enumerate}
\item $\mathrm{d}_{\mathcal{C}}(\mathit{c}_1, \mathit{c}_2) < \delta$ \hspace{-2mm}  $\implies$ \hspace{-2mm}  $\mathrm{d}_{\mathcal{M}}( \mathrm{Dec}_k(\mathit{c}_1), \mathrm{Dec}_k(\mathit{c}_2)) < \tau\delta$.
\item $\tau\delta \ll \max\limits_{c_i,c_j \in \mathcal{C}} \{\mathrm{d}_{\mathcal{M}}(\mathrm{Dec}_k(\mathit{c}_i), \mathrm{Dec}_k(\mathit{c}_j)) \}$.
\end{enumerate}
\label{distortion_tolerance}
\end{definition}

Unlike in distance-preserving encryption, we use two different metrics over the message and the ciphertext spaces. Furthermore, we allow some distance expansion $\tau$ between the decrypted plaintexts $\mathrm{Dec}_k(\mathit{c}_1)$ and $\mathrm{Dec}_k(\mathit{c}_2)$, which is still small compared to the maximum distance between plaintexts. Finally, the distortion-tolerant property applies locally in a ciphertext neighborhood.

Every encryption scheme with a distortion-tolerant property is malleable by design.\footnote{An encryption algorithm is `malleable' if it is possible to transform a ciphertext into another ciphertext which decrypts to a bona fide plaintext.} Without additional data-integrity mechanisms, an active attacker can modify the decrypted plaintext by carefully changing the encrypted data. Moreover, given a pair $(m,\mathrm{Enc}_k(m))$, the attacker may easily guess or approximate the decryption result of all ciphertexts close to $\mathrm{Enc}_k(m)$. On the other hand, the malleability does not necessarily compromise the secrecy of encryptions if the enciphering algorithm uses a fresh cryptographic key $k \in \mathcal{K}$ for every encryption.

In this section, we describe a distortion-tolerant technique for scrambling unit vectors on hyperspheres in even dimensions. The technique could be used for enciphering the spectral envelopes of speech signals represented as spherical vectors on the hypersphere $S^n$ (how it is done will be detailed in Section~\ref{speech_encryption}). The scrambling procedure relies on a spherical commutative (abelian) group code \cite{slepian1968group} and uses a secure PRNG with a secret seed (a cryptographic key). The result of scrambling is a new spherical vector on $S^{2n-1}$ that may represent another spectral envelope. Due to the distortion-tolerant property of the enciphering technique, this vector can be efficiently descrambled despite some small noise.

\subsection{Hypersphere mappings}

\begin{figure}[h!]
\centering
\includegraphics[width=0.45\textwidth]{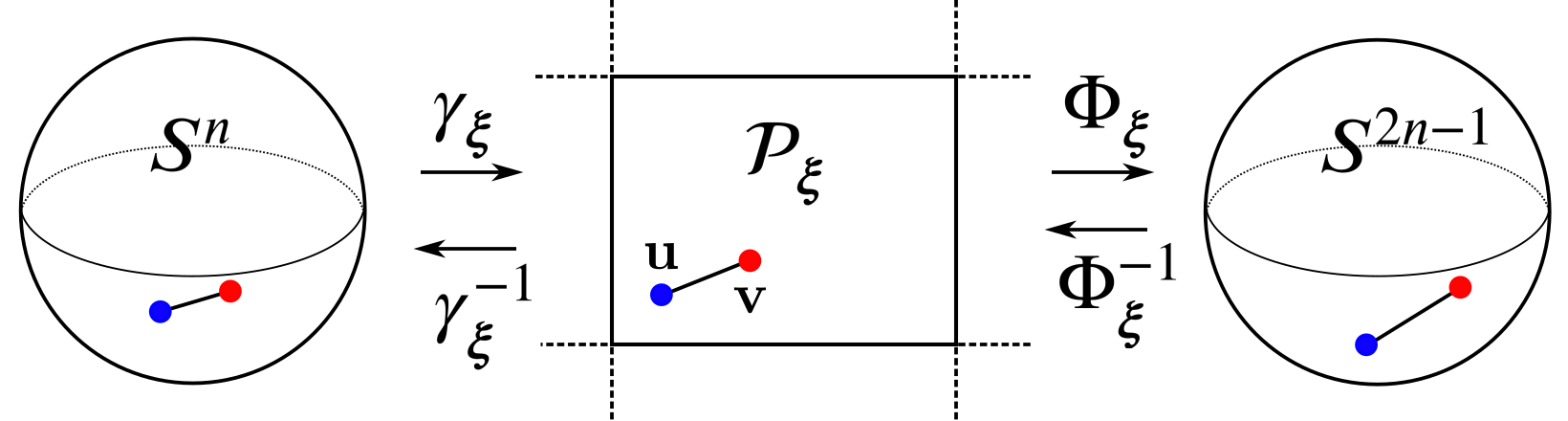}
\caption{Mapping unit vectors on $S^{n}$ to vectors on $S^{2n-1}$ using two mappings $\gamma_{\boldsymbol \xi}$ and $\Phi_{\boldsymbol \xi}$.}
\label{mappings}
\end{figure}

For $n>1$, let $ \boldsymbol\xi = [\xi_1,...,\xi_n]^T$ be a unit vector such that $\xi_i \geq 0$. We describe two mappings, ${\gamma_{\boldsymbol \xi}: S^{n} \rightarrow \mathbb{R}^n}$ and ${\Phi_{\boldsymbol \xi}:  \mathbb{R}^n \rightarrow S^{2n-1}}$, which preserve the Euclidean distance approximately. The composition $\Phi_{\boldsymbol \xi} \circ \gamma_{\boldsymbol \xi}$ transforms vectors on $S^n$ to vectors on $S^{2n-1}$, as shown in Fig. \ref{mappings}. In our scheme, enciphering is performed over the intermediate domain $\mathcal{P}_{\boldsymbol \xi}$ in $\mathbb{R}^n$.

Let $\mathbf{x} \in S^n$. Then, $\gamma_{\boldsymbol \xi}(\mathbf{x})$ is defined as $[\xi_1\varphi_1,...,\xi_n\varphi_n]^T$, where $\varphi_1,...,\varphi_{n-1} \in [0,\pi)$ and $\varphi_n \in [0,2\pi)$ are the spherical coordinates of $\mathbf{x}$:
\begin{equation}
\begin{aligned}
x_1 &= \cos(\varphi_1),  \\
x_2 &= \sin(\varphi_1)\cos(\varphi_2),  \\
&\vdots  \\
x_{n} &= \sin(\varphi_1) \cdot ... \cdot \sin(\varphi_{n-1})\cos(\varphi_n),  \\
x_{n+1} &= \sin(\varphi_1) \cdot ... \cdot \sin(\varphi_{n-1})\sin(\varphi_n), &&
\end{aligned}
\end{equation}

\noindent For a vector $\mathbf{u} \in \mathbb{R}^n$, let $\Phi_{\boldsymbol \xi}(\mathbf{u})$ be a torus mapping defined as in \cite{siqueira2008flat}: 
\begin{equation}
\begin{aligned}
\Phi_{\boldsymbol\xi}(\mathbf{u}) = [ &\xi_1\cos(u_1/\xi_1), \ \xi_1\sin(u_1/\xi_1), ..., \\
 &\xi_n\cos(u_n/\xi_n), \ \xi_n\sin(u_n/\xi_n) ]^T.
\end{aligned}
\end{equation}

The image of the mapping $\Phi_{\boldsymbol\xi}$ is a composition of orbits on $S^{2n-1}$, which can be viewed as a flat torus \cite{lavor2018advances}. The whole family of flat tori with $\|{\boldsymbol\xi}\| = 1$ and $\xi_i \geq 0$ foliates $S^{2n-1}$, meaning that every point on the hypersphere belongs to one and only one flat torus \cite{lawson1974foliations,candel2000foliations}. The pre-image of the mapping $\Phi_{\boldsymbol\xi}$ is $\mathcal{P}_{\boldsymbol \xi} = \Pi_{i=1}^{n}[0,2\pi \xi_i)$.

The mappings preserve the Euclidean distance between vectors approximately. Let $\mathbf{u},\mathbf{v} \in \Pi_{i=1}^{n-1}[0,\pi \xi_i) \times [0,2\pi \xi_n)$. The distance $\| \gamma_{\boldsymbol\xi}^{-1}(\mathbf{u})-\gamma_{\boldsymbol\xi}^{-1}(\mathbf{v}) \|$ is bounded by:
\begin{align}
0 &\leq \| \gamma_{\boldsymbol\xi}^{-1}(\mathbf{u})-\gamma_{\boldsymbol\xi}^{-1}(\mathbf{v}) \| \leq \frac{\|\mathbf{u}-\mathbf{v}\|}{\xi_{min}}, 
\label{distances1}
\end{align}
where $\xi_{min} = \min\limits_{1 \leq i \leq n}\xi_i \neq 0$. On the other hand, by assuming $\|\mathbf{u}-\mathbf{v}\| \leq \xi_{min}$, we have \cite{torezzan2015optimum}:
\begin{equation}
\frac{2}{\pi}\|\mathbf{u}-\mathbf{v}\| \leq \| \Phi_{\boldsymbol\xi}(\mathbf{u}) - \Phi_{\boldsymbol\xi}(\mathbf{v})  \| \leq \|\mathbf{u}-\mathbf{v}\|.
\label{distances2}
\end{equation}
\noindent By combining Eq. (\ref{distances1}) and Eq. (\ref{distances2}) we finally obtain:
\begin{equation} 
0 \leq \| \gamma_{\boldsymbol\xi}^{-1}(\mathbf{u}) \hspace{-1mm}-\hspace{-1mm}\gamma_{\boldsymbol\xi}^{-1}(\mathbf{v}) \| \leq \frac{\pi}{2\xi_{min}}\| \Phi_{\boldsymbol\xi}(\mathbf{u})\hspace{-1mm}-\hspace{-1mm} \Phi_{\boldsymbol\xi}(\mathbf{v})\|. 
\label{distances3}
\end{equation}
\noindent From Eq. (\ref{distances3}), we may conclude that the composed mapping $\gamma^{-1}_{\boldsymbol \xi} \circ  \ \Phi^{-1}_{\boldsymbol \xi}$ is robust against some small channel noise, since the error introduced to $\Phi_{\boldsymbol\xi}(\mathbf{u})$ will be mapped to $\gamma_{\boldsymbol\xi}^{-1}(\mathbf{u})$ with the expansion factor up to $\pi/2\xi_{min}$. 

\subsection{Spherical group codes from lattices}

The spherical mappings $\gamma_{\boldsymbol \xi}$ and $\Phi_{\boldsymbol \xi}$ provide a framework for a distortion-tolerant scrambling of spectral envelopes. Another challenge is to find a suitable model for enciphering vectors over $\mathcal{P}_{\boldsymbol \xi}$. In this work, we propose a discrete model based on a pair of nested lattices in $\mathbb{R}^n$ and a spherical commutative group code on $S^{2n-1}$. Spherical commutative group codes were introduced in \cite{slepian1968group} as a new encoding method for sending non-binary information over non-binary Gaussian channel. However, the useful literature on using these codes for enciphering remains scarce.

\begin{definition}{\cite{slepian1968group}}
A spherical commutative group code $\mathscr{C}$ of order $M$ is a set of $M$ unit vectors $\mathscr{C} = \{G\boldsymbol\sigma : \ G \in \mathcal{G}\}$, where $\boldsymbol\sigma$ lies on the unit hypersphere $S^{n-1} \subset \mathbb{R}^n$ and $\mathcal{G}$ is a finite group of order $M$ of $n \times n$ orthogonal matrices. 
\end{definition}

Commutative spherical group codes are geometrically uniform, i.e., for any two elements $\mathbf{p},\mathbf{q} \in \mathscr{C}$, there exists an isometry $\mathrm{f}_{\mathbf{p},\mathbf{q}}$ such that $\mathrm{f}_{\mathbf{p},\mathbf{q}}(\mathbf{p})=\mathbf{q}$ and $\mathrm{f}_{\mathbf{p},\mathbf{q}}(\mathscr{C})=\mathscr{C}$~\cite{forney91geometrically}. Moreover, they have congruent Voronoi regions \cite{tessellations}, the same detection probability in the presence of transmission noise, and a distribution of codewords invariant to multiplication by matrices from $\mathcal{G}$. Although spherical commutative group codes do not offer packing densities as high as general spherical codes, this shortcoming is compensated by their simple structure and the easiness of encoding and decoding~\cite{costa2017lattices}. 

Every element in $\mathcal{G}$ can be uniquely represented as a product of powers of generator matrices $\{ G_1, ..., G_k \}$, such that $G_i \in \mathcal{G}$ for $i=1,...,k$, and $G_i$ generate $\mathcal{G}$:
\begin{equation}
\mathcal{G} = \{ G_1^{w_1} \cdot G_2^{w_2} \cdot ... \cdot G_k^{w_k} : 0 \leq w_i < d_i, \ i=1,...,k \}.
\end{equation} 
Furthermore, $\mathcal{G}$ is isomorphic to $\mathbb{Z}_{d_1} \oplus ... \oplus \mathbb{Z}_{d_k}$ where $d_1 \cdot d_2 \cdot...\cdot d_k=M$ and $d_i | d_{i+1}$ for $i=1,...,k-1$ \cite{cohen1993course}. We can thus conveniently index $G \in \mathcal{G}$ (and $G\boldsymbol\sigma \in \mathscr{C}$) by a vector $[w_1, ..., w_k]^T \in \mathbb{Z}_{d_1} \oplus ... \oplus \mathbb{Z}_{d_k}$. 

\begin{figure}[h!]
\centering
\includegraphics[width=0.45\textwidth]{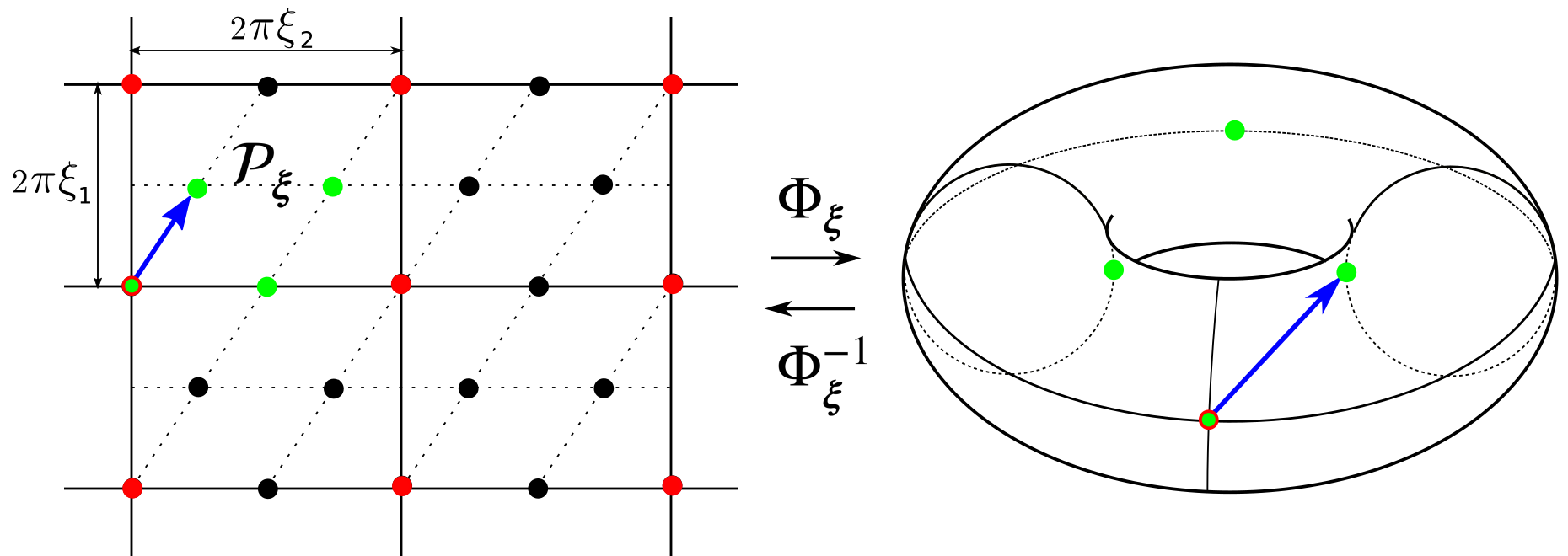}
\caption{Example of a construction of a spherical code on $S^3$ from nested lattices in $\mathbb{R}^2$. The flat torus on the right side lies on the hypersphere in $\mathbb{R}^4$.}
\label{spherical_code}
\end{figure}

Costa et al. \cite{costa2017lattices} proposed a very efficient method for constructing some spherical group codes in even dimensions using lattices. Figure \ref{spherical_code} presents a simple example of two nested lattices in $\mathbb{R}^2$ associated with a spherical group code on $S^3$. The red dots on the left side of the picture belong to an orthogonal lattice $\Lambda_{(\beta)} = 2\pi \xi_1 \mathbb{Z} \times 2 \pi \xi_2 \mathbb{Z}$, where $(\xi_1,\xi_2) = (0.8, 0.6)$. Since $\boldsymbol \xi = [\xi_1, \xi_2]^T$ is nonnegative with a unit norm, the points of $\Lambda_{(\beta)}$ can be viewed as vertices of frames that are the pre-images of~$\Phi_{\boldsymbol \xi}$.

Besides, the red and black dots combined form another lattice $\Lambda_{(\alpha)}$ such that $\Lambda_{(\beta)} \subset \Lambda_{(\alpha)}$. It can be noticed that the quotient $\Lambda_{(\alpha)}/\Lambda_{(\beta)}$ of order 4 can be mapped by $\Phi_{\boldsymbol \xi}$ to some spherical code $\mathscr{C}$ of order 4 on $S^3$. Moreover, since $\Lambda_{(\alpha)}/\Lambda_{(\beta)}$ is closed under translation by a single basis vector (the blue arrow), the code should be closed under an associated rotation on $S^3$. Consequently, the code $\mathscr{C}$ is a commutative group code of order 4 with a single generator matrix. A generalized approach for constructing spherical codes from lattices is given by Corollary~\ref{corollary2}.

\begin{corollary}{\cite{costa2017lattices}}
Let $\Lambda_{(\beta)} \subset \Lambda_{(\alpha)}$ be a pair of full rank lattices with respective generator matrices $A_{\alpha}$ and $A_{\beta}=[\boldsymbol\beta_1,..., \boldsymbol\beta_n]$, where $\{ \boldsymbol\beta_1,..., \boldsymbol\beta_n \}$ is an orthogonal basis of $\Lambda_{(\beta)}$. There exists an integer matrix $H$ such that $A_{\beta}$ = $A_{\alpha}H$. Matrix $H$ has a Smith normal form $H = PDQ$ where $P$ and $Q$ are integer matrices with determinant -1 or 1, and $D$ is a diagonal matrix with $\mathrm{diag}(D) = [d_1, ..., d_n]^T$, $d_{i} \in \mathbb{N}$ and $d_{i}|d_{i+1}$ for $i = 1,..., n -1$. 

Let us define $b_i = \|\boldsymbol\beta_i\|$, $b =  \sqrt{\sum_{j=1}^{n} \|\boldsymbol\beta_j\|^2}$, $\xi_i = b_i/b$, $\boldsymbol\xi = [\xi_1, ..., \xi_n]^T$ and the torus mapping $\Phi_{\boldsymbol \xi}$. Then, the quotient of the lattices $(2\pi b^{-1}\Lambda_{(\alpha)})/(2\pi b^{-1}\Lambda_{(\beta)})$ is associated with a spherical code $\mathscr{C} \subset S^{2n-1}$ with the initial vector $\boldsymbol\sigma = [\xi_1, 0, \xi_2, 0, ..., \xi_n, 0]^T$, and a generator group of matrices determined by the Smith normal decomposition of $H$.
\label{corollary2}
\end{corollary}

Corollary \ref{corollary2} states that for a given pair of nested lattices ${\Lambda_{(\beta)} \subset \Lambda_{(\alpha)}}$,  $\Lambda_{(\beta)}$ being orthogonal, and an integer matrix $H$ such that $A_{\beta} = A_{\alpha}H$, with Smith normal form $H=PDQ$, one can easily get generator matrices of $\mathcal{G}$. The task is to correctly associate basis vectors of the group $\Lambda_{(\alpha)}/\Lambda_{(\beta)}$ with appropriate rotations on the hypersphere. Firstly, from $\mathrm{diag}(D)$ we may conclude that $\Lambda_{(\alpha)}/\Lambda_{(\beta)}$ (and hence $\mathcal{G}$) is isomorphic to $\mathbb{Z}_{d_{m}} \oplus ... \oplus \mathbb{Z}_{d_n}$, where $d_{m}$ is the first element of $\mathrm{Diag}(D)$ larger than 1. In consequence, we get that $\mathcal{G}$ is of order $M=d_{m} \cdot ... \cdot  d_n$ and has $k=n-m+1$ generator matrices. We may express the $2n \times 2n$ generator matrices $\{G_1, G_2, ..., G_k\}$ in block-diagonal form:
\[
G_{i+1} = \begin{bmatrix}
\mathrm{rot}(2\pi r_{i+m,1}) & 0 & ... & 0\\
0 & \mathrm{rot}(2\pi r_{i+m,2}) & ... & 0\\
\vdots & \vdots & \ddots & \vdots\\
0 & 0 & ... & \mathrm{rot}(2\pi r_{i+m,n})
\end{bmatrix}
\]
where $i = 0,...,k-1$, $r_{i+m,j}$ are elements of the matrix $R=A_{\beta}^{-1}A_{\alpha}P$  and $\mathrm{rot}(x)$ are $2 \times 2$ rotation matrices:
\begin{equation*}
\qquad \mathrm{rot}(x) =
\begin{bmatrix}
\cos(x) & -\sin(x)\\
\sin(x) & \phantom{-}\cos(x)
\end{bmatrix}.
\end{equation*}
For a relevant discussion, the reader is referred to \cite{costa2017lattices}.

It can be noticed that performing basic rotations on the codewords of $\mathscr{C}$ is equivalent to translating points in the pre-image of the flat torus. Thus, we can impose some design rules that improve the properties of the constructed spherical code.

Firstly, from the bounds in Eq. (\ref{distances2}) we obtain that for any vectors $\mathbf{u},\mathbf{v} \in (2\pi b^{-1}\Lambda_{(\alpha)})/(2\pi b^{-1}\Lambda_{(\beta)})$ the Euclidean distance ${\|\Phi_{\boldsymbol \xi}(\mathbf{u}) - \Phi_{\boldsymbol \xi}(\mathbf{v}) \|}$ is larger than $2/\pi \|\mathbf{u} - \mathbf{v} \|$. Consequently, the maximization of the minimum distance between vectors of $(2\pi b^{-1}\Lambda_{(\alpha)})$ would improve the distance distribution between codewords on the hypersphere $S^{2n-1}$. In particular, selecting dense lattices like the cubic lattice $\mathrm{Z}^n$, the checkerboard lattice $\mathrm{D}_n$ or the Gosset lattice $\Gamma_8$ \cite{conway2013sphere} in the construction will lead to a larger minimum distance between codewords in $\mathscr{C}$. Another advantage of these mentioned lattices is the existence of very efficient algorithms \cite{conway1982fast} for solving the closest vector problem (CVP)~\cite{micciancio2012complexity}.

The distribution of codewords on $S^{2n-1}$ can be further improved by selecting a proper vector $\boldsymbol \xi = [\xi_1,...,\xi_n]^T$. From Eq. (\ref{distances3}), we should maximize the minimum component $\xi_{min}$. This is achieved by taking $\xi_1 = ... = \xi_n$.

\subsection{Scrambling}

Finally, we describe the scrambling algorithm of vectors on $S^n$. The procedure is presented in Algorithm~\ref{procedure}. The parameters of the scrambling model is a nonnegative vector $\boldsymbol \xi = [\xi_1,...,\xi_n]$ and a quotient group $\Lambda_{(\alpha)} / \Lambda_{(\beta)}$ of order $M$, where $\Lambda_{(\beta)} = \prod_{i=1}^n 2\pi\xi_i\mathbb{Z}$ and $\Lambda_{(\alpha)}$ has an efficient algorithm for solving the CVP. In addition, the scrambler uses a secure PRNG with a secret, binary random seed $s$ of length $\lambda \geq 128$. The role of the PRNG (here, the subroutine `SelectRandom') is to output pseudo-randomly vectors from $\Lambda_{(\alpha)} / \Lambda_{(\beta)}$ with a distribution computationally indistinguishable from the uniform distribution by any probabilistic polynomial-time (PPT) \cite{katz2014introduction} algorithm up to the security level given by $\lambda$.

The algorithm starts by mapping a single vector $\mathbf{x} \in S^n$ to $\gamma_{\boldsymbol \xi}(\mathbf{x})$ in $\mathcal{P}_{\boldsymbol \xi}$ and searching the closest vector $\boldsymbol\chi$ in $\Lambda_{(\alpha)} / \Lambda_{(\beta)}$ in terms of the Euclidean distance. This process may be viewed as quantization. The resulting quantization error could be mitigated by increasing the resolution of $\Lambda_{(\alpha)}$. In the next step, the PRNG selects an element $\boldsymbol \nu$ from $\Lambda_{(\alpha)} / \Lambda_{(\beta)}$. Finally, the scrambler outputs a codeword $\Phi_{\boldsymbol \xi}(\boldsymbol \chi + \boldsymbol \nu)$ that belongs to a spherical commutative group code $\mathscr{C}$ associated with $\Lambda_{(\alpha)} / \Lambda_{(\beta)}$ through the mapping~$\Phi_{\boldsymbol \xi}$.

If a sequence of plaintext vectors on $S^n$ is given, the scrambler processes them sequentially, each time using a freshly generated vector produced by `SelectRandom' with the same seed $s$.

\begin{algorithm2e}
\caption{ }
\SetAlgoLined
\KwData{initial vector $x \in S^{n}$, security parameter $\lambda \geq 128$, random seed $s \in \{0,1\}^\lambda$;}\
\KwResult{scrambled codeword $\mathbf{p} \in \mathscr{C} \subset S^{2n-1}$;}\
\tcp*[h]{find the closest vector in $\Lambda_{(\alpha)} / \Lambda_{(\beta)}$}\
$\boldsymbol\chi \longleftarrow \mathrm{FindClosestVector}(\gamma_{\boldsymbol \xi}(\mathbf{x});\Lambda_{(\alpha)} / \Lambda_{(\beta)})$\;
\tcp*[h]{select a random element from $\Lambda_{(\alpha)} / \Lambda_{(\beta)}$}\
$\boldsymbol \nu \longleftarrow \mathrm{SelectRandom}(\Lambda_{(\alpha)} / \Lambda_{(\beta)}; s)$\;
\tcp*[h]{encipher and map on the flat torus}\
$\mathbf{p} \longleftarrow \Phi_{\boldsymbol \xi}(\boldsymbol \chi + \boldsymbol \nu)$\; 
\label{procedure}
\end{algorithm2e}

The security of the scrambling algorithm depends on the PRNG and the quality of entropy used to produce the seed $s$. It can be noticed that for the scrambling vector $\boldsymbol \nu \in \Lambda_{(\alpha)} / \Lambda_{(\beta)}$ taken perfectly at random (i.e., a `one-time-pad' scenario), $(\boldsymbol \chi + \mathbf{v}) \in \Lambda_{(\alpha)} / \Lambda_{(\beta)}$ follows a uniform distribution over $\Lambda_{(\alpha)} / \Lambda_{(\beta)}$ for any distribution of $\boldsymbol \chi$. Thus, $\Phi_{\boldsymbol \xi}(\boldsymbol \chi + \boldsymbol \nu)$ can be any codeword of $\mathscr{C}$ with equal probability. Furthermore, the same result can be extended to a sequence of vectors $(\boldsymbol \nu_1, ..., \boldsymbol \nu_N)$ selected independently and uniformly at random over $\Lambda_{(\alpha)} / \Lambda_{(\beta)}$. In such a case, the scrambling procedure would give a sequence $(\Phi_{\boldsymbol \xi}(\boldsymbol \chi_1 + \boldsymbol \nu_1), ...,\Phi_{\boldsymbol \xi}(\boldsymbol \chi_N + \boldsymbol \nu_N))$ of statistically independent codewords with the uniform distribution over $\mathscr{C}$. Consequently, the adversary trying to extract any relevant information from the ciphertext could not perform better than by randomly guessing.

From the construction of the scrambling algorithm, the existence of any PPT adversary who breaks the secrecy of ciphertexts (with non-negligible probability) implies the existence of efficient attacks on the PRNG used in scrambling. In effect, the selection of a proper sequence generator is the crucial factor from the security standpoint. Some propositions of suitable PRNGs are discussed in Section \ref{discussion}.

\subsection{Transmission and descrambling}

Spherical group codes discussed here can be viewed as equal-energy block codes adapted for transmission over noisy channels \cite{slepian1968group}. Nonetheless, in contrast to classic digital communications, our scheme tolerates some detection error that is supposed to be proportionally mapped to the plaintext as some deciphering error. Optimally, a growing noise level should gradually increase detection error and hence deciphering error. However, a significant channel noise may break this desired continuity between transmission and deciphering errors, severely disrupting the communication.

Let $\mathbf{p} \in \mathscr{C}$ be an encrypted codeword sent over the Gaussian channel. Upon reception, the recipient observes $ \mathbf{\hat{q}} = \mathbf{p} + \mathbf{n}$, where $\mathbf{n}$ represents channel noise sampled from Gaussian distribution with zero mean. Provided that the enciphered vector $\mathbf{p}$ can be any codeword of $\mathscr{C}$ with equal probability, the maximum likelihood detector selects the closest codeword in $\mathscr{C}$ in terms of the Euclidean metric. Moreover, due to the uniform geometrical distribution of codewords on the hypersphere $S^{2n-1}$, the error probability of the optimal detector is the same for every sent codeword. This property turns out to be very useful in the context of this work because the deciphering error caused by Gaussian noise is statistically independent of the ciphertext.

Before decryption, the received vector $\mathbf{\hat{q}} = [\hat{q}_1,...,$ $\hat{q}_{2n}]$ should be projected to $\mathbf{q} =G\boldsymbol\sigma$ on the flat torus associated with $\Phi_{\boldsymbol \xi}$. The rotation matrix $G$ is of the form:

\[ G =
\begin{bmatrix}
\mathrm{rot}(\varphi_1) & 0 & ... & 0\\
0 & \mathrm{rot}(\varphi_2) & ... & 0\\
\vdots & \vdots & \ddots & \vdots\\
0 & 0 & ... & \mathrm{rot}(\varphi_n)
\end{bmatrix}_{2n \times 2n}
\]
where $\varphi_1,\varphi_2,...,\varphi_n$ are some unknown rotation angles. Assuming Gaussian noise, the vector $\mathbf{q} = [q_1,...,q_{2n}]^T$ can be found by projecting the coordinates of $\mathbf{\hat{q}}$ onto the respective circles of radius $\xi_{i}$:
\begin{equation}
[q_{2i-1},q_{2i}] = \xi_{i}\frac{[\hat{q}_{2i-1},\hat{q}_{2i}]}{\|[\hat{q}_{2i-1},\hat{q}_{2i}]\|}, \ i=1,...,n.
\label{projection_formula}
\end{equation}

\begin{figure}[h!]
\centering
\includegraphics[width=0.45\textwidth]{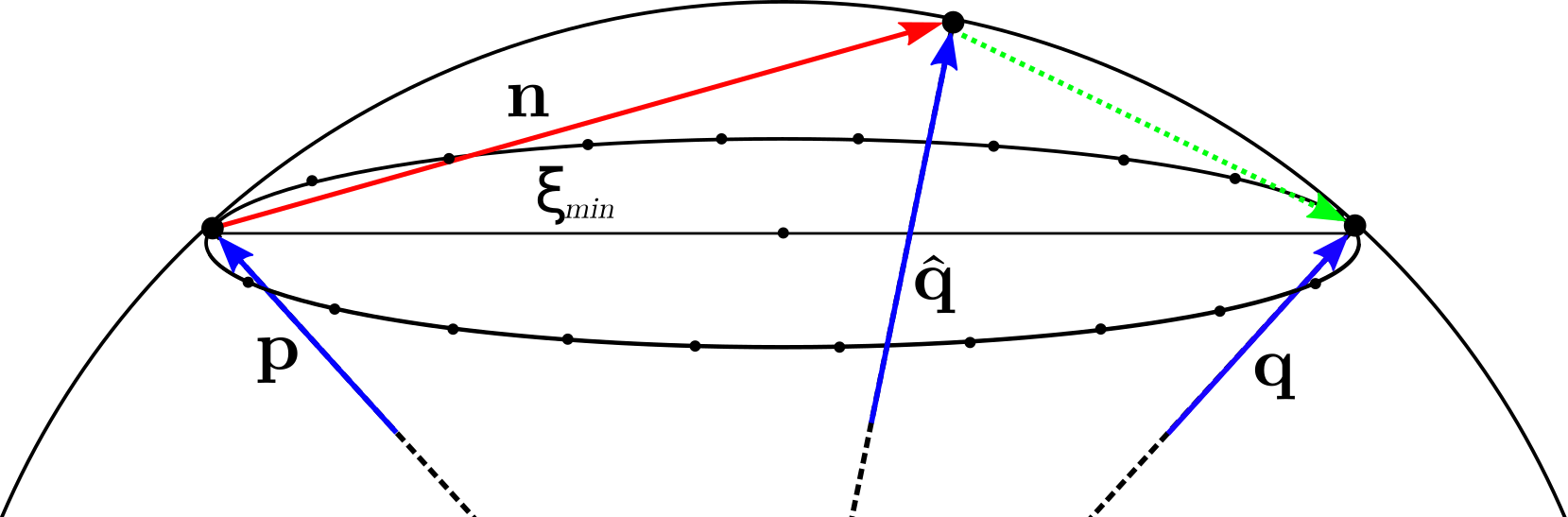}
\caption{Projection of vectors to the flat torus in presence of excessive noise. The initial vector $\mathbf{p} \in \mathscr{C}$ on the orbit of radius $\xi_{min}$ is transmitted over a transmission channel and received as $\mathbf{\hat{q}} = \mathbf{p}+\mathbf{n}$. The vector $\mathbf{\hat{q}}$ is projected to the vector $\mathbf{q}$ on the opposite side of the orbit, far from $\mathbf{p}$. A possible solution to this problem is to increase the energy of the transmitted vectors.}
\label{jump}
\end{figure}

The projection of $\mathbf{\hat{q}}$ onto $\mathbf{q}$ is an additional source of error that is difficult to tackle. When the noise level is small, however, the overall distortion caused by the operation is limited. Another situation may occur, when the distance $\|\mathbf{\hat{q}}-\mathbf{p}\|$ overreaches $\xi_{min}$. In such a case, $\mathbf{q}$ may be projected on the opposite side of the torus and cause a large error as illustrated in Fig.~\ref{jump}.

Algorithm~\ref{procedure2} details the descrambling procedure which essentially reverses scrambling operations in Algorithm~\ref{procedure}. It is assumed that both participants securely shared the random seed $s$. The subroutine `Project' transforms the received vector $\mathbf{\hat{q}}$ according to Eq. (\ref{projection_formula}). However, the projected vector  $\mathbf{q}$ is not quantized by $\Lambda_{(\alpha)} / \Lambda_{(\beta)}$. Additionally, if a descrambled vector $\boldsymbol{\hat{\chi}}$ exceeds $\Pi_{i=1}^{n-1}[0,\pi \xi_i) \times [0,2\pi \xi_n)$, respective coordinates of $\boldsymbol{\hat{\chi}}$ are reflected around $\pi\xi_i$. Finally, it is essential to maintain synchronization of the PRNGs on both communication sides. 

\begin{algorithm2e}
\caption{ }
\SetAlgoLined
\KwData{received $\mathbf{\hat{q}} \in S^{2n-1}$, seed $s \in \{0,1\}^\lambda$;}\
\KwResult{descrambled vector $\mathbf{y} \in S^{n}$;}\
\tcp*[h]{project to the flat torus}\\
$\mathbf{q} \longleftarrow \mathrm{Project}(\mathbf{\hat{q}};\boldsymbol \xi)$\; 
\tcp*[h]{select a random element from $\Lambda_{(\alpha)} / \Lambda_{(\beta)}$}\
$\boldsymbol \nu \longleftarrow \mathrm{SelectRandom}(\Lambda_{(\alpha)} / \Lambda_{(\beta)}; s)$\;
\tcp*[h]{decipher}\\
$\boldsymbol{\hat{\chi}} \longleftarrow \Phi_{\boldsymbol \xi}^{-1}(\mathbf{q})-\boldsymbol \nu \ \mod \ \Lambda_{(\alpha)} / \Lambda_{(\beta)}$\;
\tcp*[h]{correct $\boldsymbol{\hat{\chi}}$}\\
\For{$i\leftarrow 1$ \KwTo $n-1$}{
\If{$\boldsymbol{\hat{\chi}}_i \geq \pi\xi_i$}{
$\boldsymbol{\hat{\chi}}_i \leftarrow 2\pi\xi_i - \boldsymbol{\hat{\chi}}_i$\;
}
}
$\mathbf{y} \longleftarrow \gamma_{\boldsymbol \xi}^{-1}(\boldsymbol{\hat{\chi}})$; \tcp*[h]{map to $S^n$} 
\label{procedure2}
\end{algorithm2e}

From Algorithm \ref{procedure2} and Eq. (\ref{distances2}), we may conclude that $\| \boldsymbol{\chi} - \boldsymbol{\hat{\chi}}\| \leq \| \mathbf{p} - \mathbf{q}\|$. By inserting this inequality to Eq. (\ref{distances3}) we obtain:
\begin{equation}
\quad 0 \leq \| \mathbf{y} - \mathbf{x}\| \leq \frac{\pi}{2\xi_{min}}\| \mathbf{p} - \mathbf{q}  \|.
\label{distances4}
\end{equation}

In accordance with Definition \ref{distortion_tolerance}, the scrambling scheme is distortion-tolerant with the expansion factor $\tau = \pi/2\xi_{min}$ when the transmission error is no larger than $\xi_{min}$. On the other hand, the distance $\| \mathbf{y} - \mathbf{x}\|$ goes down to $0$ when approaching the poles of the hypersphere $S^{n}$. Consequently, we may notice that the vectors near the poles of $S^n$ should be relatively less distorted by channel noise than the vectors close to the equator.

\section{Speech encryption scheme}
\label{speech_encryption}

\begin{figure*}[h]
\centering
    \includegraphics[width=0.8\textwidth]{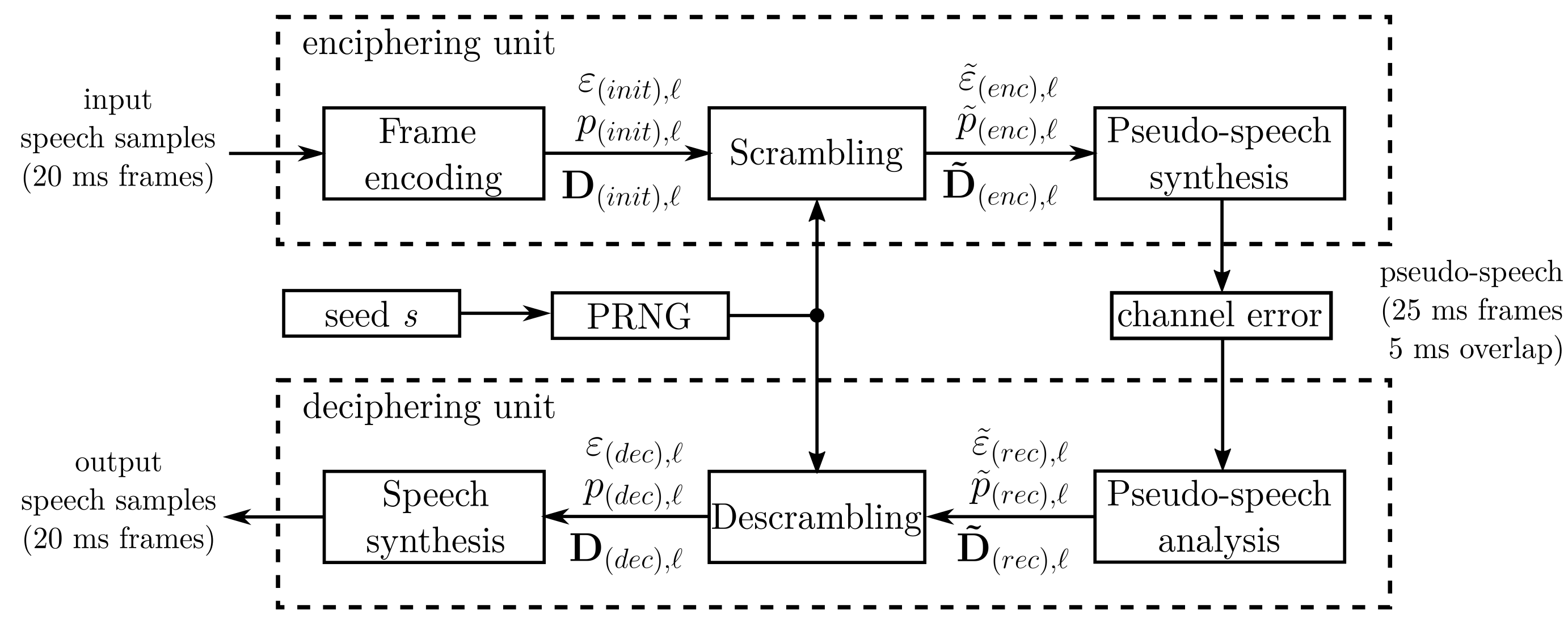}
    \caption{Simplified diagram of the distortion-tolerant speech encryption scheme.}
    \label{speech_diagram}
\end{figure*}

Figure \ref{speech_diagram} illustrates a simplified model of a distortion-tolerant speech encryption scheme, consisting of a speech enciphering unit and a complementary deciphering unit. The enciphering unit takes as an input a binary key-stream produced by a PRNG with a secret seed $s$ of length at least 128 bits, and samples of a narrowband speech signal. In the first processing step, the speech encoder maps 20 ms speech frames indexed by $\ell = 0,1,2,...$ into a sequence of vocal parameters $(\varepsilon_{(init),\ell},$ $p_{(init),\ell},\mathbf{D}_{(init),\ell})$, where $\varepsilon_{(init),\ell}$ corresponds to the frame's energy, $p_{(init),\ell}$ is a pitch period, and $\mathbf{D}_{(init),\ell}$ is a vector representing the shape of a spectral envelope.

The encoding process is followed by enciphering using randomness produced by the PRNG. Vocal parameters of every frame are independently scrambled into a new set of parameters $(\tilde{\varepsilon}_{(enc),\ell}, \tilde{p}_{(enc),\ell},\mathbf{\tilde{D}}_{(enc),\ell})$ defined over a new space of pseudo-speech parameters (tagged by a tilde). Finally, the scrambled sequence is forwarded to the pseudo-speech synthesizer, which produces a harmonic, wideband signal resembling pseudo-speech. The synthetic signal is a concatenation of 25 ms frames with a 5 ms overlap, where every frame carries one set of enciphered parameters. Consequently, the encrypted signal duration is the same as the duration of the initial speech, which is an essential requirement in real-time operation.
  
Due to the speech-like properties of the synthetic signal, it can be transmitted over a wideband digital voice channel without much risk of suppression by a Voice Activity Detector (VAD). Upon reception of the signal samples, the paired deciphering unit extracts distorted copies of the sent parameters $(\tilde{\varepsilon}_{(rec),\ell}, $ $ \tilde{p}_{(rec),\ell},$ $\mathbf{\tilde{D}}_{(rec),\ell})$ and performs descrambling using the same binary key-stream produced by its PRNG. In the last step, restored parameters $(\varepsilon_{(dec),\ell}, p_{(dec),\ell}, \mathbf{D}_{(dec),\ell})$ are decoded into narrowband speech, perceptually similar to the input speech signal.

The crucial property of the presented speech encryption system is its ability to descramble enciphered parameters $(\tilde{\varepsilon}_{(rec),\ell}, \tilde{p}_{(rec),\ell},\mathbf{\tilde{D}}_{(rec),\ell})$ distorted by channel noise. As the amount of channel distortion goes up, so does the distortion of the resynthesized speech. As a result, we obtain a progressive and controlled speech quality degradation without significant loss of intelligibility. Preservation of intelligibility comes from the remarkable human tolerance to understanding distorted speech signals.

\subsection{Speech encoding}

The speech encoder in the presented encryption scheme is essentially a harmonic speech encoder that models speech signals as a combination of amplitude-modulated harmonics \cite{mcaulay1986speech}. The perceived fundamental frequency of a harmonic speech signal is usually referred to as \textit{pitch}, signal energy is perceived as \textit{loudness}, whereas the spectral envelope is related to \textit{speech timbre}. 

\begin{figure}[h!]
 	\centering
 	    \includegraphics{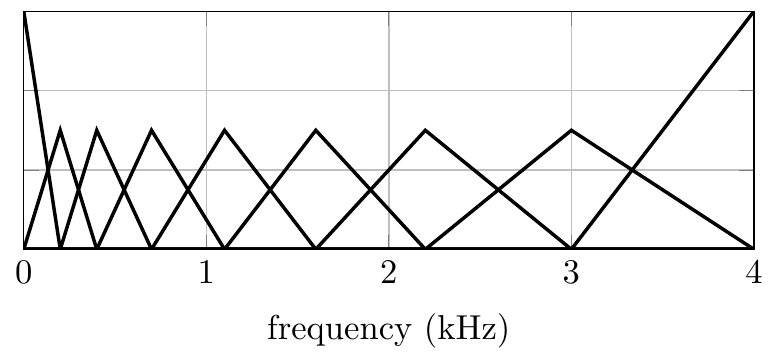}
 	\caption{Nine mel-scaled triangular spectral windows used in speech encoding. The amplitude of two side filters is doubled to compensate their missing halves. \vspace{-4mm}}
 	\label{mel_bands_scheme}
 \end{figure}

The encoder operates sequentially on 20 ms speech frames of 160 samples with a 10 ms look-ahead. Every frame is processed in the same manner, so we skip the frame indexation $\ell$ for simplicity. A speech frame is firstly pre-emphasized with a first-order filter $I(z) = 1-0.85z^{-1}$ to boost high-frequency signal components, and encoded into a set of 10 basic parameters: its pitch period and an approximation of the spectral envelope expressed by 9 coefficients. The pitch period expressed in signal samples per cycle is defined as: 
\begin{equation}
p_{(init)} \coloneqq \frac{f_s}{f_0},
\end{equation}
where $f_0$ is the estimated fundamental frequency of the harmonic structure of the speech signal and $f_s = 8000$ is the sampling frequency. The spectral envelope is obtained from the Power Spectral Density (PSD) on a moving window of 40 ms speech signal with 20-ms offset and 50\% overlap. The PSD is windowed using 9 mel-scaled triangular filters shown in Fig. \ref{mel_bands_scheme}, resulting in 9 band-limited energies $\mathrm{E}_1$, ..., $\mathrm{E}_9$ such that their sum is close enough to the frame energy:
\begin{equation}
\varepsilon_{(init)} \coloneqq \sum_{i=1}^9 \mathrm{E}_i.
\end{equation}

It may be noticed that the vector of square roots of energy coefficients $[\sqrt{\mathrm{E}_1}$, ..., $\sqrt{\mathrm{E}_9}]^T$ is identified as a point on the non-negative part of the 9-dimensional hypersphere centered at 0. The radius $\sqrt{\varepsilon_{(init)}}$ of the 8-sphere is related to the frame energy, whereas the normalized vector:
\begin{equation}
\mathbf{D}_{(init)} \coloneqq \left[\sqrt{\mathrm{E}_1/\varepsilon_{(init)}}, ..., \sqrt{\mathrm{E}_9/\varepsilon_{(init)}}\right]^T
\end{equation}
corresponds to the shape of the spectral envelope, i.e. speech timbre. Since a typical spectral envelope consists of about 4 formants \cite{rabiner2011theory}, it is a reasonable assumption that $\mathbf{D}_{(init)}$ should capture the most relevant features in the speech spectrum.

The enciphering procedure requires the encoded pitch period and the signal energy to be bounded by some predefined intervals $[p_{min}, \ p_{max}]$ and $[\varepsilon_{min}, \ \varepsilon_{max}]$. Thus, if $p_{(init)}$ or $\varepsilon_{(init)}$ exceed these intervals, they are thresholded to the closest bound. A selection of bounds is a compromise between the dynamic range required for proper speech representation and its sensitivity to distortion. Moreover, the lower energy bound $\varepsilon_{min}$ is slightly larger than 0, meaning that the scheme could be unable to register some very low-amplitude sounds.

\subsection{Enciphering}

A blockwise scrambling is applied on the input parameters $(p_{(init)}, \varepsilon_{(init)}, \mathbf{D}_{(init)})$ defined over the space of speech parameters into a new set $(\tilde{p}_{(enc)}, \tilde{\varepsilon}_{(enc)}, \mathbf{\tilde{D}}_{(enc)})$ defined over the space of pseudo-speech parameters. Each of these parameters is crucial for maintaining speech intelligibility \cite{huang2001spoken}, and hence contains information that could be exploited by a cryptanalyst to reconstruct the vocal message. For this reason, we consider them as equally salient. Consequently, $(p_{(init)}, \varepsilon_{(init)}, \mathbf{D}_{(init)})$ are enciphered using a single, shared PRNG.

The enciphering of each frame requires a vector of 10 freshly-generated random integers $\boldsymbol \nu = [\nu_1,\ ...\ \nu_{10}]$, where $\nu_3$ belongs to the additive ring $\mathbb{Z}_{2^{15}}$ with $2^{15}$ elements, $\nu_{10}$ belongs to $\mathbb{Z}_{2^{17}}$, and the remaining coefficients belong to $\mathbb{Z}_{2^{16}}$. These non-uniform ranges of values determine the quantization resolution of the input parameters: $p_{(init)}$ and $\varepsilon_{(init)}$ are quantized using $2^{16}$ levels, and the vector $\mathbf{D}_{(init)}$ is encoded by one of the $2^{128}$ possible values. Consequently, we obtain a 16-bit quantization per encoded coefficient, which is a reasonable resolution for encoding vocal parameters. The vector $\boldsymbol \nu$ can be efficiently computed from a sequence of 160 bits produced by the PRNG. Given the random bits, the scrambling block splits the binary sequence into chunks of length 15, 16 and 17 bits, and reads them as unsigned integers.  

\begin{figure}[h]
\centering
    \includegraphics[width=0.49\textwidth]{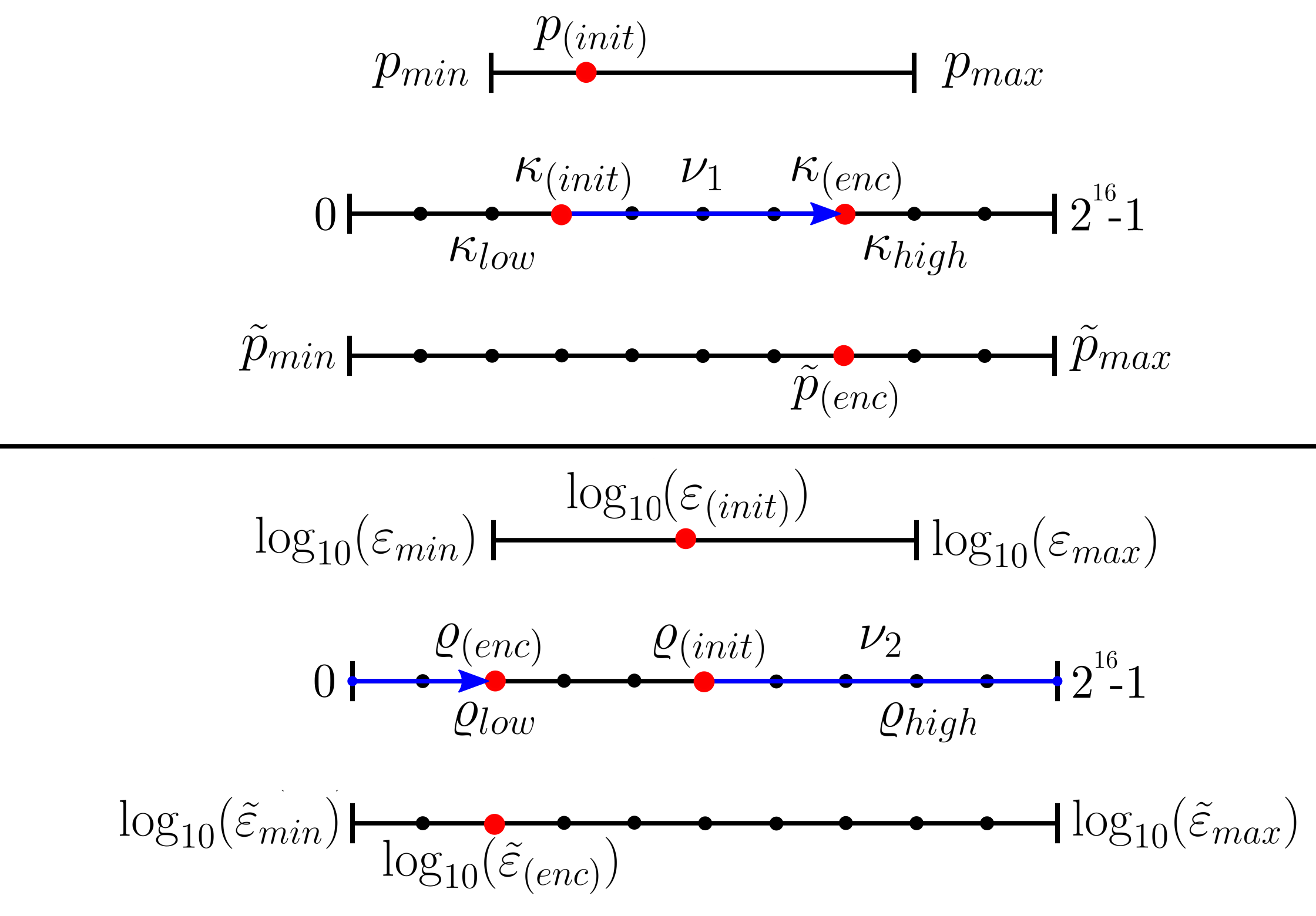}
    \caption{Enciphering of the frame pitch period $p_{(init)}$ (top) and the frame energy $\varepsilon_{(init)}$ (bottom). \textbf{Step~1}: $p_{(init)}$ and $\log_{10}(\varepsilon_{(init)})$ are linearly scaled and rounded to $\kappa_{(init)}$ and $\varrho_{(init)}$ in $\mathbb{Z}_{2^{16}}$. \textbf{Step~2}: $\kappa_{(init)}$ and $\varrho_{(init)}$ are translated by random $\nu_1$ and $\nu_2$ over $\mathbb{Z}_{2^{16}}$ to $\kappa_{(enc)}$ and $\varrho_{(enc)}$. \textbf{Step~3}: $\kappa_{(enc)}$ and $\varrho_{(enc)}$ are linearly scaled to $\tilde{p}_{(enc)}$ and $\tilde{\varepsilon}_{(enc)}$   defined over the space of pseudo-speech parameters.}
    \label{pitch_energy_enc}
\end{figure}

Enciphering of pitch and energy is illustrated in Fig.~\ref{pitch_energy_enc}. The input pitch period $p_{(init)}$ is linearly scaled into an interval $[\kappa_{low},\kappa_{high}]$ such that $0 < \kappa_{low} < \kappa_{high} < 2^{16}-1$, and rounded to the closest integer $\kappa_{(init)} \in \mathbb{Z}_{2^{16}}$. Similarly, the frame energy in logarithmic scale $\log_{10}(\varepsilon_{(init)})$ is transformed to  $\varrho_{(init)} \in [\varrho_{low},\varrho_{high}]$. Then, the obtained integers $\kappa_{(init)}$ and $\varrho_{(init)}$ are translated respectively by $\nu_1$ and $\nu_2$ over the additive ring $\mathbb{Z}_{2^{16}}$:
\begin{align}
\kappa_{(enc)} &= (\kappa_{(init)}+\nu_1) \mod 2^{16} \\
\varrho_{(enc)} &= (\varrho_{(init)}+\nu_2) \mod 2^{16}.
\end{align}
Finally, the enciphered integers $\kappa_{(enc)}$ and $\varrho_{(enc)}$ are scaled to $\tilde{p}_{(enc)} \in [\tilde{p}_{min}, \ \tilde{p}_{max}]$ and $\tilde{\varepsilon}_{(enc)} \in [\tilde{\varepsilon}_{min}, \ \tilde{\varepsilon}_{max}]$. 

The unit vector $\mathbf{D}_{(init)} \in S^8$ is scrambled using a technique described in Section \ref{section-sphericalcodes}. Let $\boldsymbol\xi = 1/\sqrt{8}[1,...,1]^T \in\mathbb{R}^8$. It may be noticed that the vector $2\gamma_{\boldsymbol \xi} (\mathbf{D}_{(init)})$ (note the factor of 2) lies inside the hyperbox $\mathcal{P}_{\boldsymbol \xi}=\prod_{i=1}^{8} [0,2\pi/\sqrt{8})$, and thus can be mapped to a flat torus associated with the mapping $\Phi_{\boldsymbol \xi}$. Additionally, the scrambling model relies on a spherical code $\mathscr{C} = \mathcal{G}\boldsymbol \sigma$, ${\boldsymbol \sigma = [\xi_1,0,...,\xi_n,0]^T}$ , associated with the quotient $\Lambda_{(\alpha)} / \Lambda_{(\beta)}$, where $\Lambda_{(\beta)} = (2\pi\mathrm{Z}^8/\sqrt{8})$ is an orthogonal lattice and $\Lambda_{(\alpha)}$ is the scaled Gosset lattice:
\begin{equation}
\Lambda_{(\alpha)} = \frac{2\pi}{2^{16}\sqrt{8}}\Gamma_8.
\end{equation}
From its design, $\mathcal{G}$ is isomorphic to ${\mathbb{Z}_{2^{15}} \oplus \mathbb{Z}^6_{2^{16}} \oplus \mathbb{Z}_{2^{17}}}$. 

\begin{figure}[h]
 \centering
    \includegraphics[width=0.49\textwidth]{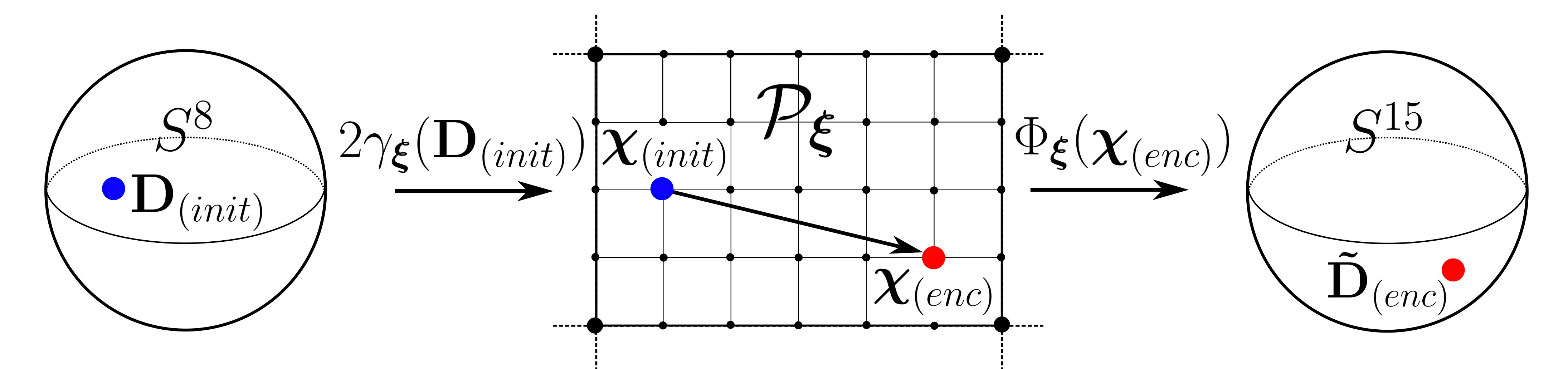}
    \caption{Enciphering of $\mathbf{D}_{(init)}$. The scaled spherical coordinates $2\gamma_{\boldsymbol \xi} (\mathbf{D}_{(init)})$ are quantized by searching the closest lattice vector from $\Lambda_{(\alpha)}$. The obtained $\boldsymbol \chi_{(init)} \in \Lambda_{(\alpha)}$ is randomly translated to a new lattice vector and projected to $\mathbf{\tilde{D}}_{(enc)} = \Phi_{\boldsymbol \xi}(\boldsymbol \chi_{(enc)})$ on the flat torus in $\mathbb{R}^{16}$. }
    \label{timbre_full_enc}
\end{figure}

Figure \ref{timbre_full_enc} illustrates an enciphering of $2\gamma_{\boldsymbol \xi} (\mathbf{D}_{(init)})$ over the pre-image $\mathcal{P}_{\boldsymbol \xi}$. Let $\boldsymbol \chi_{(init)} \in \Lambda_{(\alpha)}$ be the closest lattice vector to $2\gamma_{\boldsymbol \xi} (\mathbf{D}_{(init)})$. The vector $\boldsymbol \chi_{(init)}$ is firstly translated by a random vector:
\begin{equation}
\boldsymbol \chi_{(enc)} = \boldsymbol \chi_{(init)} + \nu_3 \boldsymbol \alpha_1 + ...+ \nu_{10}\boldsymbol \alpha_8 \mod  \Lambda_{(\alpha)} / \Lambda_{(\beta)},
\end{equation}
where $\boldsymbol \alpha_1$, ..., $\boldsymbol \alpha_8$ are the basis vectors of the lattice $\Lambda_{(\alpha)}$:
\begin{flalign*}
\boldsymbol \alpha_1 &= 2\pi/2^{15}[\phantom{-}1,\phantom{-}0,\phantom{-}0,\phantom{-}0,\phantom{-}0,\phantom{-}0,\phantom{-}0,\phantom{-}0]^T, \\
\boldsymbol \alpha_2 &= 2\pi/2^{16}[-1,\phantom{-}1,\phantom{-}0,\phantom{-}0,\phantom{-}0,\phantom{-}0,\phantom{-}0,\phantom{-}0]^T, \\
\boldsymbol \alpha_3 &= 2\pi/2^{16}[\phantom{-}0,-1,\phantom{-}1,\phantom{-}0,\phantom{-}0,\phantom{-}0,\phantom{-}0,\phantom{-}0]^T, \\
\boldsymbol \alpha_4 &= 2\pi/2^{16}[\phantom{-}0,\phantom{-}0,-1,\phantom{-}1,\phantom{-}0,\phantom{-}0,\phantom{-}0,\phantom{-}0]^T ,\\
\boldsymbol \alpha_5 &= 2\pi/2^{16}[\phantom{-}0,\phantom{-}0,\phantom{-}0,-1,\phantom{-}1,\phantom{-}0,\phantom{-}0,\phantom{-}0]^T, \\
\boldsymbol \alpha_6 &= 2\pi/2^{16}[\phantom{-}0,\phantom{-}0,\phantom{-}0,\phantom{-}0,-1,\phantom{-}1,\phantom{-}0,\phantom{-}0]^T ,\\
\boldsymbol \alpha_7 &= 2\pi/2^{16}[\phantom{-}0,\phantom{-}0,\phantom{-}0,\phantom{-}0,\phantom{-}0,-1,\phantom{-}1,\phantom{-}0]^T ,\\
\boldsymbol \alpha_8 &= 2\pi/2^{17}[\phantom{-}1,\phantom{-}1,\phantom{-}1,\phantom{-}1,\phantom{-}1,\phantom{-}1,\phantom{-}1,\phantom{-}1]^T. \\
\end{flalign*}

Provided that the PRNG produces bits with a distribution indistinguishable from the uniformly random distribution and $\nu_3,...,\nu_{10}$ are obtained by reading respective bitstrings as unsigned integers, then $\nu_3 \boldsymbol \alpha_1 + ...+ \nu_{10}\boldsymbol \alpha_8$ can be any vector of $\Lambda_{(\alpha)} / \Lambda_{(\beta)}$ with distribution indistinguishable from the uniform distribution over $\Lambda_{(\alpha)} / \Lambda_{(\beta)}$. This operation is thus equivalent to the subroutine `SelectRandom' in Algorithm \ref{procedure} in Section~\ref{section-sphericalcodes}. 

Finally, the enciphered vector is mapped to the flat torus $\mathbf{\tilde{D}}_{(enc)} = \Phi_{\boldsymbol \xi}(\boldsymbol \chi_{(enc)}) \in S^{15}$.
 
\subsection{Pseudo-speech synthesis}

The last stage of speech encryption involves the synthesis of an analog audio signal. The role of audio synthesis is to enable an efficient transmission of the enciphered values $(\tilde{p}_{(enc)}, \tilde{\varepsilon}_{(enc)}, \mathbf{\tilde{D}}_{(enc)})$ over a digital voice channel, and to prevent signal blockage by a VAD. Robust operation requires finding a trade-off between producing a signal sufficiently speech-like yet simple to encode and decode. Furthermore, the encoding procedure should comply with a typical signal distortion characteristic introduced by a particular channel to benefit from distortion-tolerant enciphering.

Since $\tilde{p}_{(enc)}$, $\tilde{\varepsilon}_{(enc)}$ and $\mathbf{\tilde{D}}_{(enc)}$ represent the enciphered pitch period, the energy and the spectral envelope of a speech frame, the natural approach is to relate these values with some homologous parameters of an encrypted signal. Then, a perceptual distortion of the signal would be proportionally mapped to the deciphered speech, to some extent reflecting the quality of the voice channel used for transmission.   

\begin{figure}[h]
 	\centering
 	    \includegraphics{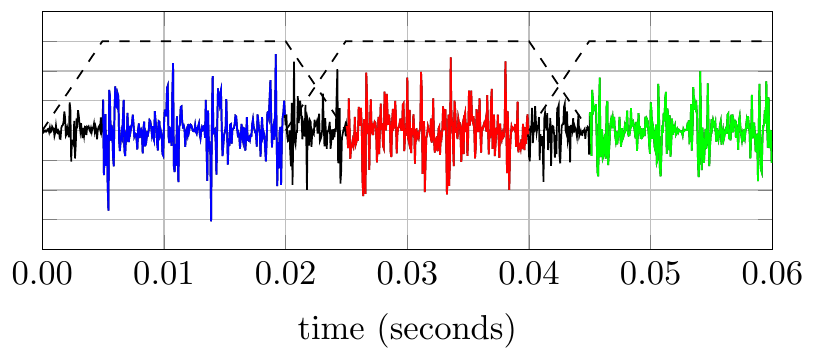}
 	\caption{Three 25 ms frames of a pseudo-speech harmonic signal. Every colored portion of the waveform encodes a different set of enciphered parameters $\tilde{p}_{(enc)}$, $\tilde{\varepsilon}_{(enc)}$ and $\mathbf{\tilde{D}}_{(enc)}$. The frames are windowed using a trapezoidal window and overlapped, forming 5 ms guard periods. }
\label{pseudospeech_symbols}
 \end{figure}

Every 25 ms frame of a pseudo-speech signal consists of three segments. The first and the last 5 ms of a frame play the role of guard periods. The remaining 15 ms is where the enciphered parameters are encoded. Once a frame is synthesized, it is windowed by a trapezoidal window and concatenated in an overlap-then-add manner, as illustrated in Fig.~\ref{pseudospeech_symbols}. 

A 25 ms signal frame $\mathbf{y}_t$ sampled at $f_s = 16$ kHz contains the samples of a harmonic waveform:
\begin{equation}
\mathbf{y}[n] = \sum_{k=1}^{{K_{(\omega_0)}}}{\eta\mathrm{A}_k}\cos(k{ \omega_0}/f_s \cdot n + {\phi_k -k\omega_0/f_s\cdot 80}),
\label{equation_yt}
\end{equation}
where $n=0,1,...,399$, ${\omega_0}$ is the fundamental frequency, ${\mathrm{A}_k}$ are the amplitudes of harmonics, $\eta$ is the energy scaling factor, ${(\phi_k -k\omega_0/f_s\cdot 80)}$ are the initial phases and ${K_{(\omega_0)}}$ is the number of harmonics depending on ${\omega_0}$. Given the harmonicity of $\mathbf{y}_t$, the encoding of $\tilde{p}_{(enc)}$, $\tilde{\varepsilon}_{(enc)}$ and $\mathbf{\tilde{D}}_{(enc)}$ essentially reduces to a careful manipulation of $\omega_0$, $\mathrm{A}_k$ and $\phi_k$. In addition, only the middle samples $\mathbf{y}[80],...,\mathbf{y}[319]$ are involved in the encoding process. Once the harmonic parameters of the frame are determined, the remaining part of $\mathbf{y}_t$ is reproduced.

The encoding of enciphered parameters into $\mathbf{y}_t$ is performed sequentially, starting from the pitch, then the envelope shape, and finally the energy. The most natural approach for encoding the pitch is to assign $\omega_0 \coloneqq 2\pi f_s/\tilde{p}_{(enc)}$. Spectral shaping involves finding a proper relation between the amplitudes $\mathrm{A}_k$ and the initial phases $\phi_k$. Finally, the spectrally shaped frame is scaled in order to match the desired frame energy $ \tilde{\varepsilon}_{(enc)} = \sum_{n=80}^{319}\mathbf{y}^2[n]$.

\begin{figure}[h]
 	\centering
 	   \includegraphics{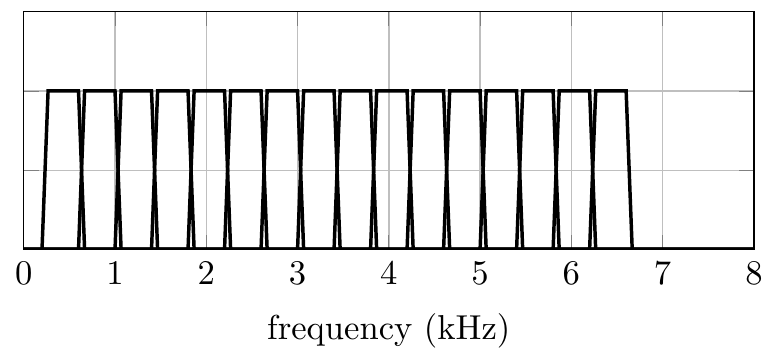}
 	\caption{Sixteen square-shaped spectral windows distributed uniformly over 300-6700 Hz.}
 	\label{square_windows}
\end{figure}

Compared to encoding the pitch and the energy, mapping the vector $\mathbf{\tilde{D}}_{(enc)}$ of length 16 into the spectrum of $\mathbf{y}_t$ appears to be less straightforward. The encoding process relies on a bank of 16 adjacent spectral windows, illustrated in Fig. \ref{square_windows}. The main idea of using these spectral windows is to encode each coordinate of $\mathbf{\tilde{D}}_{(enc)}$ into a frequency band associated to its respective spectral window. Unlike in speech encoding, the windows are square-shaped, and linearly distributed between the 300-6700 Hz range. The proposed selection of spectral windows aims to improve transmission robustness over a voice channel rather than to capture the perceptually relevant spectral features. As a result, the proposed framework is similar to using Frequency Division Multiplexing (FDM) \cite{1090705fdm} for mitigating frequency fading.

Another difference is related to how the spectral windows are applied. Instead of windowing the signal PSD as is the case in speech analysis, the windows are directly applied on the Discrete Fourier Transform (DFT) of sampled $\mathbf{y}_t$. As will be explained later in the section, this change significantly simplifies the encoding process. Besides, it seems better suited for data transmission over channels with an additive, independent noise such as AWGN.

Shaping the spectrum of the harmonic frame $\mathbf{y}_t$ can be achieved by a simultaneous manipulation of the amplitudes and the initial phases of the harmonics. Thus, it is advantageous to consider a complex-domain rewriting of $\mathbf{y}_t$, in which the amplitude $\mathrm{A}_k$ and the initial phase $\phi_k$ are merged into a single complex term $\check{\mathrm{A}}_k = \mathrm{A}_k\exp(j\phi_k)$:

\begin{equation}
\mathbf{z}[n] = \sum_{k=1}^{K_{(\omega_0)}}\check{\mathrm{A}}_k\exp(jk\omega_0/f_s \cdot n), \ n=0,1,...,239.
\label{equation_ak}
\end{equation}
The complex samples $\mathbf{z}[0],...,\mathbf{z}[239]$ correspond to respective samples $\mathbf{y}[80],...,\mathbf{y}[319]$ that encode the enciphered parameters. 

Let $\mathbf{Z}$ be the column vector representing the DFT of samples $\mathbf{z}[0],...,\mathbf{z}[239]$. The vector $\mathbf{Z}$ is a sum of harmonic components:
\begin{equation}
\mathbf{Z} = \sum_{k=1}^{K_{(\omega_0)}}\check{\mathrm{A}}_k \mathbf{B}^T_{(\omega_0),k},
\label{matrix1}
\end{equation}
where $\mathbf{B}_{(\omega_0),k}$ is a row vector representing the DFT of complex sinusoidal samples $\exp (j2\pi \omega_0/f_sn)$ for $n = 0,1,...,239$. The goal of the encoding process is to find $\mathbf{Z} = [\mathrm{Z}_1,...,\mathrm{Z}_{240}]^T$ such that:
\begin{equation}
\tilde{\mathrm{D}}_{(enc),k} \cdot \mathrm{e}^{j\frac{2\pi}{16}k} = \sum_{n=1}^{240} \mathrm{Z}_n \mathrm{H}_{k,n}, \ k=1,...,16,
\label{matrix2}
\end{equation}
where $\mathbf{H}_k = [\mathrm{H}_{k,1}, ..., \mathrm{H}_{k,240}]$ is a row vector representing the $k$-th spectral window in Fig. \ref{square_windows} sampled at frequencies $\frac{n}{240} \cdot f_s$ for $n = 0,1,...,239$. As a result, each element of $\mathbf{\tilde{D}}_{(enc)}$ is represented by a complex sum of windowed DFT samples. The predefined component $\exp (j2k\pi/16)$ is inserted to prevent the result of summation from being purely real and improve the time-domain waveform shape of the synthesized pseudo-speech frame.

The summations in Eq. (\ref{matrix1}) and Eq. (\ref{matrix2}) can be expressed conveniently in a matrix form:
\begin{equation}
\mathbf{\tilde{D}}_{(enc)} \odot \bold{W}_{16} = \mathrm{H}\mathbf{Z} \qquad \text{and} \qquad \mathbf{Z} = \mathrm{B}_{(\omega_0)}\mathbf{\check{A}},  
\end{equation}
where $\bold{W}_{16} = [\mathrm{e}^{j\frac{2\pi}{16}1}, \mathrm{e}^{j\frac{2\pi}{16}2}, ..., \mathrm{e}^{j\frac{2\pi}{16}16}]^T$ is the vector of the 16 roots of the unity, $\mathrm{H}$ is a $16 \times 240$ matrix representing 16 spectral windows sampled over the frequency domain of $\mathbf{Z}$, $\mathrm{B}_{(\omega_0)}$ is a $240 \times K_{(\omega_0)}$ matrix with columns $\mathbf{B}_{(\omega_0),k}$, $\mathbf{\check{A}}$ is the column vector of $K_{(\omega_0)}$ complex amplitudes $\check{\mathrm{A}}_k$ as defined by Eq. (\ref{equation_ak}), and $\odot$ denotes the Hadamard product. As a result, we obtain a simple, linear relation between $\mathbf{\tilde{D}}_{(enc)}$ and the amplitudes of harmonics:
\begin{equation}
\mathbf{\tilde{D}}_{(enc)} \odot \bold{W}_{16} = \mathrm{H}\mathrm{B}_{(\omega_0)}\mathbf{\check{A}}
\label{matrix_combined}
\end{equation}

The problem of finding $\mathbf{\check{A}}$ such that Eq. (\ref{matrix_combined}) holds is under-determined, because $K_{(\omega_0)}$ is larger than 16 by design. Instead, we can compute the least-square solution using the Moore-Penrose pseudo-inverse:
\begin{equation}
\mathbf{\breve{A}} = (\mathrm{H}\mathrm{B}_{ \omega_0})^{\dagger}(\mathbf{\tilde{D}}_{(enc)} \odot \bold{W}_{16})
\label{matrix_penrose}
\end{equation}
\noindent where $(\bullet)^{\dagger}$ denotes the pseudo-inverse operation. In order to improve computational efficiency, the pseudo-inverse matrix $(\mathrm{H}\mathrm{B}_{\omega_0})^{\dagger}$ can be precomputed and kept in memory. 

The least-square solutions obtained by the Moore-Penrose pseudo-inverse imply that the computed magnitudes $|\breve{\mathrm{A}}_k|$ are small. It has a positive impact on the time-domain waveform shape, minimizing the risk of producing high-amplitude peaks that are likely to be clipped during transmission. Another advantage of the pseudo-inverse is its fast computation, suitable for real-time processing.

Finally, we assign $\mathrm{A}_k \coloneqq |\breve{\mathrm{A}}_k|$ and $\phi_k \coloneqq \mathrm{Arg}(\breve{\mathrm{A}}_k)$, and set the scaling factor $\eta$ in Eq. (\ref{equation_yt}) to match the energy $\tilde{\varepsilon}_{(enc)}$. 

The remaining issue is extracting $\mathbf{\tilde{D}}_{(enc)}$ from $\mathbf{y}_t$. The previously described encoding process involved complex samples $\mathbf{z}[0],...,\mathbf{z}[239]$, where $\mathbf{y}[80+n] = \eta \Re (\mathbf{z}[n])$. Let $\mathbf{Y}$ be a column vector of length 240 representing the DFT of samples $\mathbf{y}[80],...,\mathbf{y}[319]$. From the general properties of Discrete Fourier Transform we have:
\begin{equation}
\mathbf{Y}_n = \frac{\eta}{2}(\mathbf{Z}_n + \bar{\mathbf{Z}}_{241-n}), \ n = 1,...,240,
\end{equation}
where $\mathbf{\bar{Z}}_{241-n}$ denotes the complex conjugate of $\mathbf{Z}_{241-n}$. Provided that $\mathbf{Z}$ is a sum of complex sinusoids of frequency no larger than 6700 Hz, the values $\mathbf{Z}_n$ for $n > 120$ are close to zero. As a result, we can approximate the vector $\mathbf{Y}$ as:
\begin{equation}
\mathbf{Y} \approx 
\begin{cases}
\frac{\eta}{2}\cdot \mathbf{Z}_n, & \text{for } n = 1,...,120, \\
\frac{\eta}{2}\cdot \mathbf{\bar{Z}}_{241-n}, & \text{for } n = 121,...,240.
\end{cases}
\end{equation}
Finally, the enciphered vector can be approximately retrieved by taking:
\begin{equation}
\mathbf{\tilde{D}}_{(enc)} \odot \bold{W}_{16} \approx \frac{2}{\eta}\mathrm{H}\mathbf{Y}.
\end{equation}

We estimated the root mean squared error (RMSE) of ${\mathbf{\tilde{D}}_{(enc)} \odot \bold{W}_{16}}$ approximations by simulating a sequence of $L = 10000$ pseudo-speech frames from parameters $(\tilde{p}_{(enc)},$ $\tilde{\varepsilon}_{(enc)},$ $\mathbf{\tilde{D}}_{(enc)})$ selected randomly in every frame. We used the following formula: 

\begin{flalign*}
\mathrm{RMSE}_{\mathbf{\tilde{D}}} &= \sqrt{\frac{1}{L}\sum\limits_{\ell=1}^L\|\mathbf{\tilde{D}}_{(enc),\ell} \odot \bold{W}_{16}-2\eta_\ell^{-1}\mathrm{H}\mathbf{Y}_\ell\|^2},&
\end{flalign*}
where $\mathbf{\tilde{D}}_{(enc),\ell}$ is the $\ell$-th encoded vector, $\mathbf{Y}_\ell$ is a vector representing the DFT of the $\ell$-th produced pseudo-speech frame, and $\eta_\ell$ is the respective scaling factor. The obtained error was 0.011, far lower than an anticipated distortion introduced by the voice channel.

\subsection{Signal transmission and analysis}
\label{subsection-transmission}

Successful decoding of a synthetic signal produced by the pseudo-speech synthesizer requires a high-precision, nearly sample-wise synchronization. Consequently, the presented speech encryption scheme is foremost suited for digital data storage and transmission over fully digital voice communication systems like VoIP, in which a high level of synchronization can be maintained. Upon reception, the signal analyzer processes sequentially the received signal frames and retrieves enciphered parameters.

Let $\mathbf{\hat{y}}[0], ..., \mathbf{\hat{y}}[399]$ be the samples of some received pseudo-speech frame $\mathbf{\hat{y}}_t$ and let $\mathbf{\hat{Y}}$ be a column vector of length 240 with DFT of the sequence $\mathbf{\hat{y}}[80], ..., \mathbf{\hat{y}}[319]$. The received parameters $\tilde{p}_{(rec)}$ and $\tilde{\varepsilon}_{(rec)}$ are defined as:
\begin{align}
\tilde{p}_{(rec)} &\coloneqq \frac{2\pi f_s}{\hat{\omega}_0}, \\
\tilde{\varepsilon}_{(rec)} &\coloneqq \sum_{n=80}^{319}\mathbf{\hat{y}}^2[n], 
\end{align}
where $\hat{\omega}_0$ is the estimated fundamental frequency of the signal frame and $f_s = 16000$ Hz is the sampling frequency. If any of the values $\tilde{p}_{(rec)}$ and $\tilde{\varepsilon}_{(rec)}$ exceed the intervals $[\tilde{p}_{min},$ $\tilde{p}_{max}]$ and $[\tilde{\varepsilon}_{min}, \ \tilde{\varepsilon}_{max}]$, they are tresholded to the closest bound.

The vector $\mathbf{\tilde{D}}_{(rec)}$ is retrieved from $\mathbf{\hat{Y}}$ in two steps. Firstly, we compute the normalized real-valued sum of the windowed DFT:
\begin{align}
\mathbf{\hat{D}}_{(rec)} \coloneqq \frac{\Re \left( 2H\mathbf{\hat{Y}} \odot \mathbf{\bold{\bar{W}}}_{16} \right)}{\left\|\Re \left( 2H\mathbf{\hat{Y}} \odot \mathbf{\bold{\bar{W}}}_{16} \right)\right\|},
\label{received_spectral_shape}
\end{align}
where $\bold{\bar{W}}_{16} = [\mathrm{e}^{-j\frac{2\pi}{16}1}, \mathrm{e}^{-j\frac{2\pi}{16}2}, ..., \mathrm{e}^{-j\frac{2\pi}{16}16}]^T$ and $\Re ( 2H\mathbf{\hat{Y}} \odot \bold{\bar{W}}_{16} )$ is the real component of $2H\mathbf{\hat{Y}} \odot \bold{\bar{W}}_{16}$. Then, the vector $\mathbf{\hat{D}}_{(rec)}$ is projected to $\mathbf{\tilde{D}}_{(rec)}$ on the flat torus associated with $\Phi_{\boldsymbol \xi}$ using Eq.~(\ref{projection_formula}) in Section~\ref{subsection-transmission}.

\begin{figure*}[h!]
 \centering
    \includegraphics[width=0.85\textwidth]{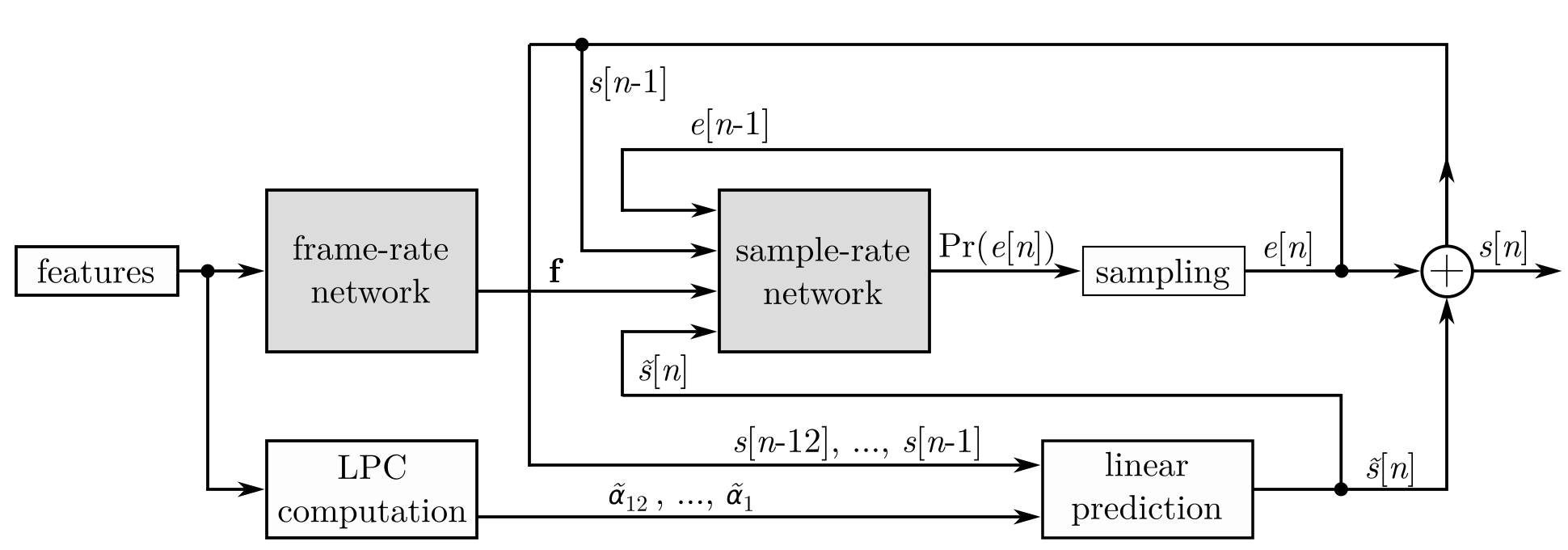}
    \caption{Overview of the narrowband LPCNet architecture.}
    \label{narrow_lpcnet}
\end{figure*}

\subsection{Deciphering}

Given the set of received parameters $(\tilde{p}_{(rec)}, \tilde{\varepsilon}_{(rec)},$ $\mathbf{\tilde{D}}_{(rec)})$, the descrambling algorithm reverses enciphering operations using the same vector $\boldsymbol \nu = [\nu_1,...,\nu_{10}]^T$ of random integers produced by the PRNG.  The values $\tilde{p}_{(rec)}$ and $\log_{10}(\tilde{\varepsilon}_{(rec)})$ are firstly linearly scaled to $\kappa_{(rec)}$ and $\varrho_{(rec)}$ over the interval $[0,2^{16}-1]$. Unlike in the enciphering stage, these values are not quantized. In the next step, $\kappa_{(rec)}$ and $\varrho_{(rec)}$ are deciphered by respective translations $-\nu_1$ and $-\nu_2$ modulo $2^{16}$: 
\begin{align}
\kappa_{(dec)} &= (\kappa_{(rec)}-\nu_1) \mod 2^{16} \\
\varrho_{(dec)} &= (\varrho_{(rec)}-\nu_2) \mod 2^{16},
\end{align}
where $\nu_1, \nu_2 \in \mathbb{Z}_{2^{16}}$ are obtained from the PRNG. If the values $\kappa_{(dec)}$ and $\varrho_{(dec)}$ exceed the respective intervals $[\kappa_{low},$ $\kappa_{high}]$ and $[\varrho_{low},\varrho_{high}]$, they are tresholded to the closest bound. In the last step, the values are transformed back into the intervals $[p_{min}, \ p_{max}]$ and $[\log_{10}(\varepsilon_{min}), \ \log_{10}(\varepsilon_{max})]$ representing the domain of speech parameters. 

The deciphering of the unit vector $\mathbf{\tilde{D}}_{(rec)}$ is done by translating $\boldsymbol \chi_{(rec)} = \Phi_{\boldsymbol \xi}^{-1}(\mathbf{\tilde{D}}_{(rec)})$:
\begin{equation}
\boldsymbol \chi_{(dec)} = \boldsymbol \chi_{(rec)} - \nu_3\boldsymbol \alpha_1 - ... - \nu_{10}\boldsymbol \alpha_8 \mod  \Lambda_{(\alpha)} / \Lambda_{(\beta)}.
\end{equation}
The coordinates of $\boldsymbol \chi_{(dec)}$ are corrected to fit the image of $\gamma^{-1}_{\boldsymbol \xi}$ as in Algorithm \ref{procedure2}. The deciphered spectral envelope vector is $\mathbf{D}_{(dec)} = \gamma^{-1}_{\boldsymbol \xi}(\boldsymbol \chi_{(dec)}/2)$. 

\subsection{Speech resynthesis}

The output of the descrambling process is a sequence $(p_{(dec),\ell}, \varepsilon_{(dec),\ell},\mathbf{D}_{(dec),\ell})$ representing harmonic parameters of 20 ms speech frames. The final speech resynthesis is possible with any adapted speech synthesizer supporting harmonic speech parametrization. An example of a suitable narrowband harmonic speech synthesizer is Codec2.\footnote{\url{https://rowetel.com}}

Unfortunately, although parametric sinusoidal speech co-ders succeed in producing intelligible speech, they often struggle to maintain satisfactory speech quality. In this work, we avoid harmonic speech synthesis by using a narrowband modification of the LPCNet, a Machine Learning (ML) based synthesizer introduced by Jean-Marc Valin (Mozilla) and Jan Skoglund (Google LLC) \cite{Valin2019lpcnet}. The narrowband LPCNet recreates the samples of a speech signal $s[n]$ from a sum of the linear prediction $\tilde{s}[n]$ and the excitation $e[n]$:
\begin{align}
s[n] &= \tilde{s}[n] + e[n] \\
\tilde{s}[n] &= \sum_{k=1}^{16} \alpha_k s[n-k] \ ,
\end{align}
where $\alpha_1$, ..., $\alpha_{12}$ are the 12-th order linear prediction coefficients (LPC) for the current frame. The excitation samples $e[n]$ are produced by two concatenated neural networks that model an excitation signal from the input vocal parameters. 

\begin{figure}[h!]
 \centering
    \includegraphics[width=0.5\textwidth]{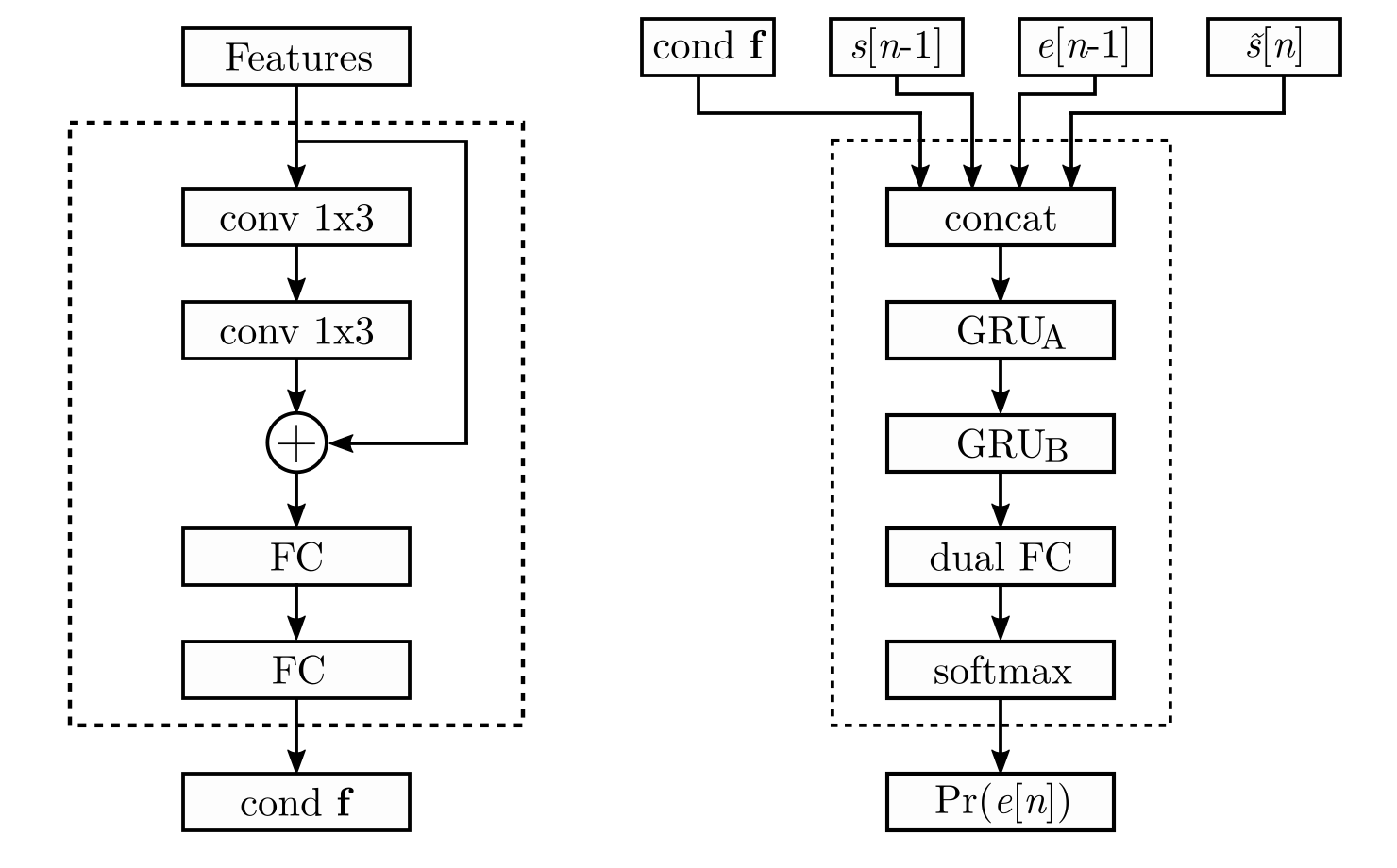}
    \caption{Architectures of the frame-rate (left) and the sample-rate (right) networks. The frame network consists of two convolutional layers with a filter of size 3, followed by two fully-connected layers. The output of the convolutional layers is added to an input connection. The sample-rate network firstly concatenates four inputs and passes the resulted combination to two gated recurrent units ($\mathrm{GRU_A}$ of size 384 and  $\mathrm{GRU_B}$ of size 16), followed by a dual fully connected layer \cite{Valin2019lpcnet}. The output of the last layer is used with a soft-max activation. }
    \label{narrow_lpcnet_networks}
\end{figure}

Figure \ref{narrow_lpcnet} depicts a simplified diagram of the modified narrowband LPCNet algorithm. The speech synthesizer combines two recurrent neural networks: a frame-rate network processing 20 ms speech frames (160 samples) and a sample-rate network operating at 8 kHz. Network architectures, shown in Fig. \ref{narrow_lpcnet_networks}, are the same as in \cite{Valin2019lpcnet}. The frame-rate network takes as input the sequence of feature vectors computed from $(p_{(dec),\ell},$ $\varepsilon_{(dec),\ell},\mathbf{D}_{(dec),\ell})$ and produces a sequence of frame-rate conditioning vectors $\mathbf{f}_\ell$ of length 128. Vectors $\mathbf{f}_\ell$ are sequentially forwarded to the sample-rate network and padded with last value to get a frame with 160 samples.

The role of the sample-rate network is to predict the multinomial probability distribution of the current excitation sample $\Pr(e[n])$, given the current conditioning vector $\mathbf{f}_\ell$, the previous signal sample $s[n-1]$, the previous excitation sample $e[n-1]$ and the current prediction $\tilde{s}[n]$. The current excitation sample $e[n]$ is obtained by randomly generating a single sample from $\Pr(e[n])$. The synthesis output of the narrowband LPCNet are pre-emphasized speech samples $s[n] = \tilde{s}[n]+e[n]$, filtered with a de-emphasis filter $J(z) = \frac{1}{1-0.85z^{-1}}$. The operation of the narrowband LPCNet algorithm stops when the last feature vector is processed, and the sample-rate network synthesizes the last speech frame.

Computing a feature vector from a set of vocal parameters $(p_{(dec),\ell}, \varepsilon_{(dec),\ell},\mathbf{D}_{(dec),\ell})$ requires few steps. Let ${\mathbf{E}_{\ell} = \varepsilon_{(dec),\ell} \cdot \mathbf{D}_{(dec),\ell} \odot \mathbf{D}_{(dec),\ell}}$ be a vector representing 9 band-limited energies of the $\ell$-th encoded speech frame. Then, the  $\ell$-th feature vector has the form $[\mathrm{C}_{\ell,0}, \mathrm{C}_{\ell,1}, ..., \mathrm{C}_{\ell,8}, \rho_{\ell}]^T$, where $\mathrm{C}_{\ell,0}$, $\mathrm{C}_{\ell,1}$, ..., $\mathrm{C}_{\ell,8}$ is the discrete cosine transform (DCT-II) of the sequence $\log_{10}(\mathrm{E}_{\ell,1})$, ..., $\log_{10}(\mathrm{E}_{\ell,9})$, and where $\rho_{\ell} = (p_{(dec),\ell}-100)/50$ is the scaled pitch period. Taking into consideration the mel-scaled distribution of spectral windows used in the speech encoder, the coefficients $\mathrm{C}_{\ell,0}$, $\mathrm{C}_{\ell,1}$, ..., $\mathrm{C}_{\ell,8}$ can be viewed as 9-band Mel-Frequency Cepstral Coefficients (MFCC).

The prediction samples $\tilde{s}[n]$ are computed from the predictor coefficients $\tilde{\alpha}_1$, ..., $\tilde{\alpha}_{12}$ obtained from $\mathbf{E}_{\ell}$ and updated for every frame. The mel-scaled windowed energies in $\mathbf{E}_{\ell}$ are firstly interpolated into a linear-frequency PSD and then converted to an autocorrelation using an inverse FFT. The LPC coefficients are obtained from the autocorrelation using the Levinson-Durbin algorithm \cite{Makhoul1975linear}.

Obtaining $\tilde{\alpha}_k$ from the low-resolution bands is different than in the classical approach, in which the autocorrelation and the predictor $\alpha_k$ are computed directly from speech samples \cite{rabiner2011theory}. As pointed out in \cite{Valin2019lpcnet}, the sample-rate network in the LPCNet learns to compensate for this difference.

\section{Discussion}
\label{discussion}

This section discusses several aspects associated with system security, tolerance to channel distortion, selection of system parameters, and the training the neural networks in the speech synthesizer.

\subsection{Security considerations}

The security of the proposed speech encryption scheme cannot be rigorously proved without an in-depth specification, which may significantly differ in particular implementations. Instead, we provide an informal justification for the asymptotic indistinguishability of encryptions in an experiment comparable with the classical adversarial indistinguishability challenge \cite{GOLDWASSER1984270,Bellare1998crypto,katz2014introduction}. By doing this, we want to emphasize that the security of our encryption scheme depends primarily on the characteristics of a real-world PRNG used and the entropy of the secret seed rather than on signal processing and scrambling operations.

In the following analysis, we will assume that the encryption scheme uses an asymptotically secure pseudo-random generator \cite{katz2014introduction} with a fresh and perfectly random seed $s$. A generator is asymptotically secure when no PPT algorithm (distinguisher) $\mathcal{D}$ can distinguish the generator's output from a perfectly uniform bitstring with a non-negligible probability. A definition of a negligible function is detailed by Definition \ref{negligibility}.

\begin{definition}{\cite{katz2014introduction}}
A function $\mathbf{f}: \mathbb{N} \rightarrow \mathbb{R}_+ \cup \{0\}$ is negligible if for every positive polynomial $p$ there is an integer $N$ such that for all integers $n > N$ it holds that $\mathbf{f}(n) < \frac{1}{p(n)}$.
\label{negligibility}
\end{definition} 

Let $\lambda$ be an integer-valued security parameter, $L$ be a polynomial s.t. $L(\lambda) > \lambda$, $\mathbf{x}_t$ be an arbitrary speech signal of finite duration $t \in [0,320L(\lambda))$, $\mathrm{RandGen}$ be an asymptotically secure binary PRNG and $s\in \{0,1\}^{\lambda}$ be a random seed. In addition, let $\Pi = (\mathrm{RandGen},$ $\mathrm{Enc},\mathrm{Dec})$ be the speech encryption scheme described in Section \ref{speech_encryption}, where $\mathrm{Enc}$ and $\mathrm{Dec}$ are respectively the encryption and decryption algorithms. The algorithm $\mathrm{Enc}$ takes as an input the speech signal $\mathbf{x}_t$ and a vector $\mathbf{r} \leftarrow \mathrm{RandGen}(s)$ such that $\mathbf{r} \in \{0,1\}^{160L(\lambda)}$, and outputs a synthetic signal $\mathbf{y}_t=\mathrm{Enc}_{\mathbf{r}}(\mathbf{x}_t)$ of the same duration as $\mathbf{x}_t$. Furthermore, let us define an adversarial indistinguishability challenge $\mathrm{PrivK}^{eav}_{\mathcal{A},\Pi}(\lambda)$:

\begin{definition} The adversarial indistinguishability challenge $\mathrm{PrivK}^{eav}_{\mathcal{A},\Pi}(\lambda)$ is defined as:
\begin{enumerate}
\item The adversary $\mathcal{A}$ is given input $1^\lambda$, and he chooses a pair of distinct signals $\mathbf{x}_{0,t}$, $\mathbf{x}_{1,t}$ of finite duration $t \in [0,320L(\lambda))$.
\item A random seed $s$ is chosen and a sequence $\mathbf{r}$ is generated by running $\mathrm{RandGen}(s)$. The challenge $\mathbf{y}_t = \mathrm{Enc}_\mathbf{r}(\mathbf{x}_{b,t})$ is given to $\mathcal{A}$, where $b \in \{0,1\}$ is chosen uniformly at random.
\item $\mathcal{A}$ outputs bit $b'$.
\item The output of the challenge is $1$ if $b = b'$ and $0$ otherwise. 
\end{enumerate}
\end{definition}

\noindent Below we present a proposition for the indistinguishability of encryptions in the presence of an eavesdropper and provide the sketch of the proof.

\begin{proposition}
\label{systemsecurity}
Let $\Pi = (\mathrm{RandGen},\mathrm{Enc},\mathrm{Dec})$ be the speech encryption scheme described in Section \ref{speech_encryption}. Then, $\Pi$ gives indistinguishable encryptions in the presence of the eavesdropper, meaning that there is a negligible function $\mathbf{negl}$ such that for any PPT adversary $\mathcal{A}$ it holds:
$$
\left|\Pr(\mathrm{PrivK}^{eav}_{\mathcal{A},\Pi}(\lambda) = 1) - 0.5 \right| \leq  \mathbf{negl}(\lambda). 
$$
\end{proposition}
\begin{proof-sketch}
Firstly, we can observe that the security of the speech encryption scheme does depend neither on the speech analysis nor the pseudo-speech synthesis algorithms. Indeed, the single output of the speech encoder is a sequence of parameters $\{(\varepsilon_{(init),\ell},p_{(init),\ell},$ $\mathbf{D}_{(init),\ell})\}_{\ell=1}^{L(\lambda)}$, which is forwarded to the scrambling block. The result of enciphering is a new sequence $\{(\tilde{\varepsilon}_{(enc),\ell},\tilde{p}_{(enc),\ell},$ $\mathbf{\tilde{D}}_{(enc),\ell})\}_{\ell=1}^{L(\lambda)}$, being the single input of the pseudo-speech synthesizer. In consequence, the indistinguishability of the synthetic pseudo-speech signal $\mathbf{y}_t$ reduces to the indistinguishability of the enciphered sequence from any sequence taken uniformly at random.

The enciphering of the initial speech parameters is done using a sequence of scrambling vectors $\{\boldsymbol \nu_\ell \}_{\ell=1}^{L(\lambda)}$, where ${\boldsymbol \nu_\ell \in \mathbb{Z}^2_{16} \oplus \mathbb{Z}_{15} \oplus \mathbb{Z}^6_{16} \oplus \mathbb{Z}_{17}}$ and $\{\boldsymbol \nu_\ell\}_{\ell=1}^{L(\lambda)}$ produced from $\mathbf{r}$ by sequentially reading short bitstrings as unsigned integers. We can easily show that if the binary pseudo-random generator $\mathrm{RandGen}$ is secure, then the resulting sequence $\{\boldsymbol \nu_\ell\}_{\ell=1}^{L(\lambda)}$ is indistinguishable from any sequence $\{\boldsymbol \nu^*_\ell\}_{\ell=1}^{L(\lambda)}$ produced from a random binary sequence $\mathbf{r}^* \in \{0,1\}^{160L(\lambda)}$ output by the true entropy collector $\mathrm{TrueRandGen}$.

The rest of the proof essentially repeats the reasoning from the proofs of Theorem~2.9 and Theorem~3.18 in \cite{katz2014introduction}. 

Let $\tilde{\Pi} = (\mathrm{TrueRandGen},\mathrm{Enc},\mathrm{Dec})$ be a new encryption scheme where $\mathrm{RandGen}$ is replaced by $\mathrm{TrueRandGen}$. In the first step, we may prove that the result of enciphering $\{(\tilde{\varepsilon}^*_{(enc),\ell},\tilde{p}^*_{(enc),\ell},\mathbf{\tilde{D}}^*_{(enc),\ell})\}_{\ell=1}^{L(\lambda)}$ with a random sequence $\{\boldsymbol \nu^*_\ell\}_{\ell=1}^{L(\lambda)}$ obtained from $\mathrm{TrueRandGen}$ is perfectly secure, i.e., the enciphered values $\{\tilde{\varepsilon}^*_{(enc),1},\tilde{p}^*_{(enc),1},$ $\mathbf{\tilde{D}}^*_{(enc),1},\tilde{\varepsilon}^*_{(enc),2},$ $...,\mathbf{\tilde{D}}^*_{(enc),L(\lambda)}\}$ are statistically independent and have uniform distributions over their respective discrete domains. Consequently, we get $\Pr(\mathrm{PrivK}^{eav}_{\mathcal{A},\tilde{\Pi}}(\lambda) = 1) = 0.5.$

Then, we can show the indistinguishability of the sequence $\{(\tilde{\varepsilon}_{(enc),\ell},\tilde{p}_{(enc),\ell},\mathbf{\tilde{D}}_{(enc),\ell})\}_{\ell=1}^{L(\lambda)}$ by contradiction: the existence of a PPT adversary $\mathcal{A}$ who distinguishes the sequence from purely random with a non-negligible advantage implies the existence of a PPT distinguisher $\mathcal{D}$ breaking the security of $\mathrm{RandGen}$. From this, we conclude that there is a negligible function $\mathbf{negl}$ such that the advantage of any PPT adversary $\mathcal{A}$ participating in the experiment $\mathrm{PrivK}^{eav}_{\mathcal{A},\Pi}$ is at most ${|\Pr(\mathrm{PrivK}^{eav}_{\mathcal{A},\Pi}(\lambda) = 1) - 0.5 |}$ $\leq  \mathbf{negl}(\lambda)$.
\end{proof-sketch}

Proposition \ref{systemsecurity} states that the encryption scheme produces indistinguishable encryptions in the presence of an eavesdropper, provided that every speech signal is enciphered using a secure pseudo-random bit generator with a fresh and uniformly distributed random seed. However, selecting a proper binary pseudo-random generator is far from being trivial and should be done very carefully. In particular, it is not evident if a `good' bit generator can always be adequately transformed into a non-binary generator and vice-versa. For instance, in \cite{baigneres2007linear} it is shown that a poorly designed non-binary sequence expanded into a bitstream could pass a randomness test by some bit-oriented distinguishers. Some statistical test suitable for checking the randomness of non-binary ciphers can be found in \cite{baigneres2007linear,epishkina2018technique}  
 
The selection of a suitable pseudo-random number generator is out of the scope of this work. Nevertheless, some promising candidates of binary pseudo-random number generators can be found in the NIST Special Publication 800-90A \cite{nist_prng}. Crucially, the presented generators are evaluated for their potential use as non-binary number generators over integer rings $\mathbb{Z}_{2^n}$, $n\in \mathbb{N}$. An example of such a generator uses Advanced Encryption Standard (AES) in the CTR mode of operation and a secret 256-bit seed. The generator is claimed to securely produce up to $2^{48}$ bitstrings of length $2^{19}$ if the input seed is taken uniformly at random. Furthermore, the input seed is updated after every request for backtracking resistance. The maximum bitstring length $2^{19}$ in a unique request is sufficient to encipher more than one minute of one-way voice communication. Finally, a parallelization of bitstring generation provided by the CTR mode is an advantage in real-time operation.

An obvious weakness of the presented scheme is the lack of mechanisms providing data integrity. Since the enciphered speech signal does not include any side information, the recipient cannot verify the source and received data correctness. Moreover, it is not clear whether a reliable data integrity mechanism even exists in this lossy framework, given that the received signal is likely to differ from the initial signal and that malleability-by-design is one of the basic features of the presented speech encryption scheme. Instead, it is important to ensure the proper authentication of the users and secure exchange of cryptographic keys (or secret seeds) before the session starts \cite{canetti2001analysis,katz2014introduction}. Some solutions include mutual authentication using public certificates, symmetric pre-shared keys or a vocal verification \cite{rfc8446,nist_AKE,pasini2006sas,krasnowski2020introducing}. 

Despite the absence of data integrity in real-time communication, an adversarial manipulation on encrypted speech giving a meaningful deciphered speech is technically challenging. Synthetic signal fragility and high synchronization requirements between the legitimate users suggest that the attacker is more likely to interrupt the communication. However, such an interruption is effectively not much different from signal blockage by a VAD. 

If the enciphered speech signal is stored, a binary representation of the signal in PCM or a compressed form should be accompanied by a message authentication code (MAC) \cite{katz2014introduction} computed with a dedicated authentication key. 
  
\subsection{Tolerance to signal distortion and large deciphering errors}

Let $\mathbf{y}_t$ be an encrypted speech signal sent over a voice channel, and $\mathbf{\hat{y}}_t$ be the signal received by the recipient. Due to channel distortion, parameters $(\tilde{\varepsilon}_{(rec),\ell},$ $ \tilde{p}_{(rec),\ell}, \mathbf{\tilde{D}}_{(rec),\ell})$ extracted from $\mathbf{\hat{y}}_t$ usually diverge from the enciphered sequence $(\tilde{\varepsilon}_{(enc),\ell}, \tilde{p}_{(enc),\ell}, \mathbf{\tilde{D}}_{(enc),\ell})$. The transmission error propagates during descrambling, causing a deciphering error between the initial $(\varepsilon_{(init),\ell},$ $p_{(init),\ell},$ $\mathbf{D}_{(init),\ell})$ and deciphered $(\varepsilon_{(dec),\ell},$ $p_{(dec),\ell},$ $\mathbf{D}_{(dec),\ell})$ values. 

When the channel noise is low, transmission and deciphering errors are related by inequalities:
\begin{flalign}
\left|\log_{10}\left(\frac{\varepsilon_{(init),\ell}}{\varepsilon_{(dec),\ell}}\right)\right| \leq &\left( \frac{2^{16}}{\varrho_{high}-\varrho_{low}}\frac{\log_{10}(\varepsilon_{max}/\varepsilon_{min})}{\log_{10}(\tilde{\varepsilon}_{max}/\tilde{\varepsilon}_{min})}\right) & \nonumber \\
&\cdot\left|\log_{10}\left(\frac{\tilde{\varepsilon}_{(enc),\ell}}{\tilde{\varepsilon}_{(rec),\ell}}\right)\right| \label{inequalities1} \\
\left|p_{(init),\ell}-p_{(dec),\ell}\right| \leq &\left(\frac{2^{16}}{\kappa_{high} - \kappa_{low}}\frac{p_{max}-p_{min}}{\tilde{p}_{max}-\tilde{p}_{min}}\right) & \nonumber \\ &\cdot\left|\tilde{p}_{(enc),\ell}-\tilde{p}_{(rec),\ell}\right| \label{inequalities2} \\
\left\|\mathbf{D}_{(init),\ell}-\mathbf{D}_{(dec),\ell}\right\| &\leq \frac{\pi}{\sqrt{2}} \left\|\mathbf{\tilde{D}}_{(enc),\ell}-\mathbf{\tilde{D}}_{(rec),\ell}\right\|
\label{inequalities3}
\end{flalign}
where $|\bullet |$ is the modulus and $\|\bullet\|$ denotes the Euclidean norm. In consequence, the deciphering procedure is distortion-tolerant with respect to parameters $\log_{10}(\tilde{\varepsilon}_{(enc),\ell})$, $\tilde{p}_{(enc),\ell}$ and $\mathbf{\tilde{D}}_{(enc),\ell}$, with three independent expansion factors. 

The distortion-tolerant property with respect to pseudo-speech parameters holds unless the amount of distortion in the received signal $\mathbf{\hat{y}}_t$ becomes too large. When the values $|\log_{10}(\tilde{\varepsilon}_{(enc),\ell}/\tilde{\varepsilon}_{(rec),\ell})|$, $|\tilde{p}_{(enc),\ell}-\tilde{p}_{(rec),\ell}|$ and the distance $\|\mathbf{\tilde{D}}_{(enc),\ell}-\mathbf{\tilde{D}}_{(rec),\ell}\|$ exceed some specific thresholds, there is a risk of a deciphering error much larger than indicated by the bounds. These large deciphering errors are perceived by the listener as unpleasant flutter degrading the overall perceived speech quality, and should be avoided.

\begin{figure}[h]
 \centering
    \includegraphics[width=0.49\textwidth]{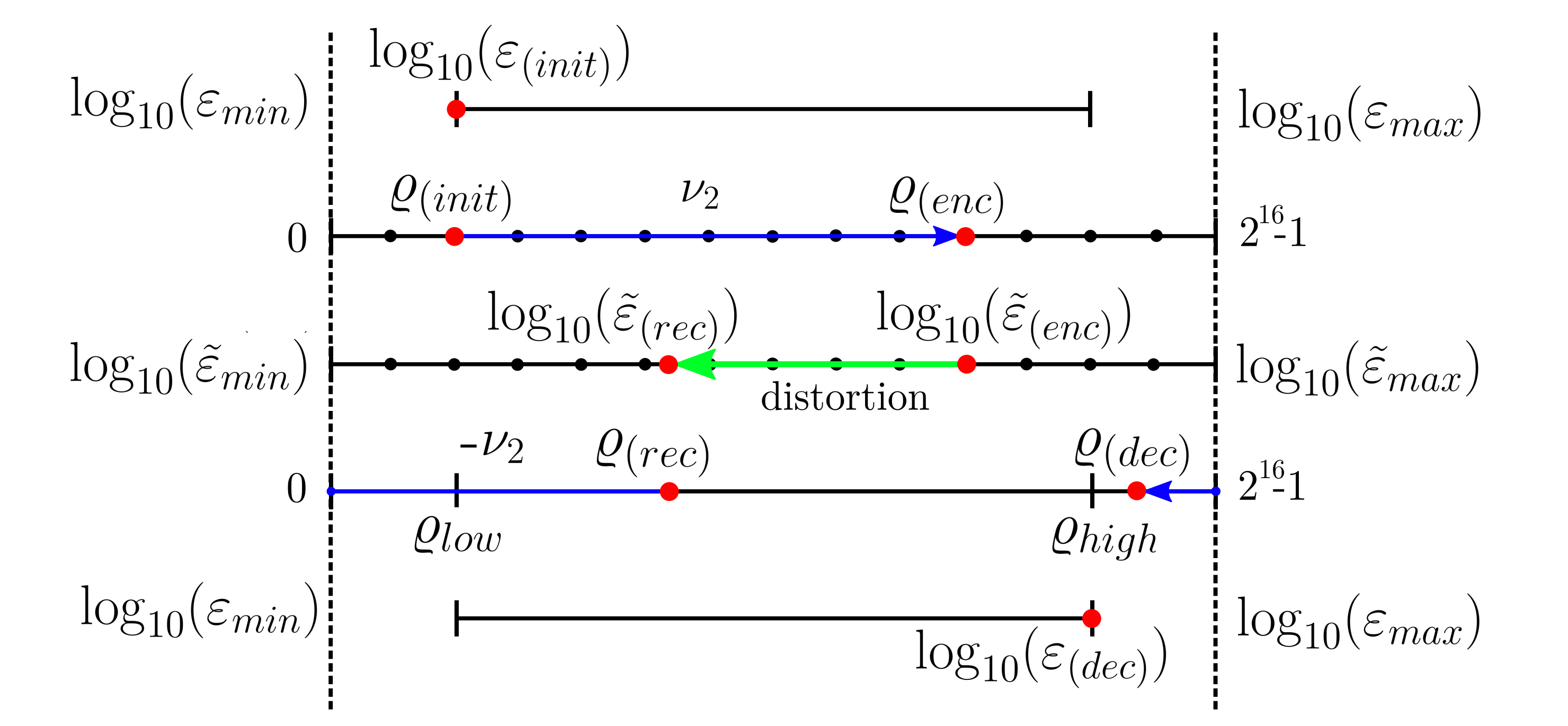}
    \caption{Large deciphering error of energy due to excessive distortion.}
    \label{energy_jump}
\end{figure}

A strong perceptual speech degradation is usually related to large deciphering errors of energy or pitch. In the example depicted in Fig. \ref{energy_jump}, a silent speech frame with the energy $\varepsilon_{(init)} = \varepsilon_{min}$ is enciphered to $\tilde{\varepsilon}_{(enc)}$ and sent over a noisy channel in a form of a pseudo-speech frame. Upon reception, the recipient observes $\tilde{\varepsilon}_{(rec)}$ such that $|\log_{10}(\tilde{\varepsilon}_{(rec)}/\tilde{\varepsilon}_{(enc)}) | > \varrho_{low}/2^{16}$. However, the deciphering result is $\varepsilon_{(dec)} = \varepsilon_{max}$, the exact opposite of the initial value.

Another factor is pitch detection accuracy in the pseudo-speech analyzer. Voice-oriented pitch estimators analyze the signal assuming small pitch variation over time \cite{szczerba2005pitch,rabiner2011theory}. However, the assumption is not valid in an encrypted signal for which the pitch period changes randomly every 20 milliseconds.

The two most common types of pitch estimation errors in noisy signals are transient errors, and octave errors \cite{beauchamp1993detection}. A transient error occurs when an abrupt change of fundamental frequency within a speech frame violates the stationarity assumption. An octave error describes a situation when the predictor incorrectly outputs a multiple $k\omega_0$ or a fraction $1/k\omega_0$ ($k\in \mathbb{N}$) of the correct fundamental frequency $\omega_0$. These errors are mitigated by pitch tracking \cite{benesty2007springer}. However, since the pitch period in the encrypted signal is uncorrelated in time, pitch tracking seems redundant if not harmful in our case. Instead, it is essential to maintain frame synchronization and ensure that neither the adjacent frames nor the guard periods damage the pitch estimation.

It may be noticed that deciphering error making a silent frame maximally loud is more damaging for perceptual quality than suppressing a loud frame into silence. A varying perceptual impact of deciphering errors is the main justification for fine-tuning the guard bounds for pitch and energy. Nevertheless, in order to maintain a robust operation of the enciphering scheme, it is important to ensure experimentally that the values $|\log_{10}(\tilde{\varepsilon}_{(rec),\ell}/\tilde{\varepsilon}_{(enc),\ell}) |$ and $|\tilde{p}_{(rec),\ell}-\tilde{p}_{(enc),\ell} |$ stay within the guard limits with the high probability. 

\begin{figure}[h!]
 \centering
    \includegraphics[width=0.45\textwidth]{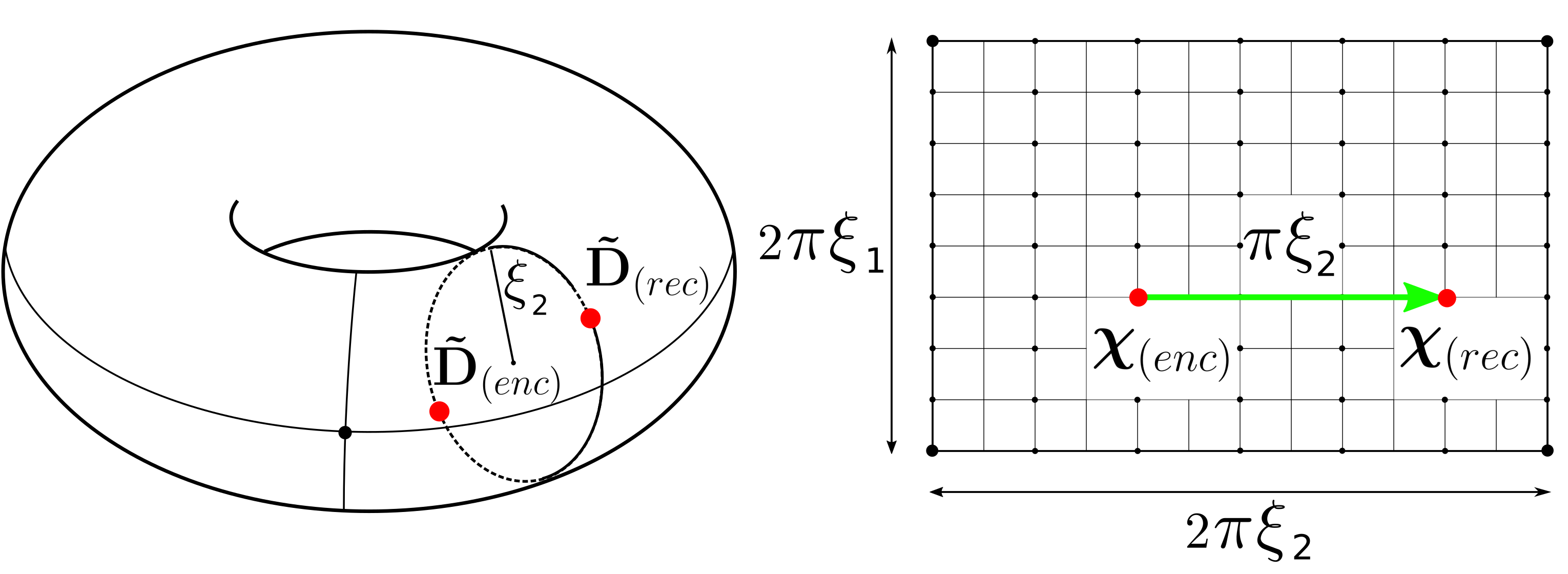}
    \caption{Projection of $\mathbf{\tilde{D}}_{(enc)}$ to the opposite side of the flat torus along single dimension, seen as translation by $\pi \xi_2$ in the pre-image of the torus.}
    \label{timbre_jump}
 \end{figure}

Another kind of a deciphering error occurs while processing the spectral shape of a pseudo-speech frame $\mathbf{\tilde{D}}_{(enc),\ell}$. As already mentioned in Section \ref{subsection-transmission}, channel distortion may cause the received $\mathbf{\hat{D}}_{(enc),\ell} \in S^{15}$ to move away from the flat torus. When the distance from the flat torus overreaches 
\begin{equation}
2\sin\left(\sin^{-1}\left(1/\sqrt{8}\right)/2\right),
\end{equation}
the vector $\mathbf{\hat{D}}_{(enc),\ell}$ could be projected to $\mathbf{\tilde{D}}_{(rec),\ell}$ on the opposite side of the torus. Figure \ref{timbre_jump} illustrates a simplified scenario of a wrong projection in $\mathbb{R}^4$. The projection of the vector $\mathbf{\tilde{D}}_{(enc)}$ to $\mathbf{\tilde{D}}_{(rec)}$ along one dimension of the torus can be viewed as translation of the corresponding coordinate of $\boldsymbol \chi_{(enc)} = \Phi_{\boldsymbol \xi}^{-1}(\mathbf{\tilde{D}}_{(enc)})$ over~$\mathcal{P}_{\boldsymbol \xi}$. A wrong projection causes an unpredictable change in the spectral envelope shape of the deciphered frame.

\vspace{-4mm}
When the channel distortion is sub-proportional to the logarithm of signal energy, the risk of a projection going on the wrong side of the torus can be mitigated by increasing the minimum pseudo-speech frame energy $\tilde{\varepsilon}_{min}$. It is because the norm in the denominator of Eq.~(\ref{received_spectral_shape}) goes up when $\tilde{\varepsilon}_{min}$ is increased, making the error $\|\mathbf{\tilde{D}}_{(enc),\ell}-\mathbf{\tilde{D}}_{(rec),\ell}\|$ relatively smaller.

\subsection{The narrowband LPCNet training data}

 The quality of synthesized speech strongly depends on the capability of the narrowband LPCNet algorithm to operate in more imperfect conditions than during the training \cite{Valin2019lpcnet}. As suggested in \cite{oord2016wavenet}, it is possible to improve the robustness of the network by adding noise during the training stage.

\begin{figure*}[h!]
 \centering
    \includegraphics[width=0.78\textwidth]{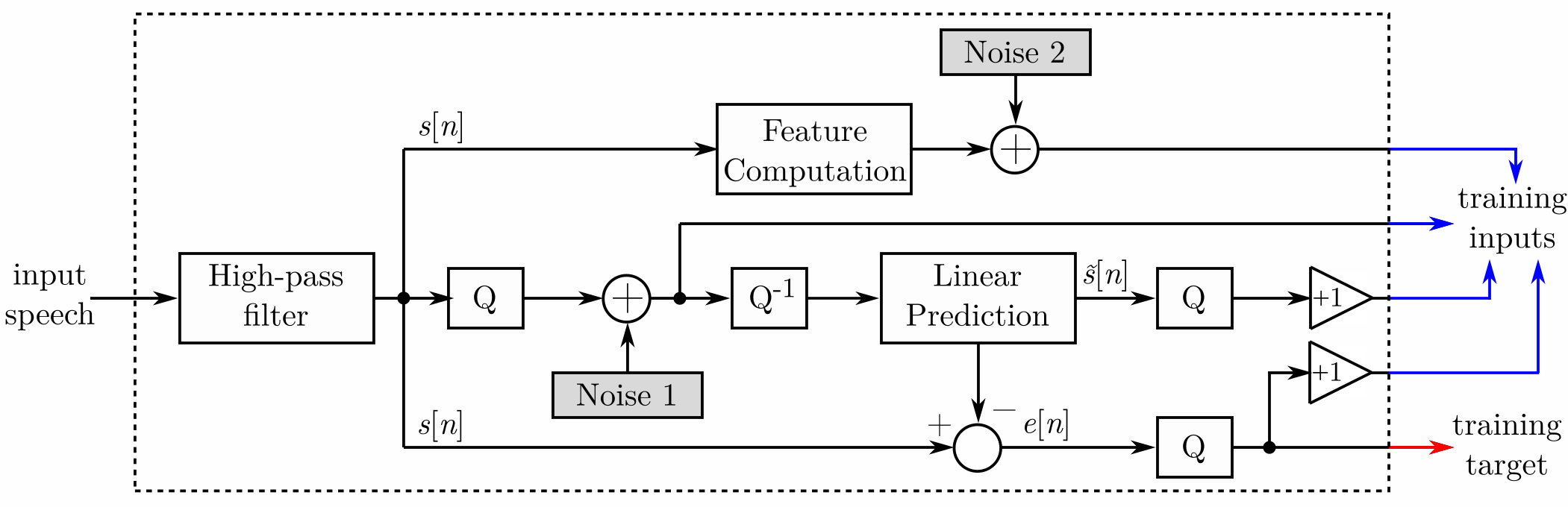}
    \caption{Computing the training data with noise injection to simulate a noisy channel (Noise 2 at the top) and a noisy speech reception (Noise 1 at the bottom). The $\mathrm{Q}$ block denotes $\mu$-law quantization and $\mathrm{Q}^{-1}$ denotes conversion from $\mu$-law to the linear domain. The prediction filter $\tilde{s}[n] = \sum_{k=1}^{12}\tilde{a}_k z^{-k}$ is applied to the noisy and quantized input. The excitation samples $e[n]$ are the difference between the clear speech samples $s[n]$ and the predicted ones. The second noise simulating a channel noise is injected directly into speech parameters. The distribution of noise simulates distortion introduced by a transmission channel and can be obtained by simulations or experiments. Triangle blocks are single sample delays.}
    \label{lpcnet_training3}
\end{figure*}

In our speech encryption scheme, there exist two independent sources of imperfections. The first source is the lower quality of real-world speech recordings taken for encryption, and the second source is channel noise. Motivated by this fact, the training process of the narrowband LPCNet was divided into two stages. During the two-step training, the ML networks consecutively learn to cope with the non-idealities of speech signals and the transmission channel. Splitting the training overcomes several typical problems with learning convergence as when the network cannot effectively compensate for both kinds of distortion at the same time. Moreover, it seems to be more practical if one considers re-training the network to different channel conditions.

A diagram for producing the training data in two steps is shown in Fig. \ref{lpcnet_training3}. The process uses two noise sources, `Noise 1' (active in both training stages) and `Noise 2' (active in the second stage only). The first training stage is nearly identical to the training process described in \cite{Valin2019lpcnet}. During the training, the network learns to predict the excitation sequence $e[n]$, using as input the previous excitation samples $e[n-1]$, the previous signal samples $s[n-1]$, the current prediction samples $\tilde{s}[n]$ and the frame-rate speech features (9-band Bark-scale cepstral coefficients, pitch). Except for the frame features, the input data is $\mu$-law quantized by the Q block \cite{ITU-T1988pulse}. The input noise is injected into the speech signal in the $\mu$-law domain to make it proportional to signal amplitude. Noise distribution varies across the training data from no noise to a uniform distribution in the [-3,3] range. It can be noticed that the injected noise propagates to all sample-rate input data, effectively imitating speech recorded in noisy environments.

After the first training stage, the network can produce intelligible speech signals from noiseless feature vectors. However, the output of the speech encryption scheme is likely to be distorted. The second stage of the training simulates a scenario when the frame-rate features are transmitted in encrypted form over a noisy channel. Since channel distortion is independent of the input signal, and the reception error linearly propagates to deciphered values, it is possible to inject noise directly into the speech parameters, as in Fig. \ref{lpcnet_training3}. The distribution of injected noise simulates channel distortion statistics obtained by simulations or measurements.

The two-stage training of the networks on a GPU card Nvidia Quadro RTX 4000 and using one hour of training speech takes approximately five days. Furthermore, we experimented with the speech quality produced by the synthesizer trained to English or Japanese\footnote{Japanese has a very simple phonology, that makes it particularly useful for experimenting with ML techniques.} language. The results obtained suggest that the synthesizer should be trained to the language used later for secure communication.

\section{Evaluation}
\label{evaluation}

This section presents the test results from some experiments with our speech enciphering algorithm. The tests verify the scheme's capability to decrypt pseudo-speech signals distorted by noise. Furthermore, the section investigates a scenario when the receiver is not fully synchronized in time and amplitude with the sender. The simulations are validated by real-world experiments.

Based on some measurements of a signal distortion introduced by FaceTime and Skype, we estimated the SNR in a typical VoIP-based voice channel to be between 10 dB and 15 dB. On the other hand, similar experiments with 3G networks revealed that signal distortion in cellular networks is much higher, and gives SNR values closer to 3-5 dB. Due to excessive noise in cellular networks, we decided to evaluate our encryption scheme for its compatibility with VoIP-based applications. The robustness of deciphering was evaluated by inserting additive white Gaussian noise (AWGN) into an encrypted signal or compressing the encrypted signal with Opus-Silk 1.3.1 \cite{opus_ietf}. Opus-Silk was chosen for experimentation because, unlike AMR \cite{amr_3gpp} or Speex \cite{speex_ietf}, its compression rate can be easily adjusted. 

The precomputed encrypted signal was successfully sent over FaceTime between two iPhones~6 running iOS~12 connected to the same WiFi network and decrypted offline. The use of FaceTime on WiFi is justified by high connection stability (limited drop-outs, constant delay) which greatly simplifies signal synchronization at the receiving end. Additionally, the selected speech excerpts reconstructed from encrypted signals were evaluated in a speech quality/intelligibility assessment on a large group of about 40 participants.

The section concludes with computational analysis in Section~\ref{section_complexity}. The system's computational complexity was estimated by measuring all floating-point operations performed during running our experimental software. The measurements suggest that the computationally optimized encryption algorithm may operate in real-time on high-end portable devices.

Selected initial, encrypted, distorted, and decrypted speech samples are available online.\footnote{\url{https://github.com/PiotrKrasnowski/Speech_Encryption}}

\subsection{Experimental setup} 

Table \ref{speech_parameters} and Table \ref{pseudospeech_parameters} present the encoding parameters of speech and pseudo-speech signals. The intervals $[\varepsilon_{min}, \ \varepsilon_{max}]$ and $[p_{min}, \ p_{max}]$ were obtained from the TSP English speech corpus.\footnote{\url{https://www-mmsp.ece.mcgill.ca/Documents/Data/}} The selection of the intervals $[\tilde{\varepsilon}_{min}, \  \tilde{\varepsilon}_{max}]$, $[\tilde{p}_{min}, $ $\tilde{p}_{max}]$ and the bounds $[\kappa_{low}, \ \kappa_{high}]$, $[\varrho_{low}, \ \varrho_{high}]$ was done based on simulations.

The speech encryption and decryption algorithms were implemented mainly in Python. The speech encoder and the speech synthesizer were obtained from the LPCNet repository\footnote{\url{https://github.com/mozilla/LPCNet/}} and adapted to the scheme. The pitch prediction with tracking for speech was based on open-loop cross-correlation search \cite{rabiner2011theory}, whereas prediction for pseudo-speech relies on a more accurate maxi-mum-likelihood estimator\footnote{\url{https://github.com/jkjaer/fastF0Nls/}} without tracking \cite{NIELSEN2017188,nielsen2013bayesian}. In the simulations, the enciphering stage takes as input a given pseudo-random bitstring produced by a built-in NumPy\footnote{\url{https://numpy.org/}} random sequence generator.

\begin{table}[h!]
\centering
\caption{Parameters used for speech encoding and synthesis.}
\begin{tabular}{|p{3cm}|p{4.5cm}|}
\hline 
 Parameter & Value \\ \hline \hline
 frame length & 20 ms \\
 sampling frequency & 8 kHz \\
 sample representation & int16 \\
 energy bounds & $(\varepsilon_{min}, \ \varepsilon_{max}) = (10, \ 10^8)$ \\
 pitch period bounds & $(p_{min}, \ p_{max}) = (16, \ 128)$ \\
 energy guard bounds & $(\varrho_{low}, \ \varrho_{high}) = (2^{13}, \ 2^{16}-2^{13}-1)$ \\
 pitch guard bounds & $(\kappa_{low}, \ \kappa_{high}) = (2^{13}, \ 2^{16}-2^{13}-1)$ \\ \hline     
\end{tabular}
\label{speech_parameters}
\end{table}
\vspace{-4mm}
\begin{table}[h!]
\centering
\caption{Parameters used for pseudo-speech encoding and synthesis.}
\begin{tabular}{|p{3cm}|p{4.5cm}|}
\hline
 Parameter & Value \\ \hline \hline
 frame length  & 25 ms \\
 guard period & 5 ms \\ 
 sampling frequency & 16 kHz \\
 sample representation & int16   \\
 energy bounds & $(\tilde{\varepsilon}_{min}, \ \tilde{\varepsilon}_{max}) = (10^9, \ 10^{10})$ \\
 pitch period bounds & $(\tilde{p}_{min}, \ \tilde{p}_{max}) = (80, \ 160)$ \\  \hline   
\end{tabular}
\label{pseudospeech_parameters}
\end{table}

The narrowband LPCNet was trained in two steps on one hour of speech from the multi-speaker TSP English corpus (12 male and 12 female speakers). In the second step of the training, inserted noise simulated channel distortion caused by a Gaussian noise at SNR = 20 dB. Each network was trained for 100 epochs per training step, with a batch consisting of 64 speech sequences of 300 ms. The training was performed on a GPU card Nvidia Quadro RTX 4000 with Keras\footnote{\url{https://keras.io/}} and Tensorflow\footnote{\url{https://www.tensorflow.org/}} using the CuDNN GRU implementation. The selected optimization method was AMSGrad \cite{Sashank2018Convergence} with a step size $\alpha = \frac{\alpha_0}{1+\delta \cdot b}$, where $\alpha_0 = 0.001$, $\delta = 5\times 10^{-5}$ and $b$ is the batch number.

\subsection{Simulations}

\begin{figure*}[h!]
\centering
  \subfloat{
\includegraphics{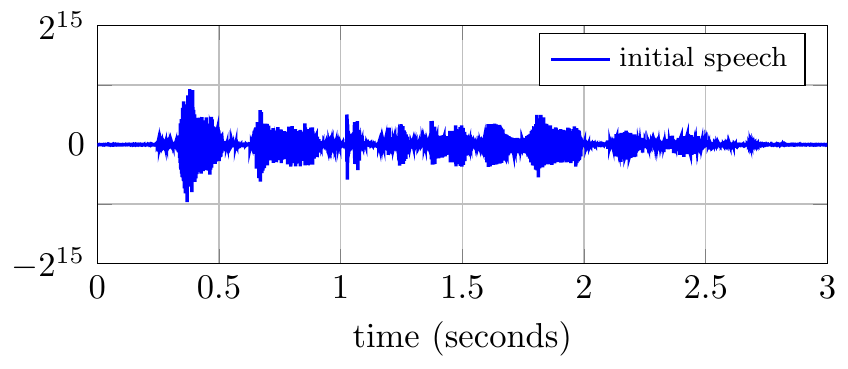}
  }
  \subfloat{
\includegraphics[height=3.8cm, width = 9cm]{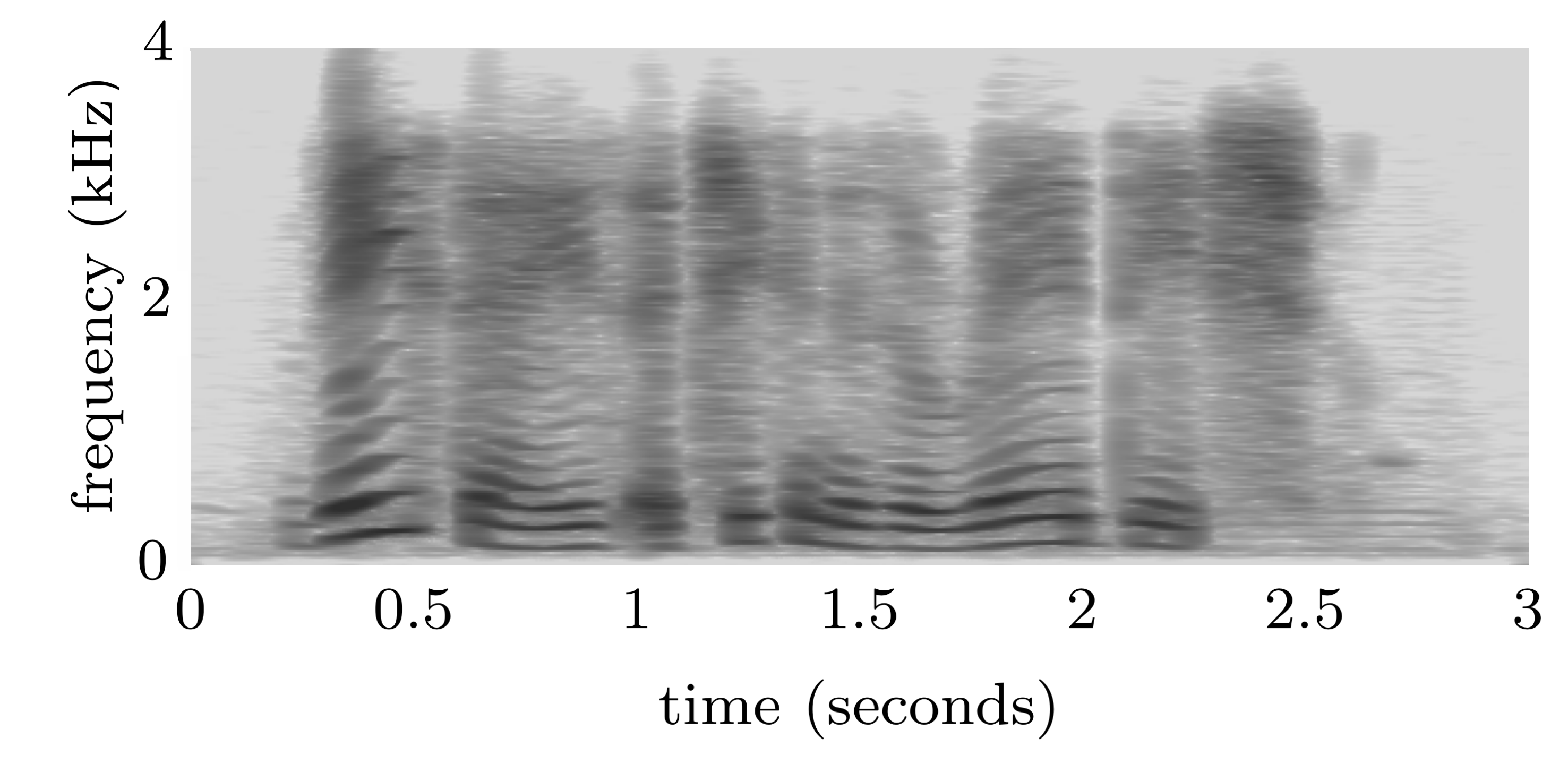}
  }
  \\
 \subfloat{
\includegraphics{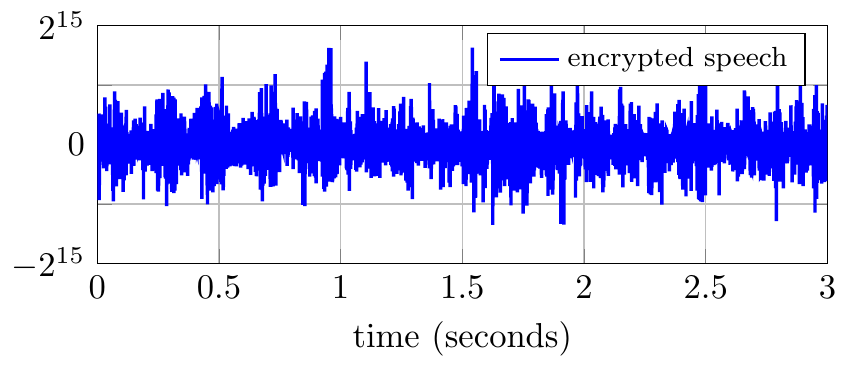}
 }
 \subfloat{
\includegraphics[height=3.8cm, width = 9cm]{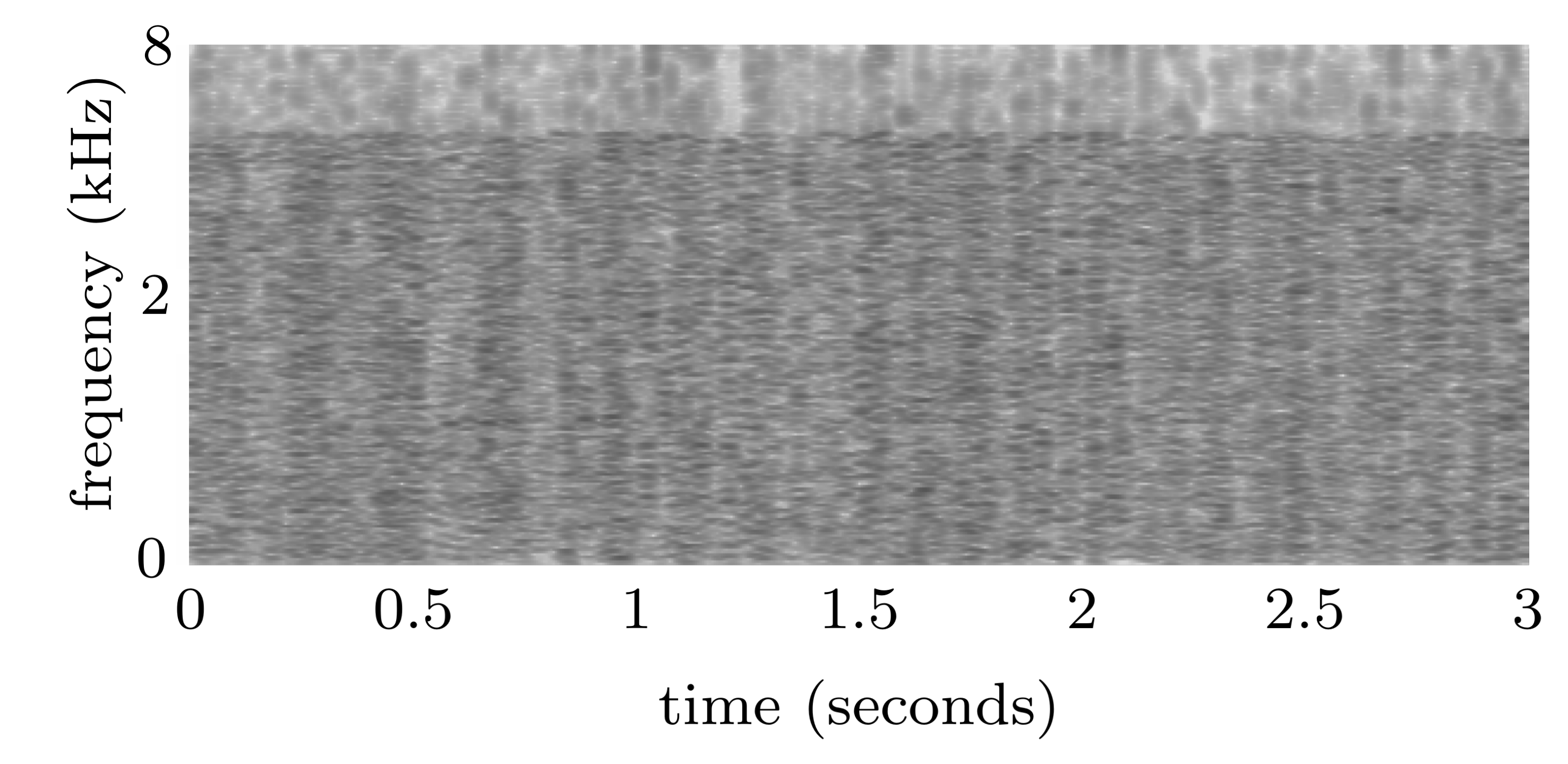}
 }
 \\
 \subfloat{
\includegraphics{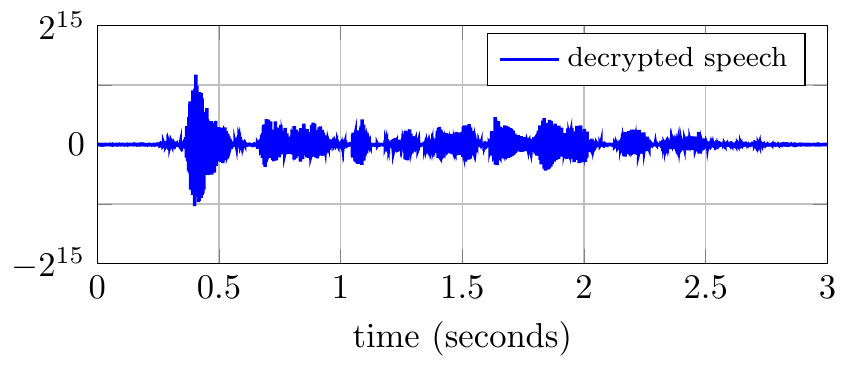}
 }
  \subfloat{
\includegraphics[height=3.8cm, width = 9cm]{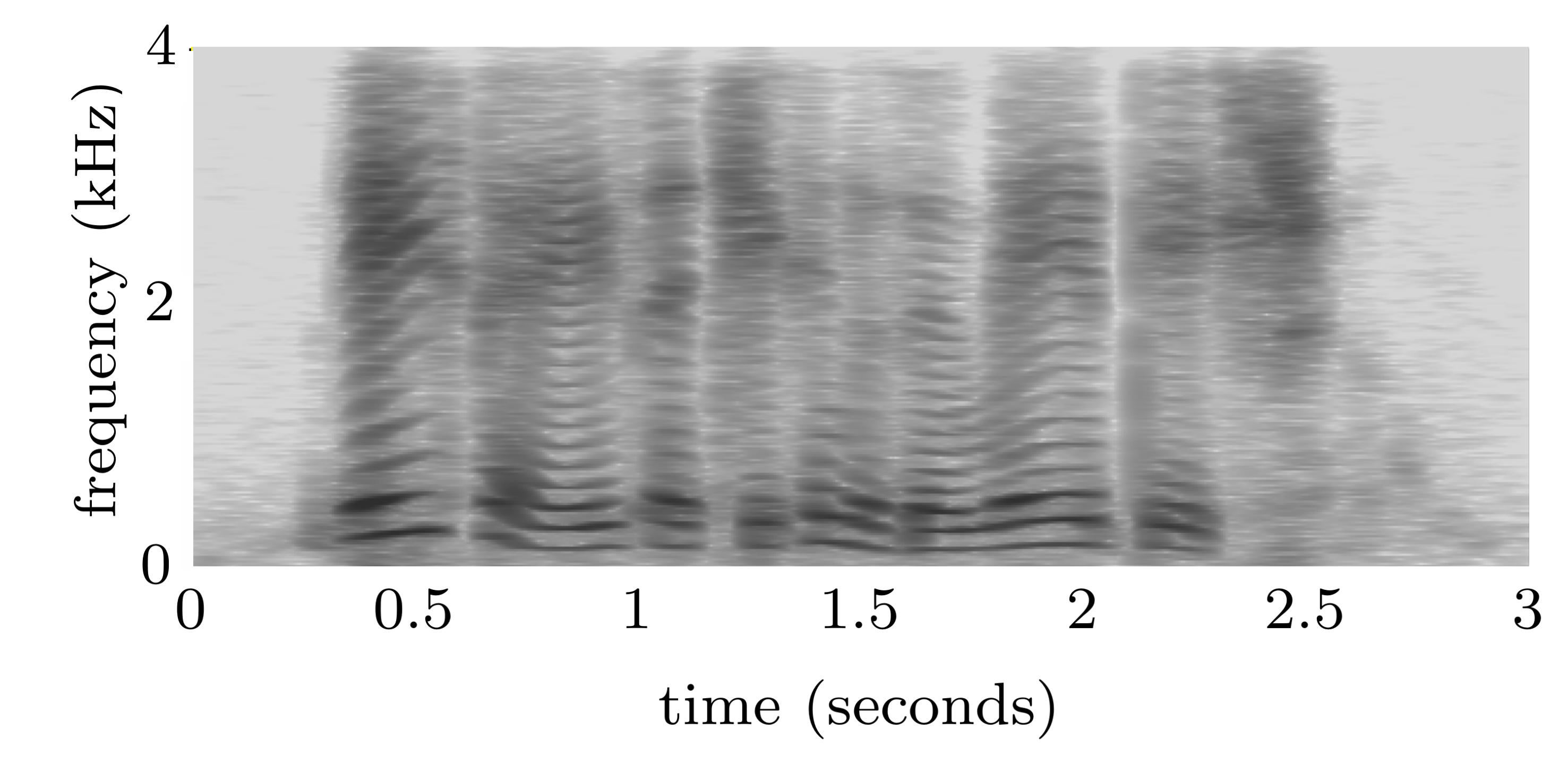}
 }
  \caption{Waveforms at different stages of signal encryption. From top to bottom: initial speech, encrypted signal and resynthesized speech. The spectrograms were obtained using a 1024-point Hann window.}
  \label{perfect_channel_time}
\begin{minipage}[t]{\textwidth}
\centering
  \subfloat{
\includegraphics{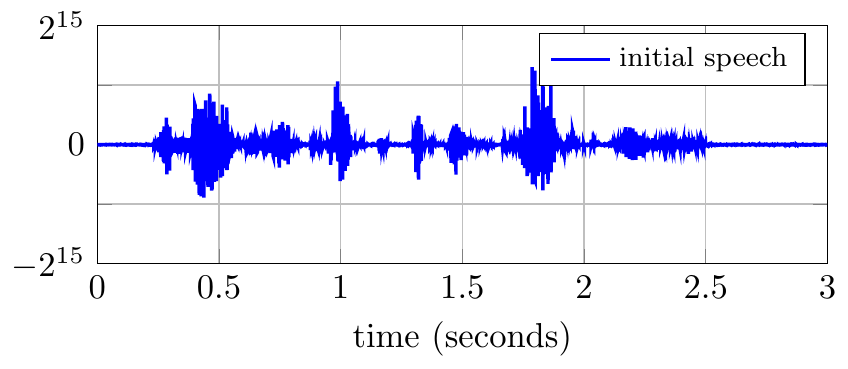}
  }
 \subfloat{
\includegraphics{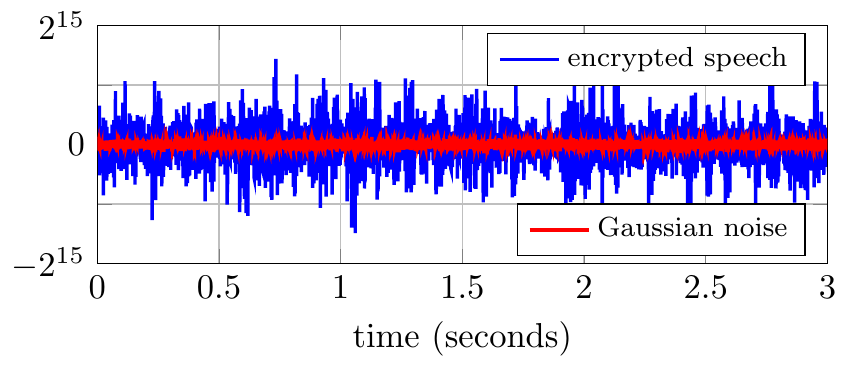}
}
 \\
 \subfloat{
\includegraphics{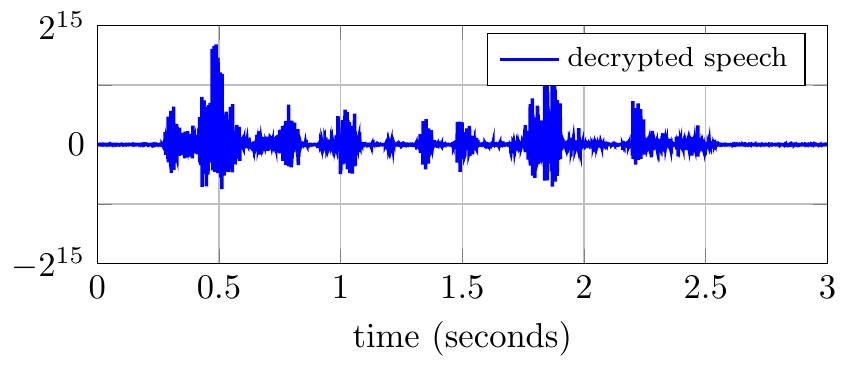}
 }
 \caption{Waveforms at different stages of signal encryption with Gaussian noise at SNR = 15~dB added to encrypted pseudo-speech.}
 \label{speech_15snr}
 \end{minipage}   
\end{figure*}

The first experiment tested the encryption and decryption operations, assuming noise-less transmission. In the example in Fig.~\ref{perfect_channel_time}, the time-domain envelopes of the initial and the reconstructed speech sentence are very similar. A high degree of similarity can also be observed in the spectrograms. It may be noticed that the trained speech synthesizer faithfully reconstructs the fundamental frequency and the formants of the initial speech. On the other hand, the encrypted signal in the time and the frequency domains resembles band-limited noise.
 
 Adding distortion into the encrypted signal degrades the decrypted speech. The time-domain envelope of the decryp-ted speech sentence in Fig.~\ref{speech_15snr} is still similar to the initial speech but not identical anymore. It may be observed that pseudo-speech decryption has a denoising effect on low-amplitude speech and silence. 

\begin{figure}[h!]
\centering
  \subfloat{
\includegraphics{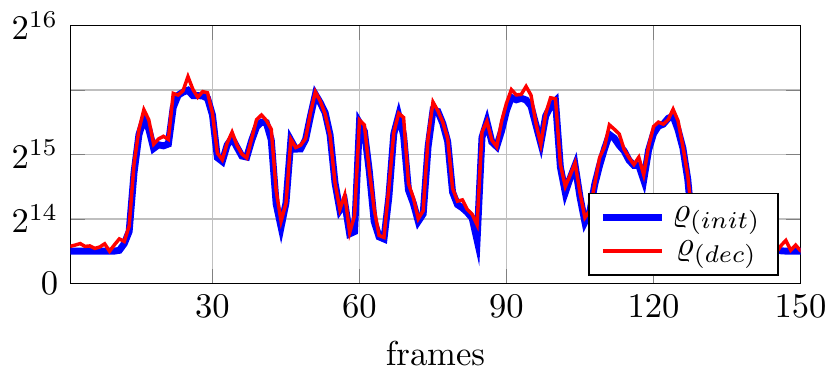}
  } \\
 \subfloat{
\includegraphics{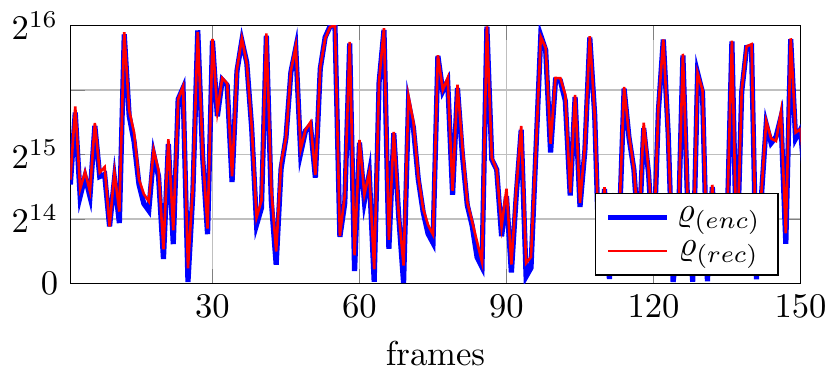}
 } \\
  \subfloat{
\includegraphics{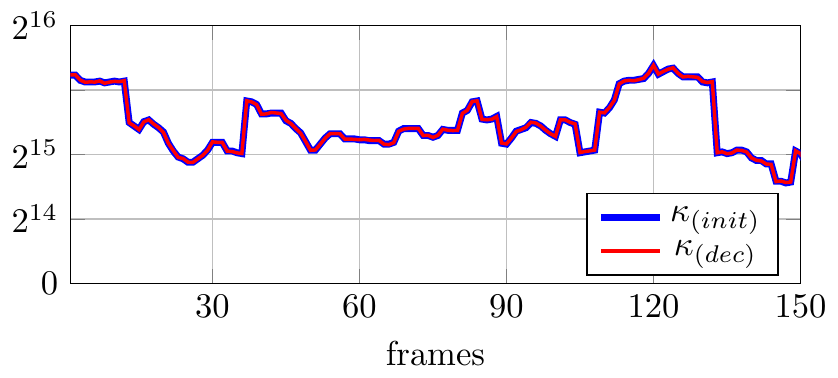}
  }
  \\
 \subfloat{
\includegraphics{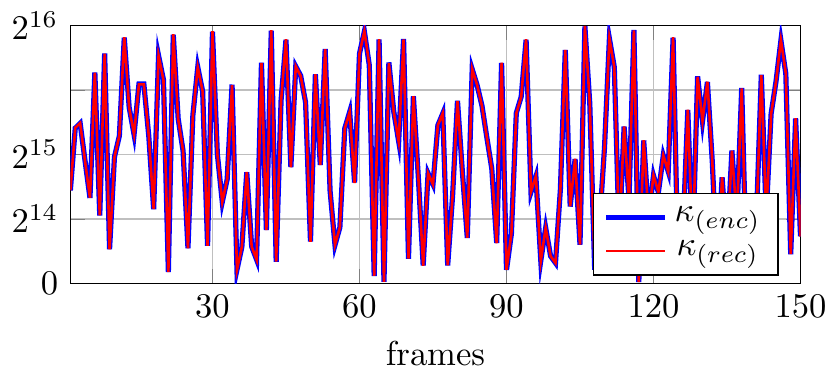}
 }
\\
\subfloat{
\includegraphics{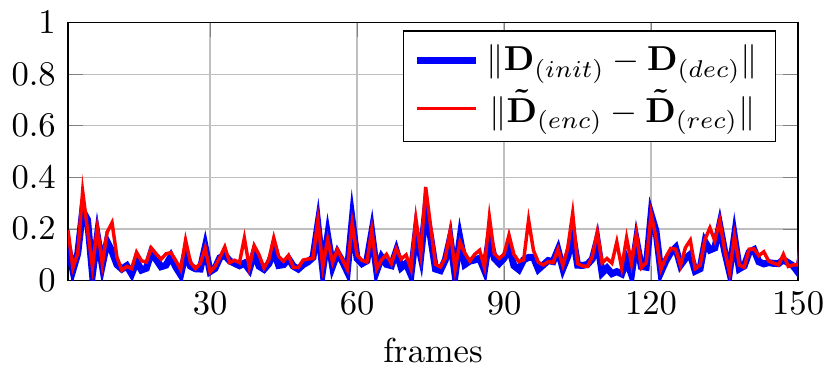}
  }
  \caption{Distortion of received and deciphered parameters caused by adding Gaussian noise at SNR = 15 dB to encrypted speech.}
  \label{parameters_15snr}
\end{figure}

The reception and deciphering errors of the same speech sentence are depicted in Fig. \ref{parameters_15snr}. As can be seen, the errors on energy and timbre are non-negligible. However, in contrast to the error $|\varrho_{(init),\ell}-\varrho_{(dec),\ell}|$, the impact of the error $\|\mathbf{D}_{(init),\ell}-\mathbf{D}_{(dec),\ell}\|$ on decrypted speech perception is more unpredictable. Unlike energy and timbre, pitch is very well preserved. 

The scheme's robustness has been tested against AWGN at SNR between 5-25 dB and Opus-Silk v1.3.1 compression at bitrates between 28-64 kbps. In each case, the error of received and deciphered parameters were expressed in terms of the RMSE defined as:
\begin{align*}
\mathrm{RMSE}_{\tilde{\varepsilon},(rec)} &= \sqrt{\frac{1}{L}\sum\limits_{\ell=1}^L|\varrho_{(enc),\ell}-\varrho_{(rec),\ell}|^2},  \\ \mathrm{RMSE}_{\varepsilon,(dec)}  &= \sqrt{\frac{1}{L}\sum\limits_{\ell=1}^L|\varrho_{(init),\ell}-\varrho_{(dec),\ell}|^2}, \\
\mathrm{RMSE}_{\tilde{p},(rec)} &= \sqrt{\frac{1}{L}\sum\limits_{\ell=1}^L|\kappa_{(enc),\ell}-\kappa_{(dec),\ell}|^2},  \\ \mathrm{RMSE}_{p,(dec)} &= \sqrt{\frac{1}{L}\sum\limits_{\ell=1}^L|\kappa_{(init),\ell}-\kappa_{(dec),\ell}|^2}, \\
\mathrm{RMSE}_{\mathbf{\tilde{D}},(rec)} &= \sqrt{\frac{1}{L}\sum\limits_{\ell=1}^L\|\mathbf{\tilde{D}}_{(enc),\ell}-\mathbf{\tilde{D}}_{(rec),\ell}\|^2}, \\ \mathrm{RMSE}_{\mathbf{D},(dec)} &= \sqrt{\frac{1}{L}\sum\limits_{\ell=1}^L\|\mathbf{D}_{(init),\ell}-\mathbf{D}_{(dec),\ell}\|^2}.
\end{align*}

\begin{figure*}[h!]
    \centering
      \subfloat{
\includegraphics{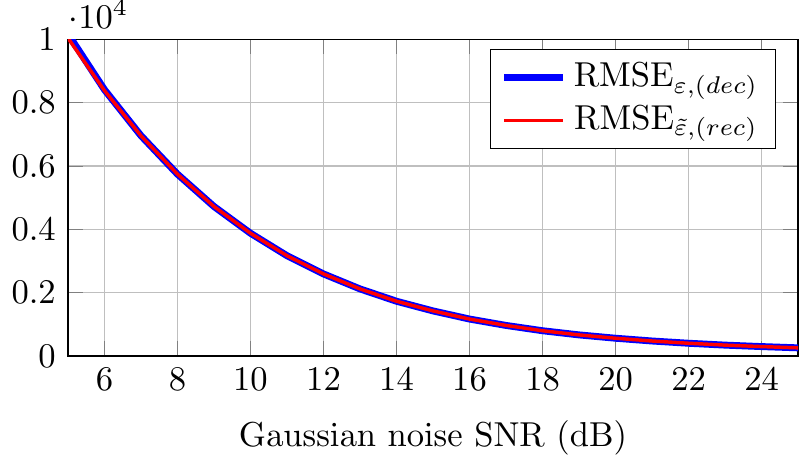}
  }
  \hfill
 \subfloat{
\includegraphics{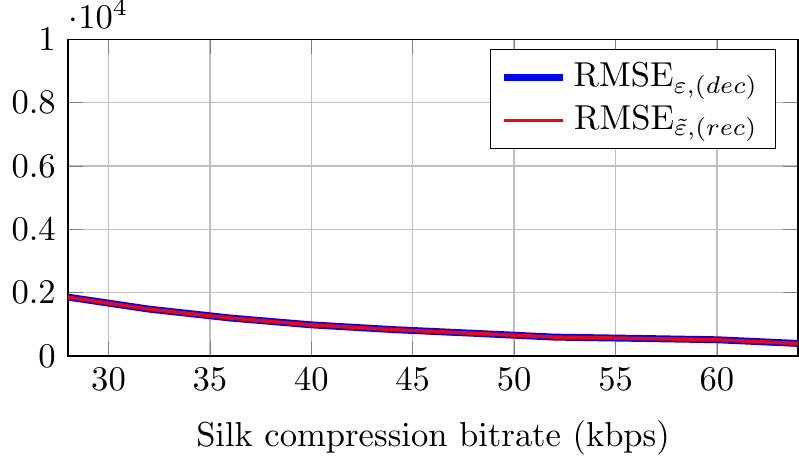}
 } \\
  \subfloat{
\includegraphics{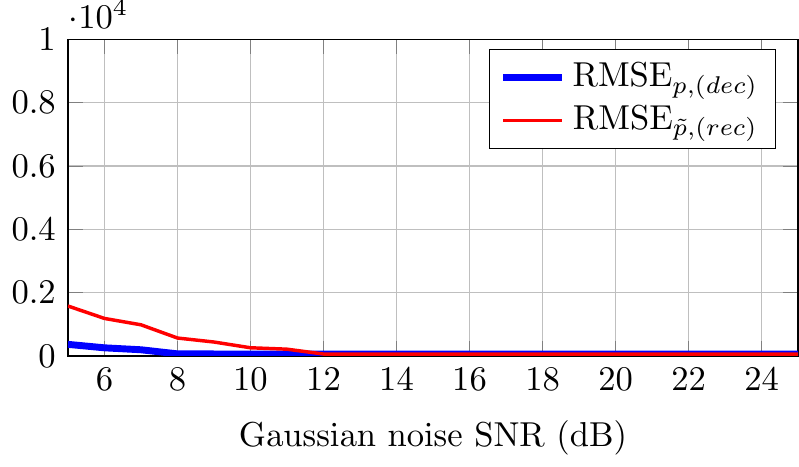}
  }
  \hfill
 \subfloat{
\includegraphics{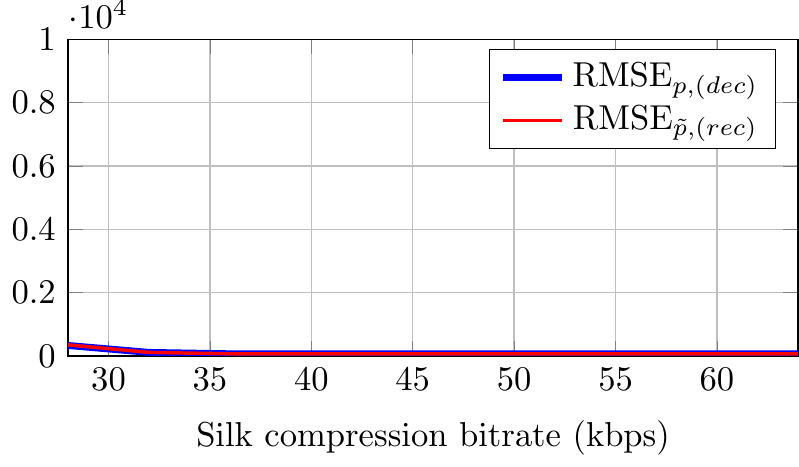}
}
\\
  \subfloat{
\includegraphics{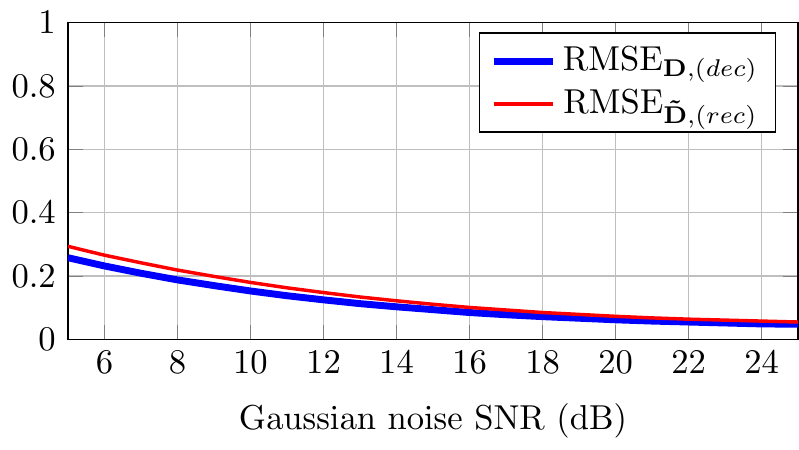}
  }
  \hfill
 \subfloat{
\includegraphics{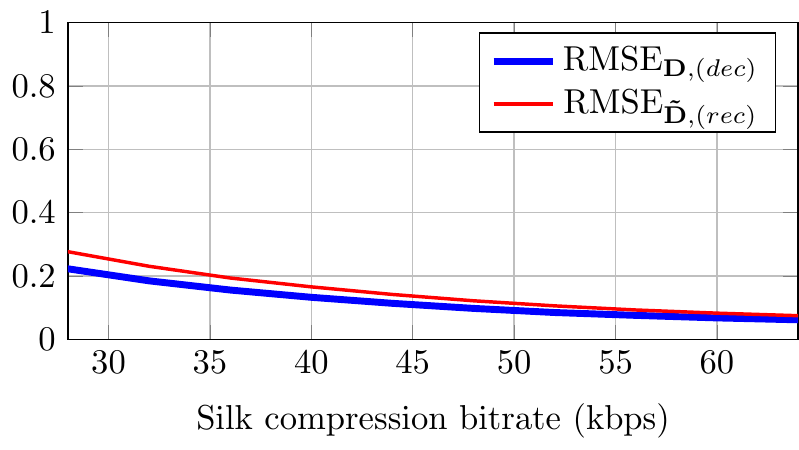}
 }
  \caption{Root mean squared error (RMSE) of deciphered speech values and received pseudo-speech values caused by adding Gaussian noise to encrypted speech (left column) or by compressing the encrypted speech with Opus-Silk (right column). Simulation based on 100000 frames.}
\label{rmse_speech}
    \begin{minipage}[t]{0.49\textwidth}
        \centering
\includegraphics{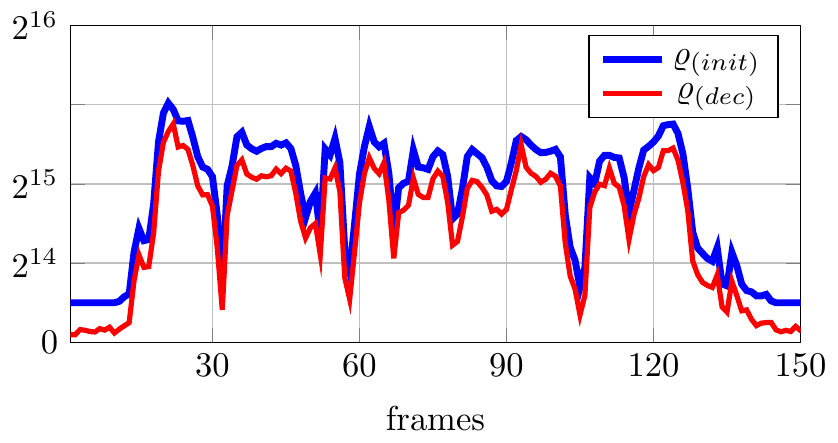}
        \caption{Deciphered speech energy from encrypted signal \\ scaled by the factor 0.85 and distorted by Gaussian noise at \\ SNR = 20~dB.}
        \label{energy_scaling}
    \end{minipage}\hfill
    \begin{minipage}[t]{0.49\textwidth}
        \centering
\includegraphics{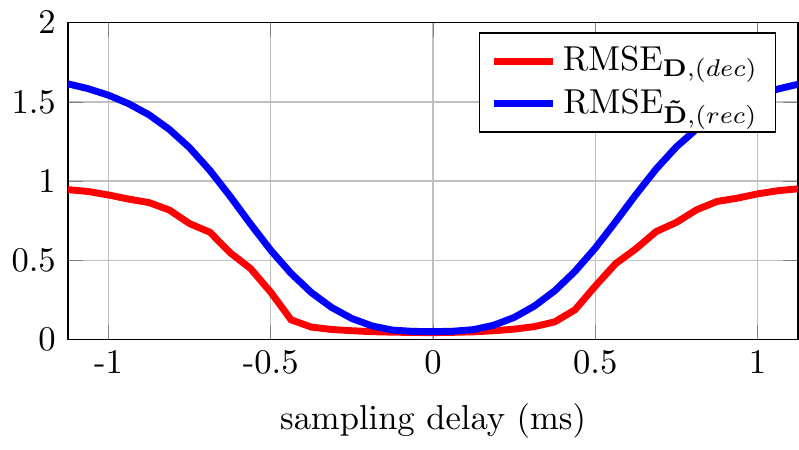}
        \caption{RMSE of received and deciphered vectors represent- \\ ing spectral envelope shape; caused by imperfect sampling syn- \\ chronization on the receiving side. Simulation based on 100000 \\ frames.}
        \label{synchro_mismatch}
    \end{minipage}
\end{figure*}

As shown in Fig. \ref{rmse_speech}, $\mathrm{RMSE}_{\varepsilon,(dec)}$ and $\mathrm{RMSE}_{\mathbf{D},(dec)}$ gradually rise when the signal distortion goes up. However, the nearly perfect alignment of $\mathrm{RMSE}_{\varepsilon,(rec)}$ and $\mathrm{RMSE}_{\varepsilon,(dec)}$ suggests that the impact of large deciphering errors on energy is statistically negligible. In consequence, the guard bounds $(\varrho_{low}, \ \varrho_{high})$ could be relaxed. Additionally, it can be noticed that the error $\mathrm{RMSE}_{\mathbf{\tilde{D}},(rec)}$ is smaller than $\mathrm{RMSE}_{\mathbf{D},(dec)}$. It is because the spherical angles $\mathbf{D}_{(dec)} = \gamma^{-1}_{\boldsymbol \xi}(\boldsymbol \chi_{(dec)}/{2})$ are divided by $2$ in the decoding stage.

The error $\mathrm{RMSE}_{p,(dec)}$ remains small for every analyzed distortion. The rarely occurring errors on pitch are usually significant and easy to detect. The observation suggests that a simple pitch tracker added at the output of the descrambling block would overperform guard bounds $(\kappa_{low}, \ \kappa_{high})$ as an error correction mechanism.

\begin{figure}[h!]
\centering
  \subfloat{
\includegraphics{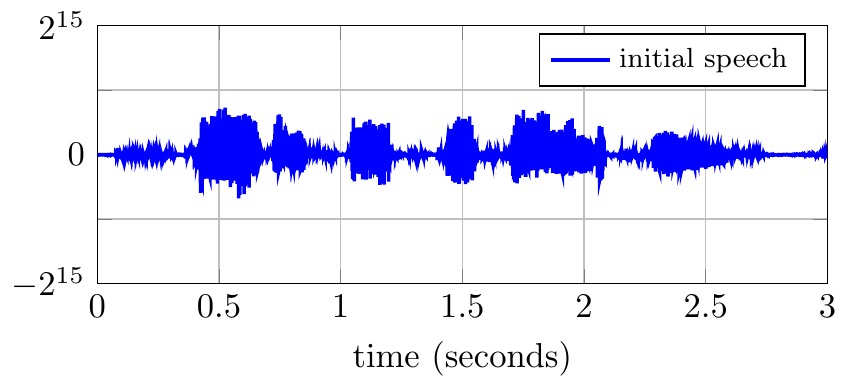}
  }
  \\
 \subfloat{
\includegraphics{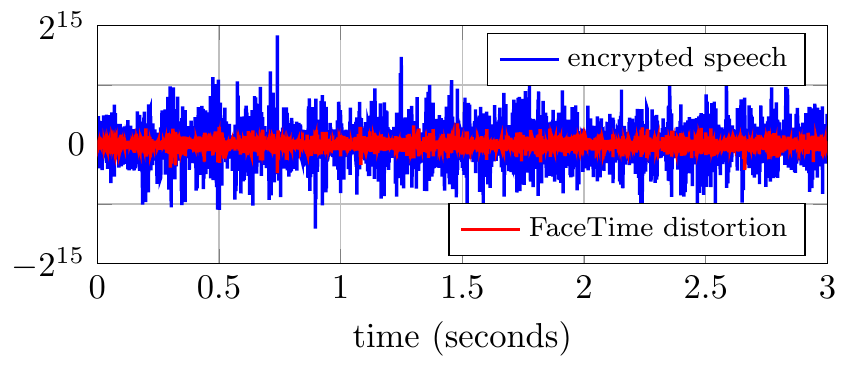}
 }
 \\
 \subfloat{
\includegraphics{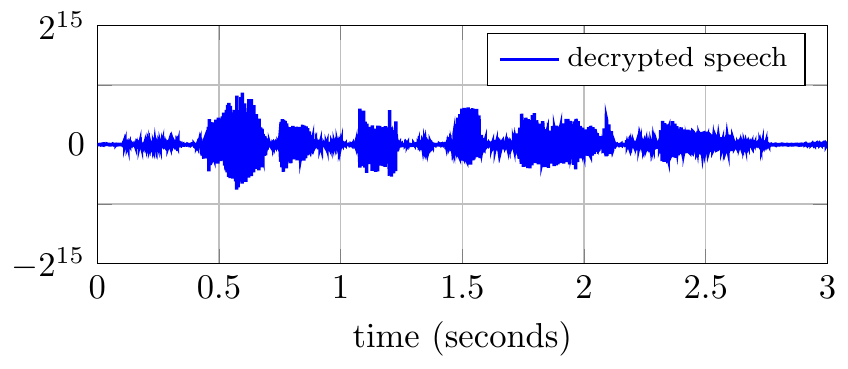}
 }
  \caption{Consecutive stages of signal encryption in communication over FaceTime between two iPhones 6. The recordings are available online at \url{https://github.com/PiotrKrasnowski/Speech_Encryption}.}
  \label{FaceTime_call}
\end{figure}

In a realistic scenario, the receiver is not always perfectly synchronized in time with the sender. Moreover, some voice channels equipped with adaptive gain control (AGC) may modify the signal amplitude. As suggested by Fig. \ref{energy_scaling}, the deciphering unit is, to some extent, tolerant of energy mismatch in the encrypted signal caused by AGC. Provided that the mismatch is no larger than the energy guard intervals, a modified signal is decrypted into an energy-scaled speech. On the other hand, the deciphering unit is very vulnerable to synchronization error larger than 0.3 ms, as shown in Fig.~\ref{synchro_mismatch}.

Finally, the scheme's robustness was checked over FaceTime on WiFi between two iPhones~6 running iOS~12. The precomputed pseudo-speech excerpts of total duration 120 seconds and enciphered with a predefined pseudo-random sequence were uploaded on one of the phones, sent over FaceTime in chunks about 10-20 second long, and recorded on the second device. Figure~\ref{FaceTime_call} illustrates an example of the recorded signal and the decrypted speech. Table  \ref{rmse_faceTime} lists the RMSE of received and deciphered parameters retrieved from 120 seconds of a recorded signal. 
\begin{table}[h]
\caption{RMSE of received and deciphered values in communication over FaceTime between two iPhones 6. Results retrieved from 120 seconds of a recorded signal.}
\centering
\begin{tabular}{|c|c||c|c|}
\hline
$\mathrm{RMSE}_{\varepsilon,(dec)}$ & 1493.30 & $\mathrm{RMSE}_{\tilde{\varepsilon},(rec)}$ & 1644.73  \\ \hline 
$\mathrm{RMSE}_{p,(dec)}$ & 525.70 & $\mathrm{RMSE}_{\tilde{p},(rec)}$  & 867.30 \\ \hline
$\mathrm{RMSE}_{\mathbf{D},(dec)}$ & 0.12 & $\mathrm{RMSE}_{\mathbf{\tilde{D}},(rec)}$ & 0.16 \\ \hline
\end{tabular}
\label{rmse_faceTime}
\end{table}

\newpage

\subsection{Speech quality evaluation}

As reported in \cite{kleijn2018wavenet}, objective measures of speech quality (i.e., PESQ \cite{ITU-T2001P862} and POLQA \cite{ITU-T2018P863}) are suboptimal for evaluating Machine Learning based, non-waveform vocoders. Consequently, we conducted a subjective listening test on a large number of participants. The tests consisted of two parts. The first part checked the subjective intelligibility of decrypted speech in perfect transmission conditions. The second part assessed the subjective quality of speech restored from an encrypted signal with different distortion types. The subset of speech samples used in the listening test has been selected from the LibriSpeech corpus \cite{panayotov2015librispeech} and is available online.

\begin{figure}[h!]
\centering
\includegraphics[width=0.49\textwidth]{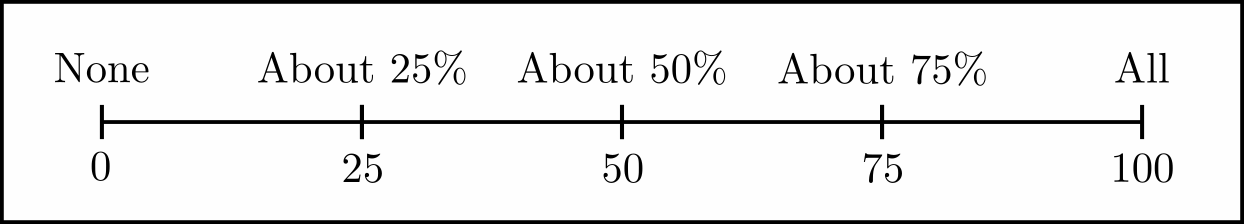}
  \caption{Rating scale used in the speech intelligibility test.}
  \label{test1_scale}
\end{figure}

The intelligibility experiment was inspired by the speech intelligibility rating (SIR) \cite{cox1989development}. During the test, the participants listened to 10 English sentences (4 female and 4 male speakers) of about 10 seconds each. In the first round, the speech utterances were consecutively encrypted and decrypted, without noise insertion. In the second round, listeners were given the initial sentences sampled at 8 kHz, which served as the reference. After listening to each speech sample, the participants were asked to estimate the percentage of recognized words in the sentence. The ratings were defined as numbers between 0 and 100, where 0 denoted no recognized word and 100 denoted that all words were recognized (Fig. \ref{test1_scale}). As opposed to rigorous, one-word or vowel/consonant intelligibility tests \cite{ITU-T2016P807}, testing the word intelligibility of a sentence allows listeners to take advantage of the context. Because the participants were non-native English speakers, they were allowed to listen to the sentences multiple times.

\begin{figure}[h!]
\centering
\includegraphics[width=0.49\textwidth]{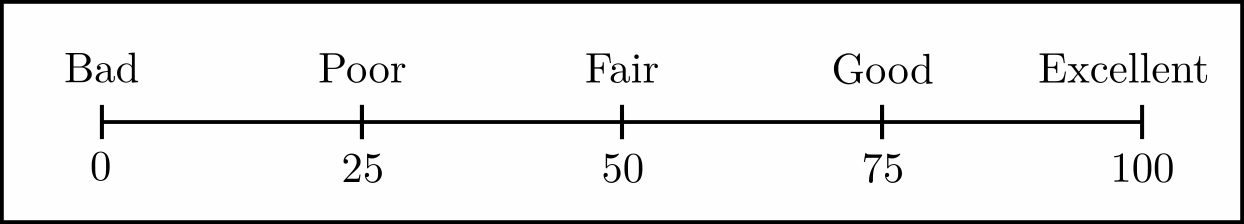}
   \caption{Rating scale used in the speech quality test.}
  \label{test2_scale}
\end{figure}

The quality assessment followed a MUSHRA methodology \cite{ITU-R2015BS} adapted for perceptual evaluation of medium quality speech signals. The method is believed to provide more reliable and reproducible results than the Mean Opinion Score (MOS) measure \cite{ITU-T2016p801}, although it is not immune to biases either \cite{zielinski2007potential}. In the MUSHRA test, a participant is given several test audio files (called excerpts) which represent the same speech utterance processed by different algorithms. To allow the participant a thorough and unbiased evaluation, these excerpts are given simultaneously and in randomized order. Among these randomized excerpts, some represent the actual speech samples under test, whereas the remaining excerpts are a hidden reference, a low-quality anchor, and a mid-quality anchor. During the quality test, the listeners were asked to rate the subjective speech quality (i.e., naturalness, fidelity) against the reference, as a number between 0 and 100 (Fig. \ref{test2_scale}). The value 100 denoted `Excellent' quality, meaning that the perceived quality of the test excerpt was identical to the reference. 

The MUSHRA tests were conducted in two rounds. The first round aimed at evaluating the quality of sentences that were consecutively encrypted by our algorithm, distorted by AWGN of varying intensity, and decrypted. In the second round, encrypted signals were compressed by Opus-Silk at varying compression rate. In addition, the test excerpts in the second round included speech utterances decrypted from the signal sent over FaceTime and recorded on iPhone 6. In both rounds, the participants had to rate 6 different sentences (3 female and 3 male) of about 10 seconds each. The reference was a wideband signal sampled at 16~kHz, the mid-anchor was a narrowband signal sampled at 8~kHz, and the low-anchor was a narrowband speech signal sampled at 8~kHz and with the MNRU distortion at SNR = 15 dB \cite{ITU-R1996P810}. In contrast to the reference signal, the mid-anchor may serve as a good benchmark to our tested signals due to the same speech bandwidth. The systems tested in the speech quality assessment are summarized in Table \ref{systems_under_test}.

\begin{table}[h!]
\centering
\caption{Hidden anchors and tested systems in the MUSHRA-based speech quality assessment.} 
\begin{tabular}{|c|c|}
\hline
\multicolumn{2}{|c|}{Reference and anchors} \\ \hline
Label  &   Description  \\ \hline 
ref       &   wideband speech sampled at 16 kHz  \\ 
mid      &   narrowband speech sampled at 8 kHz \\ 
low      &   8 kHz speech with MNRU at SNR = 15 dB \\ \hline \hline
\multicolumn{2}{|c|}{Systems under test in the assessment 1} \\ \hline
Label  &   Description  \\  \hline 
a-I  & speech decrypted from undistorted signal \\ 
b-I  & signal distorted by AWGN at SNR = 20 dB \\ 
c-I & signal distorted by AWGN at SNR = 15 dB \\ 
d-I  & signal distorted by  AWGN at SNR = 10 dB \\ \hline \hline
\multicolumn{2}{|c|}{Systems under test in the assessment 2} \\ \hline
Label  &   Description  \\ \hline 
a-II & signal compressed by Silk at 64 kbps  \\ 
b-II & signal compressed by Silk at 48 kbps  \\ 
c-II & signal sent over FaceTime between iPhones 6   \\ 
d-II & signal compressed by Silk at 32 kbps \\ \hline
\end{tabular}
\label{systems_under_test}
\end{table}

The assessment was carried out entirely online using webMUSHRA, the framework for Web-based listening tests \cite{schoeffler2015towards,schoeffler2018webmushra}. Table \ref{participants} lists the number of participants taking part in each test. All listeners were non-native English speakers with unreported hearing impairments. The participants were asked to wear headphones or earphones and were allowed to adjust the sound volume. Few participants were excluded from aggregated responses because of rating the hidden reference below 90 more than once in a single test (mostly accidentally).

Table \ref{intelligibility_results} presents sample mean and sample standard deviation of the intelligibility test. On average, the participants recognized about 12\% fewer words in synthesized speech samples than in the reference. The average rating of particular sentences varied slightly from 82\% to 89\%. On the other hand, a speaker-level average ranged from 58\% to 99\%. This high variability of average ratings explains a considerable standard deviation of aggregated responses. 

\begin{table}[h!]
\centering
\caption{Number of participants in the listening test.\newline}
\begin{tabular}{|p{4.5cm}|p{3cm}|}
\hline
\hspace{1.8cm} Test & \hspace{6mm} Participants \\ \hline \hline
\hspace{0.85cm} Intelligibility test & \hspace{1.2cm} $44\phantom{^{**}}$ \\
\hspace{1.2cm} Quality test 1 & \hspace{1.2cm} $40^*\phantom{^{*}}$ \\
\hspace{1.2cm} Quality test 2 & \hspace{1.2cm} $37^{**}$ \\ \hline 
\multicolumn{2}{|l|}{$^*\phantom{^{*}}$ 18 listeners rated 5 utterances instead of 6} \\
\multicolumn{2}{|l|}{$\phantom{^{**}}$ 4 listeners excluded for reference underrating} \\
\multicolumn{2}{|l|}{$^{**}$ 3 listeners excluded for reference underrating}\\ \hline
\end{tabular}
\label{participants}
\end{table}

\begin{table}[h!]
\centering
\caption{Intelligibility test results.}
\begin{tabular}{|c|c|c|}
\hline
System & Sample mean & Sample standard deviation \\ \hline \hline
Reference & 97.5 & 6.6 \\
Decrypted & 86.0 & 14.7 \\ \hline
\end{tabular}
\label{intelligibility_results}
\end{table}

\begin{figure}[h!]
\centering
\includegraphics{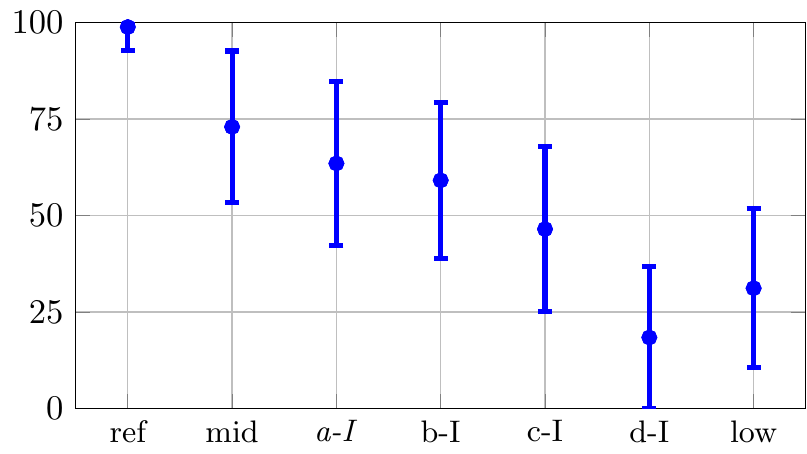}
  \caption{Results of the MUSHRA-based subjective quality assessment of speech decrypted from signals with added Gaussian noise of different intensity. Bars mark standard deviation.}
  \label{quality_test1}
\end{figure}

\begin{figure}[h!]
\centering
\includegraphics{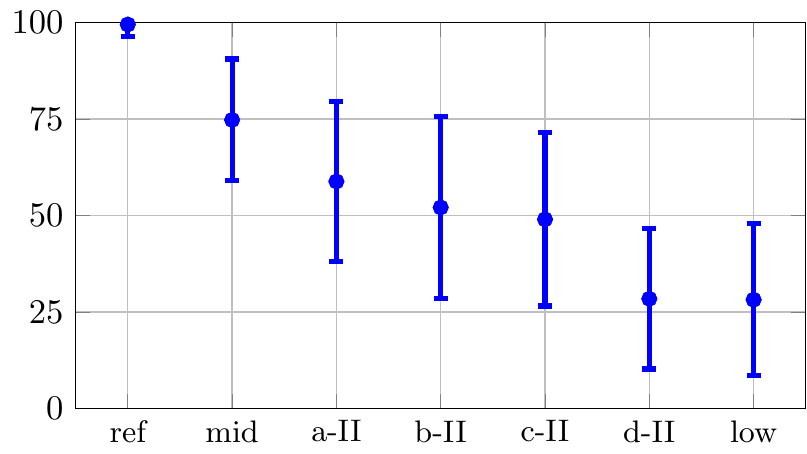}
  \caption{Results of the MUSHRA-based subjective quality assessment of speech decrypted from signals compressed by Opus-Silk vocoder at different compression rates and from signals sent over FaceTime between two iPhones 6. Bars mark standard deviation.}
  \label{quality_test2}
\end{figure}

The results of the MUSHRA-based quality assessment are depicted in Figures \ref{quality_test1} and \ref{quality_test2}. In both test rounds, the hidden reference was rated correctly as `Excellent.' The average rating of the mid-anchors given by the participants was about 75\% (`Good'), and the average rating of the low-anchors was about 30\% (`Poor'). 

The average rating of test excerpts decrypted from undistorted signals (a-I) was 64\% (`Good'/`Fair'). Compared with the average rating of mid-anchors, our algorithm reduced the speech quality by about 10\%. It may be noticed that this difference in speech quality between the mid-anchors and the excerpts labeled a-I is similar to the intelligibility loss in the SIR-based intelligibility assessment. 

The introduction of distortion into encrypted signals resulted in degraded speech quality. Gaussian noise at SNR equal to 20 dB, 15 dB, and 10 dB lowered the average ratings of speech quality respectively to 59\% (`Fair'), 46\% (`Fair'), and 19\% (`Poor'/`Bad'). It can be noticed that a small channel distortion, like the one introduced by AWGN at SNR = 20 dB, has a relatively minor impact on perceived speech quality. On the contrary, the quality becomes bad when SNR reaches 10~dB. A similar observation can be made in the case of signal compression by Opus-Silk. The compression of encrypted signals at 64 kbps, 48~kbps, and 32 kbps reduces the rated speech quality respectively to 59\% (`Fair'), 52\% (`Fair'), and 28\% (`Poor'). The excerpts decrypted from signals sent over FaceTime were rated at 49\% (`Fair').

The statistical similarity of given ratings was evaluated by the non-parametric Kruskall-Wallis test \cite{william1952ranks}, which is more suitable for ordinal scales \cite{mendona2018statistical}. The ratings of speech signals labeled a-I come from the same statistical distribution as speech signals labeled b-I with the 0.09 confidence. Additionally, the ratings of speech labeled b-II are similar to speech labeled c-I with 0.25 confidence, and the ratings of speech labeled d-II come from the same distribution as the low-anchor with 0.59 confidence. The ratings of the remaining systems were similar, with the confidence much lower than 0.05.

The obtained results suggest that the speech encryption scheme described in this study can produce intelligible speech. Moreover, the average speech quality of excerpts restored from signals sent over FaceTime hints about the possibility of making our system compatible with VoIP. However, high variability in listeners' responses indicates that the quality of decrypted speech is insufficient for having a casual conversation. Thus, some progress has to be made to improve the system's robustness to distortion and the quality of speech synthesis. 

\subsection{Algorithmic latency and computational complexity}
\label{section_complexity}

The minimum algorithmic latency in our encryption scheme is the sum of delays introduced respectively by the enciphering and deciphering algorithms. The speech encoder introduces 30 ms of delay (20 ms frame and 10 ms look-ahead), and the pseudo-speech analyzer introduces an additional 20 ms delay. Finally, two 1x3 convolutional layers in the speech synthesizer use a 40 ms look-ahead (2 frames). The combined 90~ms of the minimum algorithmic latency is significant and may reduce the perceived quality of conversation. A possible solution is to reduce the analysis look-ahead to 5 ms, and the synthesis look-ahead to 20 ms, like in the wideband LPCNet \cite{Valin2019Vocoder}. 

The speech encoder implemented in the encryption scheme is reported to have a complexity of 16 MFLOPS, where about 8 MFLOPS are used for pitch prediction \cite{Valin2019Vocoder}. Moreover, the authors hint at the possibility of significant optimizations. These given values relate to the scenario when a speech signal is sampled at 16 kHz. Thus, we roughly estimate our 8~kHz speech encoder's complexity to about 8 MFLOPS, including spherical mapping transformations.  

Enciphering (and deciphering) is relatively lightweight, as it requires only ten additions modulo per 20 ms frame. However, a higher computational load is associated with producing secure bitstrings by the pseudo-random generator at a rate 8 kbps. For this reason, it is especially important to select a PRNG based on well-established ciphers adapted for real-time applications, such as AES-CTR \cite{kasper2009faster, park2018face}.

Pseudo-speech synthesis consists of two steps: computing the harmonic parameters of a frame and producing the signal samples. The complexity of the first step is dominated by deriving the complex amplitudes of harmonics $\mathbf{\check{A}}$ of length $K$ using Eq.~(\ref{matrix_penrose}), where $K$ is the number of harmonics in a particular frame. Provided that all complex matrices $(\mathrm{H}\mathrm{\tilde{B}}_{\omega_0})^{\dagger}$ are precomputed and stored in the memory, the vector $\mathbf{\check{A}}$ can be obtained by searching the appropriate matrix from the look-up table, by element-wise complex vector multiplication, and finally by one complex matrix $16\times K$ multiplication. On the other hand, frame synthesis requires $\mathcal{O}(400K)$  floating-point operations, where 400 is the number of samples within a frame with two guard periods.

The pseudo-speech analyzer is mostly occupied by estimating the received fundamental frequency. The maximum-likelihood estimator implemented in the scheme has complexity $\mathcal{O}(N\log N)+\mathcal{O}(NK)$, where $N=2^{16}$ is the number of possible pitch values \cite{NIELSEN2017188}. Consequently, lowering the resolution of estimations or replacing the pitch predictor with a more efficient version will give considerable computational gains. 

The most computationally involving element in the encryption scheme is the final speech reconstruction. The LPCNet model implemented in the scheme has complexity:
\begin{equation}
C = (3dN_{A}^2+3N_B(N_A+N_B)+2N_BQ)\cdot 2f_s,
\end{equation}
where $N_A = 384$, $N_B = 16$, $d = 10 \% $, $Q = 256$ is the number of $\mu$-law levels, and $f_s$ is the sampling frequency. For $f_s = 16$ kHz, the estimated complexity of the synthesizer is 3 GFLOPS \cite{Valin2019Vocoder}. Additionally, it is reported that a C implementation of the synthesizer requires 20$\%$ computing power of a 2.4 GHz Intel Broadwell core, 68$\%$ of a one 2.5 GHz Snapdragon 845 core (Google Pixel 3), and 31$\%$ of a one 2.84 GHz Snapdragon 855 core. From this, we estimate that the complexity of the lightweight, narrowband implementation of LPCNet is about 2~GFLOPS, and it could operate in real-time on portable devices.

Table \ref{complexity_measurements} lists the computational complexity of various parts of our algorithm estimated using PyPaPi\footnote{\url{https://flozz.github.io/pypapi/}} library \cite{terpstra2010collecting} when processing 60 minutes of a recorded speech in Python. The measurements were done under Ubuntu kernel 5.8.0-25 and using Intel Core i7 2.9 GHz without multi-threading. The pseudo-random bitstring used for enciphering and deciphering was precomputed and stored in the memory. 

The listed results suggest that every tested part has a complexity low enough to be carried by a portable device, especially if one considers migrating the experimental Python code to a compiled code. Moreover, a replacement of the pitch predictor in the pseudo-speech analyzer would lead to significant optimization gains. On the other hand, the computational analysis does not include other essential elements of the system, such as keeping signal synchronization or adaptive energy equalization.

\begin{table}[h!]
\centering
\caption{Estimated complexity using PyPaPi library.}
\begin{tabular}{|p{5cm}|r|}
\hline
\multicolumn{1}{|c|}{Process} & \multicolumn{1}{c|}{MFLOPS} \\ \hline \hline
speech encoding         & 8 \cite{Valin2019Vocoder} \\ 
enciphering             & 2 $\phantom{^{****.}}$  \\
pseudo-speech synthesis & 1032 $\phantom{^{****.}}$   \\
pseudo-speech analysis  & 2756 $\phantom{^{****.}}$   \\
$\qquad \bullet$ pitch prediction  & 2123 $\phantom{^{****.}}$   \\
$\qquad \bullet$ remaining  & 634 $\phantom{^{****.}}$   \\
deciphering             & 2 $\phantom{^{****.}}$    \\
speech synthesis        & 2000 \cite{Valin2019Vocoder}  \\ \hline
\end{tabular}
\label{complexity_measurements}
\end{table}

\section{Conclusions}
\label{conclusions}

In this work, we proposed a new speech encryption scheme for secure voice communications over voice channels. The lossy speech encoding technique implemented in the system preserves and protects only basic vocal parameters: fundamental frequency (pitch), energy (loudness), and spectral envelope (timbre). The vocal parameters are enciphered and then encoded to a synthetic audio signal adapted for transmission over wideband voice channels. Speech is reconstructed by the narrowband vocoder based on the LPCNet architecture.

Enciphering of vocal parameters is done using norm-preserving techniques: pitch and fundamental frequency are enciphered by translations, whereas spectral envelope by rotations on the hypersphere in 16 dimensions. These techniques enable successful decryption of signals distorted by moderate transmission noise, like AWGN, or processed by some wideband VoIP applications such as FaceTime. However, the enciphering mechanism does not provide any data integrity. Instead, it is crucial to ensure strong identity authentication in the initial cryptographic key exchange.

The robustness of the speech encryption scheme a-gainst channel noise was verified experimentally. Simulations showed that the system could correctly decrypt pseudo-speech with additive Gaussian noise at SNR = 15 dB or compressed by the Opus-Silk codec at the 48 kbps rate. On the other hand, an encrypted signal is sensitive to synchronization error larger than 0.3 milliseconds. Furthermore, the results of the speech quality assessment indicated that the proposed encryption scheme could produce intelligible speech with the quality depending on channel distortion.

The preliminary complexity evaluation and the successful transmission of encrypted signals between two mobile phones hint that the proposed encryption scheme may work in real-time on high-end portable devices. However, secure communication is susceptible to short signal dropouts or de-synchronization. Consequently, robust communication is possible only over a stable vocal link between the users. Additionally, adaptive voice-enhancing algorithms implemented in commercial mobile phones (such as voice detection and noise suppression) usually lead to considerable degradation of the speech quality. This problem can be tackled using dedicated CryptoPhones or stand-alone devices connected with mobile phones in tandem. 

Speech quality could be improved by replacing our narrowband speech synthesizer exploiting the 4 kHz bandwidth with a wideband synthesizer with the 8~kHz signal bandwidth. The biggest challenge is to find a new representation for the spectral envelope that is compatible with the enciphering technique. The presented algorithm uses 9 mel-scaled frequency windows that are insufficient for encoding the wideband spectrum. A possible solution is to increase the number of mel-scaled windows to 18 and apply a dimensionality reduction technique, such as Principal Component Analysis (PCA) \cite{PCA_article} or autoencoding \cite{kramer1991nonlinear}. Dimensionality reduction may increase encoding efficiency because the coefficients within a single speech frame tend to be highly correlated.

Other improvements can be obtained in the pseudo-speech synthesis. The proposed synthesis technique, while computationally efficient, is very phase-sensitive and not enough speech-like. Instead of encoding the enciphered vector $\mathbf{\tilde{D}}_{(enc)}$ into the real part of the complex frame spectrum, it would be advantageous to encode $\mathbf{\tilde{D}}_{(enc)}$ into the power spectral density (PSD). The main limitation is that the vector $\mathbf{\tilde{D}}_{(enc)}$ contains both positive and negative values, whereas PSD is always non-negative. For this reason, envelope encoding could be performed in the cepstral domain.

Furthermore, it may be worth adding a correction unit at the deciphering output for detecting and smoothing deciphering errors. Since the vocal parameters in natural speech do not change quickly over time, the detection of large errors should be relatively straightforward. For example, the correction unit could use machine learning techniques to correct errors on a particular channel. A clear separation between the correction unit and the speech synthesizer could improve the quality of synthesized speech and simplify the two-step network training. 

Communication performance strongly depends on the stability of a vocal link. The problem with fading channels could be mitigated by combining distortion-tolerant speech encryption and multiple description coding (MDC) \cite{goyal2001multiple,Venkataramani2003Multiple, Wah2005LSP}. Multiple description coding is a technique that fragments one media stream into several substreams. Each substream is decodable into the initial stream, and decoding more substreams improves the quality. The MDC could be used to split encrypted speech into multiple audio streams and thus to increase communication reliability. 

\section*{Acknowledgements}

This document is the results of the research project funded by the AID program No SED0456JE75.

\normalsize
\bibliography{bibliography}

\end{document}